%% file: neurips_2026.tex
\documentclass{article}

\usepackage[preprint]{neurips_2026}

\usepackage[utf8]{inputenc} % allow utf-8 input
\usepackage[T1]{fontenc}    % use 8-bit T1 fonts
\usepackage{hyperref}       % hyperlinks
\usepackage{url}            % simple URL typesetting
\usepackage{booktabs}       % professional-quality tables
\usepackage{amsfonts}       % blackboard math symbols
\usepackage{nicefrac}       % compact symbols for 1/2, etc.
\usepackage{microtype}      % microtypography
\usepackage[table]{xcolor}  % colors

\usepackage{graphicx}
\usepackage{multirow}
\usepackage{wrapfig}
\usepackage{amsmath}
\usepackage{caption}
\usepackage{subcaption}
\usepackage{enumitem}
\usepackage{siunitx}
\usepackage{algorithm,algpseudocode}
\usepackage{textcomp}
\usepackage{longtable}
\usepackage{listings}
\DeclareUnicodeCharacter{2014}{\textemdash}

\def\ModelName{MACU}

\definecolor{improvegreen}{RGB}{220,245,220}
\definecolor{worsenred}{RGB}{255,225,225}
\definecolor{grey}{RGB}{230,230,230}

\title{Multi-Agent Computer Use}

\author{%
  Jing Yu Koh \\
  Carnegie Mellon University\\
  \texttt{jingyuk@cs.cmu.edu} \\
  \And
  Ruslan Salakhutdinov \\
  Carnegie Mellon University\\
  \texttt{rsalakhu@cs.cmu.edu} \\
  \And
  Daniel Fried \\
  Carnegie Mellon University\\
  \texttt{dfried@cs.cmu.edu} \\
}

\begin{document}

\maketitle

\begin{abstract}
Computer use agents (CUAs) today are primarily deployed as single serial agents. This setup is suboptimal for complex long-horizon tasks that benefit from task decomposition, parallel execution, and consistent re-planning based on new information.
In this paper, we argue that we should instead move towards evaluating and building \emph{multi-agent computer use} (\ModelName{}) systems. These systems, which emphasize planning and parallel execution, alleviate many of the shortcomings of single-agent CUAs. 
We propose a general multi-agent setup in which a manager model decomposes computer use tasks as a directed acyclic graph (DAG) of subtasks, encoding relevant dependencies and goals for subagents.
At each iteration, the manager dispatches parallel CUA subagents to carry out nodes on the ready frontier of the DAG, and continuously revises the DAG (adding, canceling, or rewriting nodes) as new findings arrive from subagents.
This design treats the partially observable environment of computer use as a first class challenge: information that downstream agents may not be able to re-observe are retained and passed forward through the manager and DAG structure. 
We demonstrate that \ModelName{} consistently improves over strong single-agent baselines by $3.4-25.5\%$ on desktop (OSWorld) and web navigation (Online-Mind2Web, WebTailBench, Odysseys) benchmarks, exhibits more favorable test-time compute scaling, and solves complex long-horizon tasks where single-agent CUAs get stuck. On Odysseys, a long-horizon web navigation benchmark, \ModelName{} also improves average task completion wall-clock time by ${\sim} 1.5 \times$, demonstrating its efficacy in speeding up traditionally slow CUA pipelines.
Our findings highlight that multi-agent coordination is a promising axis for scaling computer use agents to work productively for longer and more effectively. 
We release all code and interactive visualizations at \url{https://jykoh.com/multi-agent-computer-use}.
\end{abstract}

\section{Introduction}
\begin{figure}[t]
  \centering
  \includegraphics[width=\textwidth]{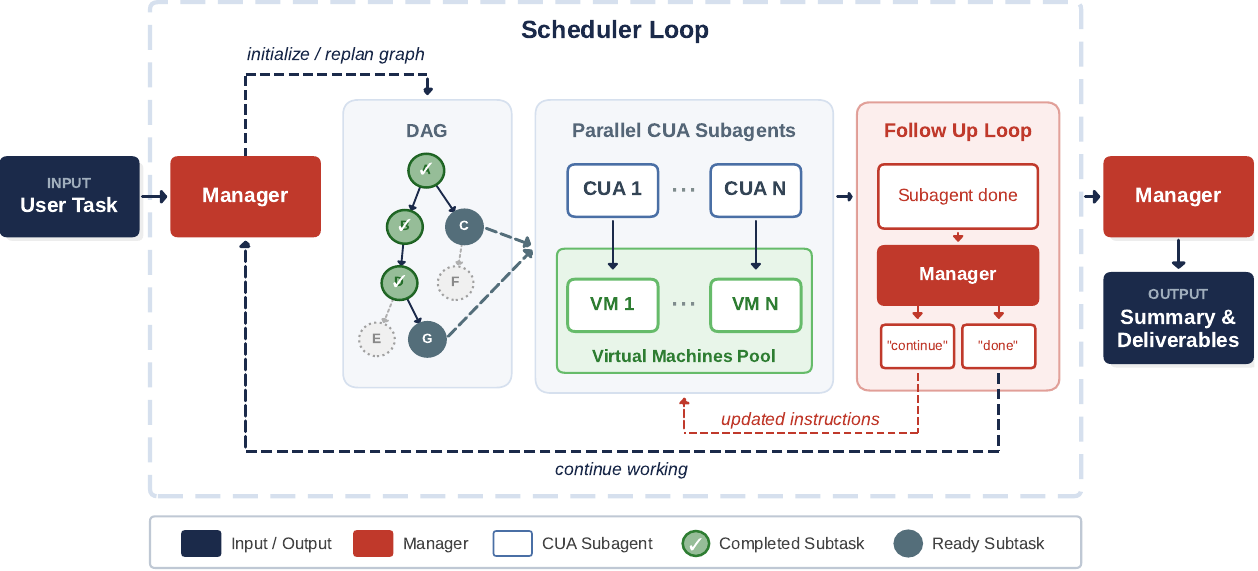}
  \caption{Overview of the setup for multi-agent computer use (\ModelName{}). At each iteration, ready subtasks (grey) are launched and handled in parallel by CUA subagents. The manager replans and updates the DAG based on information it receives from completed subagents.} \label{fig:architecture}
\end{figure}

Computer use is the task of interacting with software through the use of Graphical User Interfaces (GUIs).
Computer use agents (CUAs) today are primarily deployed and evaluated as single serial agents, but this is suboptimal for tackling complex long-horizon tasks, which benefit from decomposition, parallel execution, and consistent re-planning based on new information. 
In this paper, we argue that we should move towards \emph{multi-agent computer use} (\ModelName{}): systems which emphasize planning and parallel execution, and are promising ways to alleviate the shortcomings of single-agent CUAs. 

Towards this goal, we present a general \ModelName{} setup (Fig.~\ref{fig:architecture}), where a manager model plans, decomposes, and continuously iterates over a directed acyclic graph (DAG) whose nodes are subtasks that are executed in parallel by computer use subagents. 
The graph decomposition schema used in \ModelName{} is open-ended and the manager operates on a \emph{graph} of subagents, with the ability to add, cancel, rewire, or modify the instructions of pending subtasks. 
This directly targets long-horizon tasks and complex but parallelizable CUA tasks, key regimes where serial CUA setups struggle. 
Our work distinguishes itself from prior work in several ways.
Most prior multi-agent work based on large language models (LLMs) explores text-based specialized subagents (e.g., tool use~\citep{chen2023agentverse,li2025flow}, coding agents~\citep{hong2023metagpt,geng2026effective}, math~\citep{qian2024scaling,zhao2025majority,qi2025learning}, and science~\citep{lu2024ai,gottweis2025towards}) in effectively fully observable settings.
%: each tool call returns the state relevant to the next step, and replanning operates on a single-agent, or simply once at the beginning of the task. 
Computer use is a qualitatively different regime. The agent acts on a live operating system, whose state: filesystem, background processes, open browser tabs, login cookies, dynamic content, is only \emph{partially observable} at any given moment, and the state relevant to a downstream subtask often cannot be re-observed once the original subtask has ended. We treat partial observability as a first-class consideration in \ModelName{}, which leads us to be explicit about the initial plan being a best-guess hypothesis, which the manager needs to continuously revise once new information is collected from pending subagents. %Each completed subagent's findings are pushed forward into its pending children's instructions through the revised DAG.
% We also demonstrate empirically that \ModelName{} is an elegant and scalable method. 
In contrast to many prior approaches on orchestrating multiple subagents~\citep{agashe2025agent,lee2026infoseeker,zhang2025agentorchestra}, our proposed \ModelName{} approach is also substantially simpler and does not rely on specialized models. All subagents are identical, based off the same CUA backbone, and carry out instructions that are produced by the manager, allowing our approach to directly benefit from future advances on CUA models by directly leveraging them as subagents in \ModelName{}.
% Our simplified setup enables us to scale with less overhead and domain and benchmark -specific concerns. 

We demonstrate empirically that \ModelName{} is a promising approach to tackling computer use tasks. \ModelName{} consistently improves over strong single-agent baselines across several computer use and web navigation benchmarks, increasing success rates by 4.7\% on OSWorld~\citep{xie2024osworld}, 3.4\% on Online-Mind2Web~\citep{xue2025illusion}, 8.7\% on WebTailBench-v2~\citep{awadallah2025fara}, and 25.5\% on Odysseys~\citep{jang2026odysseys}, and also substantially improves wall-clock time on OSWorld and Odysseys.
\ModelName{} also demonstrates more promising scaling trends with increased test-time compute, solving long-horizon tasks that single-agent CUAs get stuck on. 
We perform detailed ablations on the \ModelName{} setup, and our analysis highlights design choices that are essential to maximize the effectiveness of \ModelName{}. We believe these findings provide a strong foundation for further work on effective scaling of multi-agent computer use systems.

\section{Related Work} \label{sec:related}

\paragraph{Computer Use Agents}
% Computer use agents (CUAs) are multimodal language model agents that interact with real software through the use of Graphical User Interfaces (GUIs), typically through a ReAct~\citep{yao2022react} loop where the model takes as input the screenshot of the current computer screen, and outputs reasoning traces and an action to be executed. 
Several benchmarks have been established to measure the capabilities of frontier models on computer use tasks. 
OSWorld~\citep{xie2024osworld} established an execution-based grading of 369 open ended Ubuntu tasks spanning various native apps and multi-application workflows. 
WindowsAgentArena~\citep{bonatti2024windows} adapts the OSWorld framework to 150+ tasks on a real Windows operating system. 
More recently, Gym-Anything~\citep{aggarwal2026gym} leveraged coding and audit agents to automatically install and configure real software. 
% They release the CUA-World benchmark, 10K+ long-horizon tasks generated across 200 applications, as well as checklists for grading results. 
Other benchmarks have focused on web navigation capabilities: WebArena~\citep{zhou2023webarena} provides a simulated environment of realistic self-hosted sites (Shopping, Reddit, GitLab, etc.) with execution-based grading, and VisualWebArena~\citep{koh2024visualwebarena} extends it with visually grounded multimodal tasks. 
WebVoyager~\citep{he2024webvoyager} evaluates CUAs directly on 15 real world websites, and WebLINX~\citep{lu2024weblinx} evaluates multi-turn dialogue for web navigation. Online-Mind2Web~\citep{xue2025illusion} evaluates agents with a more robust judge, across 300 tasks on 136 live websites, and WebTailBench~\citep{awadallah2025fara} proposes 609 tasks that include multi-step and cross-site tasks, drawn from high-traffic webpages.
% , introducing an improved LLM-as-a-judge mechanism for scalable grading. 
Odysseys~\citep{jang2026odysseys} recently proposed 200 long-horizon web navigation tasks based on real human browsing patterns. 
On the capabilities front, frontier models have quickly improved on computer use, saturating many of these benchmarks. GPT-5.4~\citep{gpt54thinkingsystemcard} achieves 75.0\% on OSWorld, Opus 4.6~\citep{claudeopus46systemcard} achieves 72.7\%, and Kimi K2.6~\citep{team2025kimi} achieves 73.1\%, which surpass human performance (72.4\%). Other smaller open weight models such as MolmoWeb~\citep{gupta2026molmoweb}, Qwen3.5~\citep{qwenteam2026qwen35omnitechnicalreport}, also achieve strong performance on these benchmarks, while being small enough (e.g., 4--9B parameter scale) to run locally. 
Most of these CUAs are evaluated and deployed as single serial actors. In this paper, we explore how we might instead coordinate and orchestrate CUAs in a multi-agent system, unlocking more effective planning and backtracking, higher parallelism, and improved test-time scaling behavior.

\paragraph{Multi-Agent LLM Systems}
Multi-agent systems are composed of multiple interacting autonomous agents~\citep{wooldridge2009introduction}, with a long history dating back as early as the 1980s~\citep{durfee1989trends,smith1979framework}. 
More recently, multi-agent systems based on LLMs have been explored across a wide range of settings, including software engineering (SWE)~\citep{hong2023metagpt,geng2026effective}, scientific discovery~\citep{lu2024ai,gottweis2025towards}, web search~\citep{lee2026infoseeker}, and more.
% CAMEL~\citep{li2023camel} introduces inception prompting, a role-playing framework in which two chat agents cooperate to solve tasks. 
% \cite{hong2023metagpt,geng2026effective} propose structured multi-agent coordination paradigms to improve performance on software engineering (SWE) tasks
% an assembly line framework to encode standardized operating procedures (SOPs) into prompt sequences and assigns roles to specialized agents, improving performance on software engineering (SWE) tasks. \cite{geng2026effective} introduce a structured multi-agent coordination paradigm that uses existing SWE primitives to improve performance on SWE benchmarks.
% Generative Agents~\citep{generativeagents} populate a sandbox with 25 agents that plan, reflect, and interact via a long-term memory architecture, yielding emergent social behavior. 
% chains idea generation, experiment execution, and paper writing into an end-to-end pipeline for scientific discovery. The AI co-scientist~\citep{gottweis2025towards} uses a multi-agent system, with a generate/debate/evolve loop, asynchronous task execution, and a tournament evolution process for biomedical hypothesis generation. 
Many of these works frame the system as a computational graph~\citep{zhuge2024language,qian2024scaling} that can be dynamically adjusted~\citep{chen2023agentverse} and scaled, or as systems that can recursively spawn subagents~\citep{zhang2025recursive,gandhi2026recursive}.
% GPTSwarm~\citep{zhuge2024language} frames LLM agents as optimizable computational graphs whose nodes (LLM queries, tools) and edges (information flow) can be automatically tuned, and MacNet~\citep{qian2024scaling} scale multi-agent collaboration in a directed-acyclic-graph topology, demonstrating performance improvements as the number of agents are scaled. AgentVerse~\citep{chen2023agentverse} allows dynamic adjustment of agents based on current progress, modifying its composition across expert-recruiting, collaborative decision-making, action, and evaluation stages. 
% GPTSwarm~\citep{zhuge2024language} frames LLM agents as optimizable computational graphs whose nodes (LLM queries, tools) and edges (information flow) can be automatically tuned, and MacNet~\citep{qian2024scaling} scale multi-agent collaboration in a directed-acyclic-graph topology, demonstrating performance improvements as the number of agents are scaled. AgentVerse~\citep{chen2023agentverse} allows dynamic adjustment of agents based on current progress, modifying its composition across expert-recruiting, collaborative decision-making, action, and evaluation stages. 
Others target decomposition and scheduling in a more structured form, such as using a planner module that delegates tasks to specialized modules~\citep{li2025flow,zhang2025agentorchestra,nielsen2025learning,xu2025trinity}.
% AgentFlow~\citep{li2025flow} proposes a trainable framework that coordinates four modules and trains the planner module via on-policy reinforcement learning. 
% AgentOrchestra~\citep{zhang2025agentorchestra} has a central planner agent which delegates tasks to specialized sub-agents. 
% InfoSeeker~\citep{lee2026infoseeker} uses a three-tier host/manager/worker hierarchy and reports a 5.7$\times$ end-to-end speedup for the web information seeking tasks in BrowseComp-zh~\citep{wei2025browsecomp,zhou2025browsecomp} by using specialized workers and managers. 
% \cite{xu2025trinity} orchestrates multiple LLMs, assigning one of three roles (thinker, worker, or verifier), and trains the coordinator with an evolutionary strategy. 
% \cite{nielsen2025learning} trains a coordinator with reinforcement learning to learn targeted communication and prompting strategies for coordinating LLMs.
Specifically, in the computer use agent literature, a parallel thread focuses on \emph{dynamic} replanning as the agent navigates. 
\citep{koh2024tree} proposes a tree search algorithm for CUAs which performs lookahead and backtracking using a value function to score each proposed action.
\cite{wang2025inducing} introduces a framework for inducing, verifying, and using program-based skills on the fly, as an agent continuously interacts with a web environment. 
Agent Workflow Memory~\citep{wang2024agent} learns reusable workflows and stores these for subsequent generations. 
\cite{aghzal2026llm} find in a controlled hierarchical-planning study that a single round of exploratory replanning substantially improves success on web navigation tasks. 
Others explore orchestrating specialized submodules: Agent~S2~\citep{agashe2025agent} introduces proactive hierarchical planning, refining plans after each subgoal success, and delegating tasks to workers and specialized expert modules, and CoAct-1~\citep{song2025coact} and InfantAgent-Next~\citep{lei2025infantagent} delegate tasks to either a GUI operator or a code agent. 
Several systems have also been deployed successfully in user-facing products for web navigation and information gathering~\citep{yutori2025scouts,anthropic2025multiagentresearch}.

In this paper we investigate how to best develop a multi-agent framework for general computer use: a partially observable problem that involves complex challenges such as grounding, computer interaction, and long-horizon planning. 
Our framework is general, simple and scalable. We only use~2 types of agents: a planner, and a homogenous set of subagent workers, while prior work typically uses specialized modules for handling low-level actions~\citep{agashe2025agent,lei2025infantagent,song2025coact}, or focus on fully observable environments such as code generation~\citep{hong2023metagpt,geng2026effective} and math~\citep{motwani2024malt,zhao2025majority,madaan2025rethinking}.
Our work aims to develop the principles for building a strong, simple, and generalizable framework for multi-agent computer use.

% \paragraph{Inference-time scaling for agents.}
% A growing body of work scales agents at inference time rather than at training time. Agent~S3's Behavior Best-of-$N$ (bBoN)~\citep{agents3} samples multiple whole-task trajectories and selects among them using compact behavior narratives, reaching 72.6\% on OSWorld; related work in deep research scales similarly by sampling and aggregating parallel rollouts~\citep{parallelmuse,widerdeep}. A complementary line applies tree search: Language Agent Tree Search (LATS)~\citep{lats} unifies reasoning, acting, and planning via MCTS-style rollouts guided by self-reflection; \citet{treesearchlmagents} extend tree search to realistic web agents with substantial gains over single-shot rollouts. Agent~Alpha~\citep{agentalpha} introduces Alpha-UCT and action/environment parallelism for MCTS over CUA tasks, WebOperator~\citep{weboperator} develops an action-aware tree search that dynamically adapts the action space based on the current observation, and \citet{treesearchrl} use tree search as an RL signal. Separately, BATS~\citep{bats} adapts stepwise effort to a continuous remaining-budget signal, and OSWorld-Human~\citep{osworldhuman} argues that step-efficiency, not success rate alone, is the right axis for CUA evaluation.

\section{Multi-Agent Computer Use} \label{sec:method}
% \begin{figure}[t]
%   \centering
%   \includegraphics[width=\textwidth]{figures/example_4c26e3f3.pdf}
%   \caption{Example \ModelName{} run for an OSWorld task on enhancing image brightness. Initially, the manager spawns a subagent to attempt the task through the GUI, which takes too long and fails. It spawns another subagent that attempts to use the CLI, which also fails. Finally it launches two parallel subagents: (1) another CLI attempt and (2) one that resumes the original GUI attempt but with slightly modified instructions. The second GUI attempt successfully exports the image, rescuing a task which would have initially failed. Additional qualitative examples are provided in Appendix.~\ref{sec:additional-qualitative}.}
%   \label{fig:graph-example}
% \end{figure}

\begin{figure}[t]
  \centering
  \includegraphics[width=\textwidth]{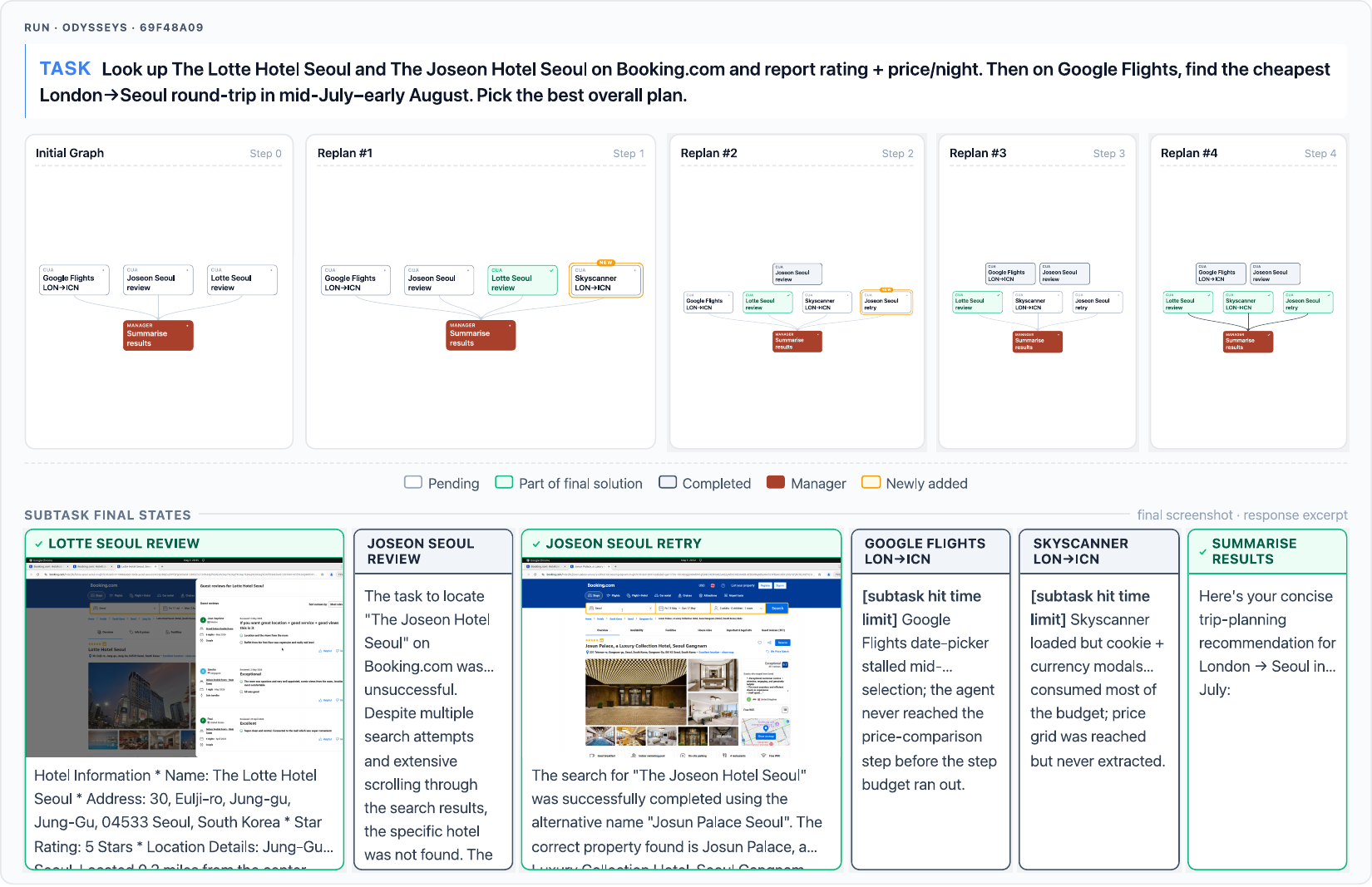}
  \caption{Example Odysseys travel planning task with \ModelName{}. The manager launches parallel hotel review and flight search subtasks, then retries blocked branches with alternate hotel and flight search routes. The final response combines the successful hotel research with the available flight-search evidence to recommend an accepted London--Seoul plan.}
  \label{fig:graph-example}
  \vspace{-1ex}
\end{figure}

\begin{algorithm}[t]
\caption{Algorithm for a \ModelName{} run on a single task.} \label{alg:macu}
\begin{algorithmic}[1]
\State \textbf{Input:} task $T$, max parallel subagents $N$, replan budget $B$
\State $G \gets \textsc{Decompose}(T)$ \Comment{manager call for initial task decomposition}
\State $b \gets 0$; $\texttt{files} \gets \emptyset$; $\texttt{running} \gets \emptyset$; $\texttt{completed} \gets \emptyset$; $\texttt{processed} \gets \emptyset$
\While{$G$ has uncompleted nodes}
  \State $R \gets \{v \in G : \text{deps}(v) \subseteq \texttt{completed},\, v \notin \texttt{running}\}$
  \State Launch $N - |\texttt{running}|$ CUA subagents from $R$  %\Comment{Spawn CUA subagents}
  \State Wait until a subagent completes or for a followup hook
  \For{$\{v \in G: v \subseteq \texttt{completed}, v \notin \texttt{processed}\}$}
    \State $\texttt{results}[v] \gets$ $v.\texttt{response}$; 
    \State $\texttt{files} \gets \texttt{files} \cup \textsc{FileMgmt}(v)$ \Comment{manager copies relevant files from subagents}
    \If{$b < B$}
      \State $G \gets$ \textsc{Replan}($G$); 
      \State $b \gets b + 1$
    \EndIf
    \State $\texttt{processed} \gets \texttt{processed} \cup v$
  \EndFor
  \If{$|\texttt{running}| < N$}
    \State $G \gets \textsc{Replan}(G, |\texttt{running}|)$ \Comment{optionally use available parallelism}
  \EndIf
  \For{$\{v \in R: v = \texttt{final\_aggregation}\}$}
  \State \Return \textsc{Aggregate}($G$, results, candidate VMs) \Comment{final manager summary}
  \EndFor
\EndWhile
\end{algorithmic}
\end{algorithm}

The overall flow of \ModelName{} is shown in Fig.~\ref{fig:architecture} and Alg.~\ref{alg:macu}. 
\ModelName{} involves a single \emph{manager} agent that coordinates a pool of \emph{CUA subagents} by mutating a directed acyclic graph (DAG) of subtasks to be completed. 
Conditioned on an instruction specified in natural language, the manager produces an initial DAG decomposition of the task (Fig.~\ref{fig:graph-example}). At each iteration, CUA subagents are dispatched in parallel to tackle subtasks on the ready frontier of the DAG (i.e., nodes whose dependencies are satisfied). 
Whenever new observations and evidence become available, the manager is prompted to revise the DAG. We describe the roles of the manager and subagents in detail below.

\paragraph{Manager} The manager, which can be a local model or an API based model, orchestrates and coordinates CUA subagents that operate in isolated virtual machines (VMs). The manager is implemented as an LLM call that takes in the task instruction, the current DAG, and the latest observations from the CUA subagent nodes. The manager is invoked at several key points:
\begin{enumerate}
    \item \textbf{Initial DAG decomposition:} The manager is prompted with the original task instruction and the screenshot of the starting VM state (if any), and is instructed to produce a DAG $G$ represented as a JSON string (which is validated for correctness). % (and we retry up to 3 times if it is not valid JSON).
    \item \textbf{Replanning:} After a CUA subagent $v$ completes its assigned task, the manager is provided with the current DAG $G$, the last $k$ screenshots and final response from $v$, as well as context from the parent nodes of $v$ in the form of their final VM screenshot and response. The manager is prompted to respond with a possible list of edits to make to $G$: \texttt{add} a node, \texttt{remove} a node, \texttt{modify} the instructions or dependencies of a node, or \texttt{cancel} a currently running node. These actions can change the structure of the graph or modify the instructions to be provided to future subagent nodes. Multiple edits can be made in a single replan call, provided they still result in a valid DAG. Each edit consumes a unit of the replan budget $B$. Once $B$ is exhausted we do not allow any additional edits to $G$. We also perform replanning intermittently if we have additional worker capacity that is not being used, to allow the manager to use spare capacity productively (e.g., exploration to find additional information, or try out different possible solutions).
    \item \textbf{File management:} A key requirement of many CUA tasks is the ability to create, maintain, and produce files. On the completion of each CUA subagent, the manager is provided with a diff of the filesystem to identify added or modified files. It is prompted to determine which files to save. When starting up a new subtask, the manager can similarly provide previously retained files as input to the new subtasks (e.g., handing off a Word document from subtask 1 to be used by subtask 2 to populate slides in PowerPoint).
    \item \textbf{Follow up:} When each CUA subagent completes its unit of work, the manager inspects it by being prompted with its latest responses and screenshot observations. The manager can optionally task the CUA subagent to continue working with a modified instruction, or get the subagent to summarize its work and terminate.
    \item \textbf{Aggregate and summarize:} The final leaf node in the DAG prompts the manager to aggregate and summarize results from the observations and responses of its parent nodes. This produces a final response for the user. %Any nodes that are not ancestors of this leaf node are also discarded (as they are unnecessary for producing the final result).
\end{enumerate}

\paragraph{Subagents} Each subagent executes a standard ReAct~\citep{yao2022react} loop used by most frontier CUA models. At each iteration, the subagent observes a screenshot from its VM, produces reasoning traces and a computer action, which is then executed on the VM. \ModelName{} treats this subagent loop as a black box: each subagent runs on its own isolated VM.
This treatment enables the manager to be free of model-specific assumptions, and agnostic to the specific CUA used, enabling \ModelName{} to work with new releases of improved CUA models (see Sec.~\ref{sec:ablations} which show improved scaling trends with stronger subagents). 
We use the same CUA backbone for each subagent. Specialization only arises from manager-written instructions and the VMs that subagents initialize from or inherit. %In future work, it would be interesting to explore allowing the manager to dynamically choose the subagent model backbone at runtime to improve latency or leverage the unique individual capabilities of different models~\citep{nielsen2025learning}.

\section{Experiments} \label{sec:experiments}

We evaluate MACU on a diverse set of computer use benchmarks:
\begin{itemize}
    \item \textbf{OSWorld}~\citep{xie2024osworld} measures the ability of models to control GUIs to complete 369 desktop tasks on an Ubuntu machine. Success rate is measured by execution-based evaluation scripts that test whether a model satisfied certain requirements.
    \item \textbf{Online-Mind2Web}~\citep{xue2025illusion} is a set of 300 tasks testing web navigation on 136 different real world websites. Success rate is measured through calling an LLM-as-a-judge on the completed trajectories to judge if the CUA accomplished the task.
    \item \textbf{WebTailBench-v2}~\citep{awadallah2025fara} is a set of 609 tasks designed to test breadth and depth of navigation, and aims to capture under-represented tasks in previous benchmarks such as Online-Mind2Web. Success rate is measured with an LLM-as-a-judge over task-specific rubrics.
    \item \textbf{Odysseys}~\citep{jang2026odysseys} is a benchmark of 200 long-horizon web navigation tasks derived from real user browsing behavior. Each task comes with predefined rubrics that measure key task requirements. We report the success rate (percentage of tasks that satisfied all rubrics), as well as the average rubric score (percentage of rubrics per task that are satisfied).
\end{itemize}

For our main experiments, we use the Qwen3.6-27B~\citep{qwen36plus} model for the CUA subagent, and Claude Opus 4.6~\citep{claudeopus46systemcard} as the manager. We also ran experiments where Qwen3.6-27B is used as both the subagent and the manager, to highlight the effectiveness of \ModelName{} even with a weaker manager model. We limit the run to maintain at most $N=4$ subagent runs in parallel, and set replan budget $B=10$ (see ablations in Sec.~\ref{sec:ablations}). More detailed experimental settings and all prompts used are provided in Appendices~\ref{appendix:experimental-details} and \ref{appendix:manager-prompts}. 
Since the general \ModelName{} setup makes no assumptions about the subagent model, it can in principle work off-the-shelf with any newer and more capable models, and we expect \ModelName{} to continue to be complementary to stronger CUA models released in the future.

\begin{table}[t]
\centering
\captionof{table}{Single vs.\ multi-agent success rate (SR) and median wall-clock time. All models use Qwen3.6-27B as the subagent backbone. $\Delta$ indicates the change when going from single-agent to multi-agent ({\setlength{\fboxsep}{1pt}\colorbox{improvegreen}{improvement}} or {\setlength{\fboxsep}{1pt}\colorbox{worsenred}{deterioration}}).}
\vspace{1ex}
\label{tab:results}
% \resizebox{\textwidth}{!}{%
\small
\begin{tabular}{lllcccc}
\toprule
\multirow{2}{*}[-2pt]{\textbf{Benchmark}} 
& \multirow{2}{*}[-2pt]{\textbf{Setup}}  
& \multirow{2}{*}[-2pt]{\textbf{Manager}}
& \multicolumn{2}{c}{\textbf{SR (\%) ($\uparrow$)}}
& \multicolumn{2}{c}{\textbf{Time (min) ($\downarrow$)}} \\
\cmidrule(lr){4-5} \cmidrule(lr){6-7}
& 
& 
& \textbf{Avg.}
& \textbf{$\Delta$}
& \textbf{Avg.}
& \textbf{$\Delta$} \\
\midrule
\multirow{3}{*}{OSWorld~\citep{xie2024osworld}} 
& Single-Agent & -- & 43.8 & -- & 26.6 & -- \\
& Multi-Agent  & Qwen3.6-27B & 44.9 & {\setlength{\fboxsep}{2pt}\colorbox{improvegreen}{+1.1}} & \textbf{14.2} & {\setlength{\fboxsep}{2pt}\colorbox{improvegreen}{-12.4}} \\
& Multi-Agent  & Opus 4.6 & \textbf{48.5} & {\setlength{\fboxsep}{2pt}\colorbox{improvegreen}{+4.7}} & 21.4 & {\setlength{\fboxsep}{2pt}\colorbox{improvegreen}{\phantom{0}-5.2}} \\
\midrule
\multirow{3}{*}{Online-Mind2Web~\citep{xue2025illusion}} 
& Single-Agent & -- & 52.2 & -- & \textbf{18.5} & -- \\
& Multi-Agent  & Qwen3.6-27B & 52.4 & {\setlength{\fboxsep}{2pt}+0.2} &  23.6 & {\setlength{\fboxsep}{2pt}\colorbox{worsenred}{\phantom{0}+5.1}} \\
& Multi-Agent  & Opus 4.6 & \textbf{55.6} & {\setlength{\fboxsep}{2pt}\colorbox{improvegreen}{+3.4}} & 33.6 & {\setlength{\fboxsep}{2pt}\colorbox{worsenred}{+15.1}} \\
\midrule
\multirow{3}{*}{WebTailBench-v2~\citep{awadallah2025fara}} 
& Single-Agent & -- & 20.8 & -- & \textbf{39.0} & -- \\
& Multi-Agent  & Qwen3.6-27B & 23.8 & {\setlength{\fboxsep}{2pt}\colorbox{improvegreen}{+3.0}} & 61.0 & {\setlength{\fboxsep}{2pt}\colorbox{worsenred}{+22.0}} \\
& Multi-Agent  & Opus 4.6 & \textbf{29.5} & {\setlength{\fboxsep}{2pt}\colorbox{improvegreen}{+8.7}} & 55.2 & {\setlength{\fboxsep}{2pt}\colorbox{worsenred}{+16.2}} \\
\midrule
\multirow{3}{*}{Odysseys~\citep{jang2026odysseys}} 
& Single-Agent & -- & 8.5 & -- & 162.4 & -- \\
& Multi-Agent  & Qwen3.6-27B & 32.0 & {\setlength{\fboxsep}{2pt}\colorbox{improvegreen}{+23.5}} & 116.3 & {\setlength{\fboxsep}{2pt}\colorbox{improvegreen}{-46.1}} \\
& Multi-Agent  & Opus 4.6 & \textbf{34.0} & {\setlength{\fboxsep}{2pt}\colorbox{improvegreen}{+25.5}} & \textbf{110.3} & {\setlength{\fboxsep}{2pt}\colorbox{improvegreen}{-52.1}} \\
% \midrule
% \multirow{3}{*}{\textbf{Average}} 
% & Single-Agent & -- & 31.3 & -- & 61.6 & -- \\
% & Multi-Agent  & Qwen3.6-27B & 38.3 & {\setlength{\fboxsep}{2pt}\colorbox{improvegreen}{+7.0}} & \textbf{53.8} & {\setlength{\fboxsep}{2pt}\colorbox{improvegreen}{-7.9}} \\
% & Multi-Agent  & Opus 4.6 & \textbf{46.1} & {\setlength{\fboxsep}{2pt}\colorbox{improvegreen}{+11.2}} & 55.1 & {\setlength{\fboxsep}{2pt}\colorbox{worsenred}{+0.1}} \\
\bottomrule
\end{tabular}
% }
\vspace{-3ex}
\end{table}

\begin{figure}[t]
    \centering
    \includegraphics[width=\linewidth]{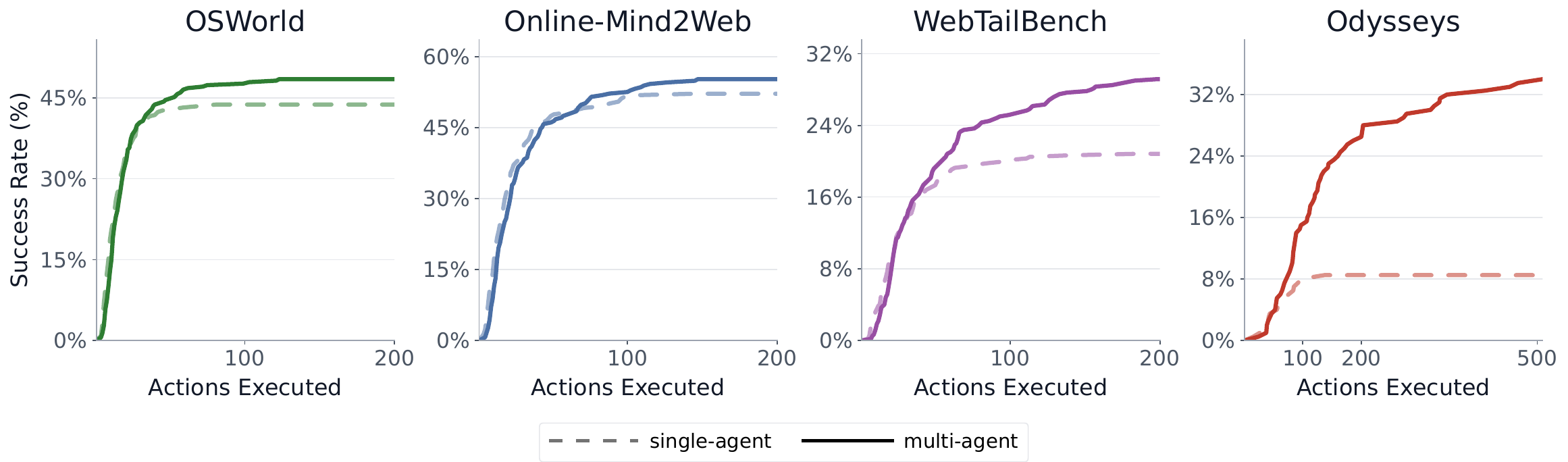}
    \vspace{-3ex}
    \caption{Success rate of \ModelName{} (Opus 4.6 manager) against number of CUA actions executed. Multi-agent unlocks more effective test-time scaling, solving tasks that single-agent setups do not.}
    \label{fig:success-rate-steps}
    \vspace{-2ex}
\end{figure}

\subsection{Results} \label{sec:results}
Our main experimental results are summarized in Tab.~\ref{tab:results}.
% These benchmarks spans multiple axes of task difficulty: long-horizon navigation, cross-site/app information gathering, and complex tasks in partially observable environments, where graph-structured multi-agent decomposition would plausibly help.
\ModelName{} consistently improves the success rate across all four benchmarks. With the Opus 4.6 manager, \ModelName{} improves the success rate on OSWorld from 43.8\% to 48.5\%, and reduces the median wall-clock time from 26.6 to 21.4 mins per task. Online-Mind2Web demonstrates similar trends, improving success rate from 52.2\% to 55.6\%.
We observed that wall-clock time on Online-Mind2Web tasks actually increase, as the tasks are largely serial, and additional manager calls also add a substantial amount of wall-clock time (see Sec.~\ref{sec:analysis} for more discussion), resulting in less benefits from multi-agent deployment and parallelization. 

Conversely, \ModelName{} achieves larger improvements on WebTailBench-v2 and Odysseys, which contain tasks that are typically more complex or require working over longer horizons. These tasks benefit more from being decomposed into smaller subproblems which can be executed in parallel.
On WebTailBench-v2, \ModelName{} improves the success rate from 20.8\% to 29.5\%, and the percentage of rubrics satisfied from 35.9\% to 46.3\%. Manager replanning also takes up significant wall time, which contributes to an overall increase in the median wall-clock time.
On Odysseys, \ModelName{} substantially improves the success rate from 8.5\% to 34.0\%, as well as the percentage of rubrics satisfied per task from 42.1\% to 62.3\%. This indicates that \ModelName{} successfully completes tasks at a higher rate, as well as achieves higher partial credit on tasks that it does not fully complete. 
We also observed a $1.47 \times$ speed up in wall-clock time on Odysseys: the median time spent per task was reduced from 162 to 110 mins, due to the effectiveness of parallel execution on these tasks. We observed similar trends across benchmarks with Qwen3.6-27B as the manager (albeit with lower absolute improvements).

\ModelName{} also demonstrates positive scaling trends as models are provided with greater inference budget. Fig.~\ref{fig:success-rate-steps} plots the success rate against the total number of CUA subagent steps that each model has executed.
Single-agent setups plateau much earlier on compared to \ModelName{}, and the success rate gap widens as more actions are executed, demonstrating that \ModelName{} can effectively leverage greater inference budget to achieve stronger results. 

As we move towards longer horizon computer use tasks, such as those in Odysseys~\citep{jang2026odysseys}, multi-agent approaches such as \ModelName{} become more valuable. \ModelName{} achieves substantially higher performance on long-horizon tasks, while maintaining its performance on shorter tasks (such as those in OSWorld and Online-Mind2Web), making it a general approach that excels on more complex tasks while preserving performance on simpler ones.
These findings strongly motivate moving towards building, researching, and evaluating multi-agent computer use systems instead of serial agents, especially as we start deploying LLM agents on more complex, long running real world tasks.

\subsection{Ablations} \label{sec:ablations}

We conduct a series of ablation experiments to justify the design choices made in the \ModelName{} setup. Unless otherwise specified, all ablations are conducted on the small subset of OSWorld (36 tasks), with the Qwen-3.5-4B model as the worker and the Opus-4.6 model as the manager.

\begin{figure}[t]
\centering

\begin{minipage}[t]{0.4\textwidth}
\vspace{0pt}
\centering
\captionof{table}{Results with varying planning budget $B$. Average success rate (SR) (\%) and cost (USD \$) are shown.}
\label{tab:replanning}

\begin{tabular}{l
                S[table-format=2.2]
                S[table-format=2.1]}
  \toprule
  \small
  \textbf{$B$} \hspace{3mm}
  & \textbf{Cost ($\downarrow$)}  \hspace{1mm}
  & \textbf{SR ($\uparrow$)} \hspace{1mm} \\
  \midrule
  0  & 0.00 & 25.0 \\
  1  & 0.08 & 27.8 \\
  5  & 0.41 & 47.2 \\
  10 & 0.46 & 58.3 \\
  \bottomrule
\end{tabular}
\end{minipage}
\hfill
\begin{minipage}[t]{0.55\textwidth}
\vspace{0pt}
\centering
\includegraphics[width=0.95\textwidth]{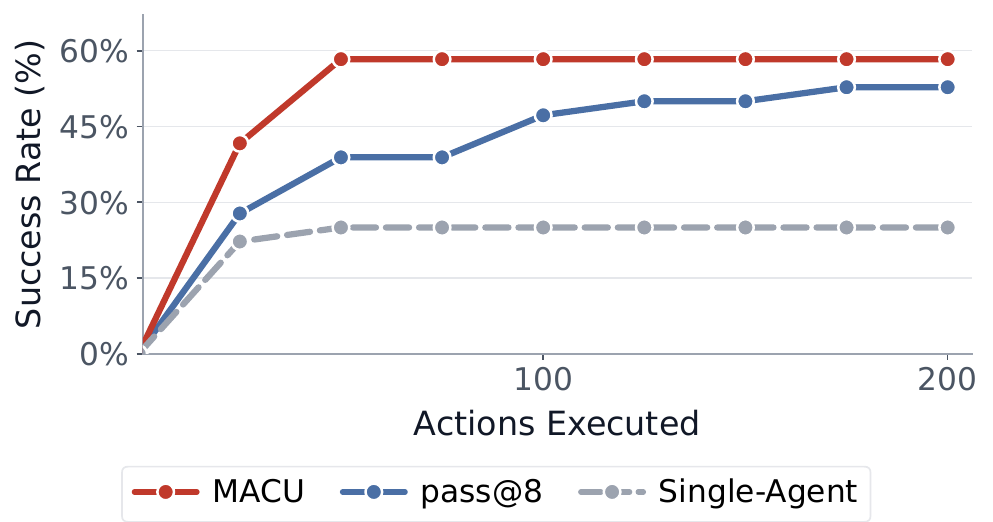}
\captionof{figure}{Comparison of \ModelName{} against pass@$8$ and single-agent baselines.}
\label{fig:pass_at_k_vs_steps}
\end{minipage}
\vspace{-2ex}
\end{figure}

% \# Steps ($\downarrow$) column values from Table~\ref{tab:replanning}: 0 $\rightarrow$ 51.2, 1 $\rightarrow$ 46.2, 5 $\rightarrow$ 76.9, 10 $\rightarrow$ 66.4.

\paragraph{Importance of planning} We ablate increasing the planning budget $B$, which represents how many times the manager is able to modify the DAG when attempting a task. The results (Tab.~\ref{tab:replanning}) highlight the importance of re-planning on the fly. 
Increasing $B$ from 0 (no planning) to 1 (generate an initial DAG, but no modifications allowed after) results in a minimal change in success rate ($25.0\% \rightarrow 27.8\%$), but enabling \ModelName{} to modify the graph during the runtime of the task ($B > 1$) improves success rate much more significantly, from 25.0\% (no planning) to 47.2\% with $B=5$ and 58.3\% with $B=10$. 
These results highlight the importance of continuous replanning when new observations are received.

\paragraph{Comparison against pass@$k$}
A natural baseline to compare \ModelName{} against is pass@$k$: we run a single-agent repeatedly up to $k=8$ times, stopping when the groundtruth evaluator reports a success. The evaluator is typically not available at test time, but pass@$k$ is still useful for analysis.
Fig.~\ref{fig:pass_at_k_vs_steps} shows the success rate given a maximum number of total actions executed (tasks which are incomplete at that point are considered failures). We find that \ModelName{} outperforms pass@$k$ up to 200 total actions executed, with both setups plateauing after. \ModelName{} is able to more effectively leverage inference-time compute, despite pass@$k$ having access to the groundtruth evaluator.
This highlights the value of the task decomposition, replanning, and error correction features of \ModelName{}. % for improved test-time scaling behavior.

\begin{table}[t]
\begin{minipage}[t]{0.4\textwidth}
\centering
\captionof{table}{Ablation results with different sized CUAs for subagents (success rate, \%) for single-agent vs. \ModelName{}.}
\vspace{1ex}
\resizebox{\textwidth}{!}{%
\begin{tabular}{lcc}
\toprule
\textbf{Model}
& \textbf{Single}
& \textbf{\ModelName{}} \\
\midrule
Qwen3.6-27B       & 47.2 & 66.7 \\
Qwen3.6-35B-A3B   & 50.0 & 58.3 \\
% Qwen3.5-27B       & 47.2 & 55.6 \\
Qwen3.5-9B        & 50.0 & 61.1 \\
Qwen3.5-4B        & 25.0 & 58.3 \\
Qwen3.5-2B        & 27.8 & 44.4 \\ 
\bottomrule
\end{tabular}
}
\label{tab:model_sweep}
\end{minipage}
\hfill
\begin{minipage}[t]{0.55\textwidth}
\centering
\captionof{table}{Manager model ablation with fixed CUA subagent model (Qwen3.5-4B). We report success rate (SR) and API cost averaged over tasks. Models we host and run locally are reported as having no API cost.}
\vspace{1ex}
\resizebox{\textwidth}{!}{%
\begin{tabular}{lcc}
\toprule
\textbf{Manager Model} 
& \textbf{SR (\%) ($\uparrow$)} 
& \textbf{Cost (\$) ($\downarrow$)} \\
\midrule
--- (single-agent)        & 25.0 & -- \\
\midrule
Opus-4.6               & \textbf{58.3} & 0.46 \\
Sonnet-4.6             & 52.8 & 0.28 \\
GPT-5.4                & 44.4 & 0.42 \\  %  38.9 & 2.39
GPT-5.4-mini           & 33.3 & 0.37 \\
gemini-3.1-pro-preview & 41.7 & 0.24 \\  % 38.9, 0.18
gemini-3.1-flash-lite-preview & 36.1 & 0.01 \\  % 44.4 & 0.01
Kimi-K2.6              & 41.7 & -- \\
Qwen3.6-27B            & 41.7 & -- \\
% Qwen3.6-35B-A3B       & 25.0 & -- \\
\bottomrule
\end{tabular}
}
\label{tab:manager_ablation}
\end{minipage}
\vspace{-2ex}
\end{table}

\paragraph{Stronger CUA subagents} We experiment with larger and more capable models as the backbone for the subagents. Tab.~\ref{tab:model_sweep} summarizes the results, which validate that using stronger subagents generally leads to improved performance on both the single-agent and multi-agent setups.
The strong results of the Qwen3.6-27B model led us to adopt it as the subagent backbone for our main experiments.

\paragraph{Stronger manager models} We ablate the choice of manager model while keeping the CUA worker fixed (Qwen3.5-4B) in Tab.~\ref{tab:manager_ablation}.
The \ModelName{} framework improves upon the single-agent with all manager models we tested, but using stronger LLMs as managers substantially improve performance. Capable managers can more than double the success rate of the fixed Qwen3.5-4B CUA worker, whereas weaker managers provide only marginal gains.
Opus 4.6 obtains the best result among the evaluated managers, reaching 58.3\% SR, a 33.3\% absolute gain over the single-agent baseline. 
Sonnet 4.6 achieves 52.8\%, the next best score, while GPT-5.4 achieves achieve 44.4\% SR and the gemini-3.1-pro-preview, Kimi-K2.6, and Qwen3.6-27B reach 41.7\% SR.

\begin{figure}[t]
\centering
\includegraphics[width=\textwidth]{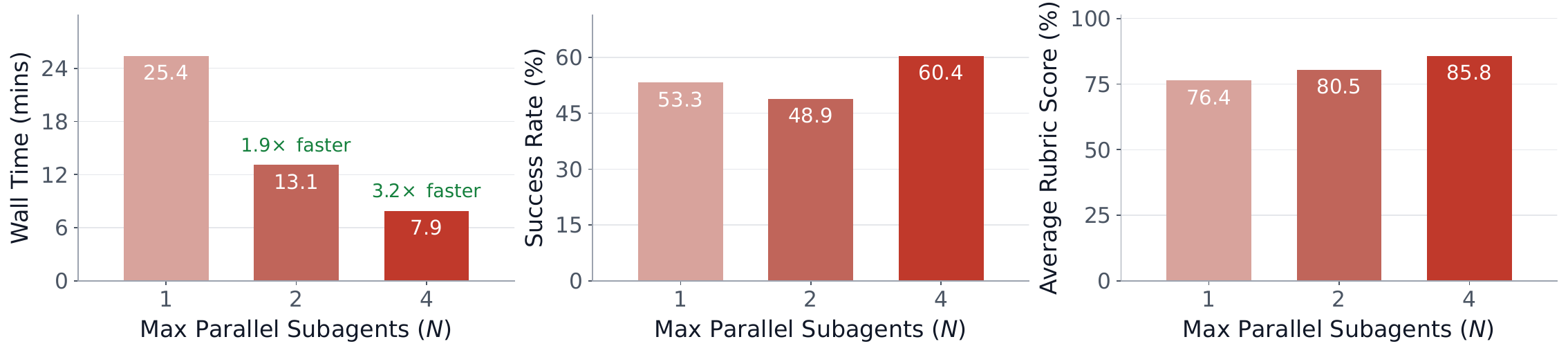}
\vspace{-3ex}
\captionof{figure}{Wall-clock time, success rate, and average rubric score on the Odysseys easy subset as a function of the maximum number of parallel subagents $N$.} \label{fig:parallel_wall_time}
\vspace{-1ex}
\end{figure}

\paragraph{Number of parallel subagents} On the Odysseys easy subset (45 tasks), we varied the maximum number of parallel subagents $N$ to examine the wall-clock speedup provided by \ModelName{}.
As visualized in Fig.~\ref{fig:parallel_wall_time}, increasing $N$ substantially reduces median wall-clock time, from 25.4 minutes ($N=1$) to 7.9 minutes ($N=4$), while improving success rate from 53.3\% to 60.4\%, and average rubric score from 76.4\% to 85.8\%. 
These results exclude inference overhead (as this is noisy based on cluster usage and number of concurrent requests). This represents a ${\sim}1.9\times$ theoretical speedup when going from $N=1$ to $N=2$ and ${\sim}3.2\times$ from $N=1$ to $N=4$ on this subset of Odysseys tasks.

\paragraph{Manager-subagent asymmetry} 
Our results using Qwen3.6-27B as both the subagent and the manager in Tab.~\ref{tab:results} demonstrate that the \ModelName{} setup has benefits beyond merely inference-time distillation of the manager model outputs. In Sec.~\ref{sec:analysis}, we also identify that \ModelName{} unlocks structural changes to task execution that go beyond knowledge distillation.

\input{figures/graph_shape_examples/table_snippet}

\subsection{Analysis} \label{sec:analysis}
\begin{figure}[t]
\centering
\includegraphics[width=\textwidth]{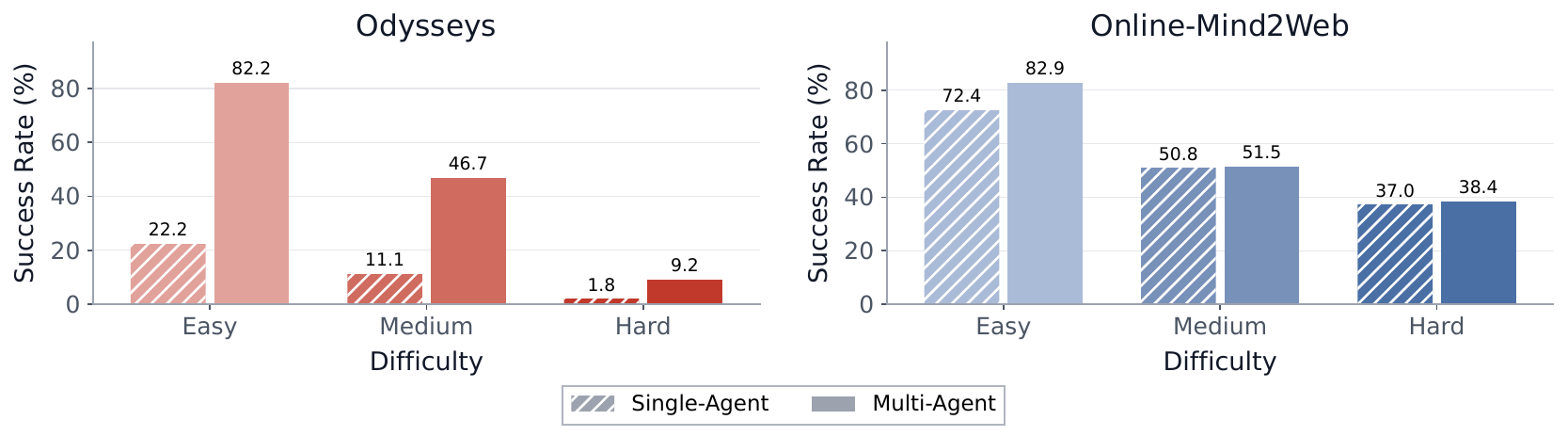}
\vspace{-3ex}
\captionof{figure}{\ModelName{} improves success rate over the single-agent baseline across all difficulty levels on Odysseys and Online-Mind2Web tasks.} \label{fig:difficulty_metrics}
\end{figure}

\paragraph{When does multi-agent help?}
We stratified performance by difficulty level (Fig.~\ref{fig:difficulty_metrics}), task domain, and also qualitatively by examining execution traces.
On Odysseys, gains from \ModelName{} were especially large on easy and medium tasks, with success rate improving from $22.2\% \rightarrow 82.2\%$ and $11.1\% \rightarrow 46.7\%$ respectively. The hard subset also improved substantially on both success rate ($1.8\% \rightarrow 9.2\%$) and rubric score ($26.5\% \rightarrow 43.1\%$). 
WebTailBench-v2 categories which required parallel information gathering and comparison benefit the most from \ModelName{}: \texttt{price\_comparison} tasks improved from $3.7\% \rightarrow 33.9\%$, \texttt{flights} from $14.0\% \rightarrow 34.0\%$, and \texttt{compositional\_tasks\_v2} from $24.0\% \rightarrow 41.8\%$.
Online-M2W showed a smaller but consistent gain primarily concentrated on the easy subset ($72.4\% \rightarrow 82.9\%$) but remained relatively similar on medium ($50.8\% \rightarrow 51.5\%$) and hard ($37.0\% \rightarrow 38.4\%$) tasks.
On OSWorld tasks (which doesn't have difficulty labels), we observed the strongest gains on OS and Ubuntu tasks ($45.8\% \rightarrow 70.8\%$), as well as tasks on the LibreOffice productivity suite: LibreOffice Writer ($47.8\% \rightarrow 60.9\%$), Calc ($34.0\% \rightarrow 46.8\%$), and Impress ($40.2\% \rightarrow 50.9\%$).
Across benchmarks, we observed a general trend that \ModelName{} is strongest on decomposable tasks that require parallel exploration and information gathering. This often requires preserving multiple pieces of the state before the final response (e.g., Odysseys tasks in Tab.~\ref{tab:graph-shape-visuals}).

\paragraph{Types of graphs} We show some examples of tasks in each benchmark and their final DAGs in Tab.~\ref{tab:graph-shape-visuals}. 
We find several common trends across tasks in the benchmarks: OSWorld and Online-M2W test mostly single-task GUI execution, resulting in smaller and more serial graphs. 
However, we did observe that on a substantial number of these tasks, \ModelName{} still improved performance by spawning retry nodes, or attempting the same task through different means (e.g., see Fig.~\ref{fig:graph-example} and Appendix~\ref{sec:additional-qualitative}).
These behaviors are similar to tree search~\citep{koh2024tree}, which can be considered a special case of \ModelName{} (which is more general), although \ModelName{} operates at the subtask level rather than at every step like \citep{koh2024tree}.
This also explains why the success rate of OSWorld and Online-Mind2Web see a relatively lower gain compared to WebTailBench-v2 and Odysseys, as tasks inherently benefit less from multi-agent decomposition and parallelization. 
On the Odysseys benchmark, tasks are generally more highly parallelizable and decomposable, and many benefit from launching parallel subagents to collect independent information. We observed these types of parallel graph structures much more frequently in Odysseys task attempts.

\begin{table*}[t]
\centering
\small
\captionof{table}{DAG statistics across benchmark runs. Replan rate is the percentage of tasks that executed at least one DAG modification during runtime.} \label{tab:graph-shapes}
\begin{tabular}{lllc}
\toprule
\textbf{Benchmark} \hspace{5mm} & \textbf{Initial DAG} & \textbf{Final DAG} & \textbf{Replan Rate} \\
\midrule
OSWorld &
2.3 nodes / 1.5 edges &
%; med. depth 2, width 1 &
2.9 nodes / 2.1 edges &
%; p90 5 nodes, width 3; max 9 nodes &
45.7\% \\
%& 165/361 tasks replanned (45.7\%); 142 added nodes; 306 applied replans total \\
%& 220 chain; 78 retry-heavy; 54 fork-join; 7 fan-in \\
Online-M2W &
2.1 nodes / 1.1 edges &
4.3 nodes / 2.1 edges &
% 2.05 nodes / 1.07 edges; 286 single-subtask plans &
% 4.31 nodes / 2.06 edges; p90 8 nodes, width 7; max 17 nodes &
68.0\% \\
%& 204 tasks with applied replans; 781 added nodes  \\
% & 134 chain; 123 retry-heavy; 18 retry-light; 25 DAG/fan-in \\
WebTailBench &
2.3 nodes / 1.5 edges &
% 2.32 nodes / 1.49 edges; mostly one CUA subtask + aggregation &
4.2 nodes / 2.6 edges &
% 4.19 nodes / 2.56 edges (last snapshot); retry-heavy fan-out &
73.5\% \\
%& 441/600 tasks with applied replans; 1436 applied replans total \\
Odysseys &
6.0 nodes / 8.0 edges &
7.6 nodes / 8.3 edges &
% 5.99 nodes / 8.02 edges; med. depth 3, width 3 &
% 7.63 nodes / 8.26 edges; med. width 4; max 23 nodes &
74.5\% \\
%149 tasks with applied replans; 89 with retry/variant nodes \\
%& 89 retry/variant; 67 fork-join; 43 fan-out/star \\
\bottomrule
\end{tabular}
\end{table*}

\paragraph{How does the graph change over time?} Tab.~\ref{tab:graph-shapes} shows the statistics of the DAG at the initial decomposition state, and at the end of the task (after potentially undergoing replanning). We observe that across all benchmarks, the graphs grow during the course of a task, due to the manager replanning and adding new subtasks as more information is acquired. On most benchmarks (with OSWorld as the exception), the majority of tasks required at least one replan.
% We also observe that OSWorld has the lowest replan rate among the four benchmarks (45.7\% of tasks were replanned). 
On Online-M2W, although replanning occurs less often (68\% of tasks) compared to WebTailBench and Odysseys, the DAG changes the most dramatically, doubling in size from 2.1 nodes to 4.3 nodes on average, due to nodes added on-the-fly for retry attempts.
Odysseys graphs are the largest, which matches our intuition that tasks are more complex and parallelizable (many tasks with the map-reduce pattern shown in Tab.~\ref{tab:graph-shape-visuals}).

\section{Conclusion}

In this paper, we argued that computer use agents should move beyond the prevailing single serial agent paradigm, and proposed a general multi-agent computer use (\ModelName{}) framework that directly targets complex and long-horizon CUA tasks.
Across four diverse computer use benchmarks, we demonstrated that \ModelName{} consistently improves the success rates of strong single-agent baselines, and exhibits more favorable test-time scaling than its single-agent counterpart, solving long-horizon tasks that serial single-agents get stuck on. 
Our ablation results highlight that continuous replanning and parallel subagent execution are key to unlocking these gains, and that multi-agent coordination is a promising axis for scaling computer use agents to work productively for longer. 
Integrating new CUA models into the \ModelName{} framework is seamless, and we expect success rates to continue to compound with future advances in stronger CUA models and capabilities.
We hypothesize that finetuning or reinforcement learning will also substantially improve the abilities of the system, as managers and workers become trained for multi-agent coordination and interaction, which is a promising direction for future work.
We release our code and interactive visualizations of \ModelName{} on selected tasks at \url{https://jykoh.com/multi-agent-computer-use} to facilitate future work.

\section*{Acknowledgments}
We thank Lawrence Jang, Pranjal Aggarwal, Apurva Gandhi, and others for feedback and helpful discussions. We thank Wendy Kua for feedback on the figures. 
Jing Yu Koh is supported by a Jane Street Graduate Research Fellowship.
This work was supported in part by ONR N000142312368 and ONR MURI N00014-25-1-2116. 

% References
\bibliographystyle{plain}
\bibliography{references}

%%%%%%%%%%%%%%%%%%%%%%%%%%%%%%%%%%%%%%%%%%%%%%%%%%%%%%%%%%%%
\pagebreak
\input{appendix}
%%%%%%%%%%%%%%%%%%%%%%%%%%%%%%%%%%%%%%%%%%%%%%%%%%%%%%%%%%%%

\end{document}

%% file: figures/graph_shape_examples/table_snippet.tex
% Requires \usepackage{xcolor}
\DeclareRobustCommand{\workernodecircle}{\textcolor[HTML]{546E7A}{\raisebox{-0.38ex}{\LARGE$\bullet$}}}
\DeclareRobustCommand{\managernodecircle}{\textcolor[HTML]{C0392B}{\raisebox{-0.32ex}{\LARGE$\bullet$}}}

\begin{table*}[t]
\centering
\small
\setlength{\tabcolsep}{4pt}
\renewcommand{\arraystretch}{1.08}
\newcommand{\graphcell}[2][1.0in]{%
  \begin{minipage}[t][#1][c]{\linewidth}%
    \raggedright\includegraphics[height=#1]{#2}%
  \end{minipage}%
}
\newcommand{\textcell}[1]{{\raggedright #1\par}}
\newcommand{\benchcell}[1]{\multirow{2}{=}[0pt]{\raggedright #1}}
\captionof{table}{Types of DAG created for different tasks from the different benchmarks. Lettered nodes \workernodecircle{} are referenced in the decomposition descriptions; red nodes \managernodecircle{} denote manager action.}
\label{tab:graph-shape-visuals}
\begin{tabular}{p{0.15\linewidth}p{0.21\linewidth}>{\raggedright\arraybackslash}p{0.25\linewidth}p{0.3\linewidth}}
\toprule
\textbf{Benchmark} & \textbf{Task} & \textbf{Graph} & \textbf{Decomposition Pattern} \\
\midrule
\benchcell{OSWorld} &
\textcell{Change GIMP's theme from dark to light.} &
\vspace{-0.12in}\graphcell[0.35in]{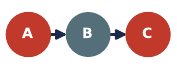} &
\textcell{{\setlength{\fboxsep}{1pt}\colorbox{grey}{\textbf{Simple chain:}}} Manager (A) creates a CUA worker (B) to change the theme, then (C) summarizes the result.} \\
\cmidrule(lr){2-4}
&
\textcell{Gather Google Maps information for five Hong Kong restaurants and enter it into a spreadsheet.} &
\vspace{-0.23in}\graphcell[0.8in]{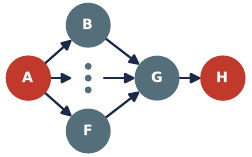} &
\textcell{\setlength{\fboxsep}{1pt}\colorbox{grey}{\textbf{Map-reduce:}} Manager (A) creates parallel restaurant lookups (B--F), which feed a spreadsheet worker (G), then a manager action (H) reports.} \\
\midrule

\benchcell{Online-M2W} &
\textcell{Find the newest digital-trends report for the Finance \& Insurance industry in China.} &
\vspace{-0.23in}\graphcell[0.9in]{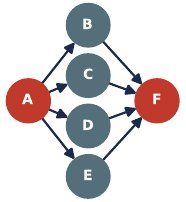} &
\textcell{\setlength{\fboxsep}{1pt}\colorbox{grey}{\textbf{Runtime retry expansion:}} Manager (A) creates the initial search (B) and alternate search variants (C--E), then a manager action (F) selects evidence.} \\
\cmidrule(lr){2-4}
&
\textcell{Compare Afghan Hound, Akita, and Azawakh dog-breed information.} &
\vspace{-0.23in}\graphcell[0.75in]{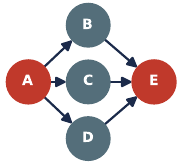} &
\textcell{\setlength{\fboxsep}{1pt}\colorbox{grey}{\textbf{Map-reduce:}} Manager (A) creates three independent breed lookups (B--D), which feed one comparative manager action (E).} \\
\midrule

\benchcell{WebTailBench-v2} &
\textcell{Research where to buy \emph{A Tale of Two Cities} and tabulate paperback and hardcover prices across retailers.} &
\vspace{-0.23in}\graphcell[0.9in]{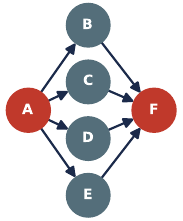} &
\textcell{\setlength{\fboxsep}{1pt}\colorbox{grey}{\textbf{Map-reduce:}} Manager (A) creates four independent retailer lookups (B--E; Amazon, Barnes \& Noble, Books-A-Million, ThriftBooks), which feed the final manager action (F).} \\
\cmidrule(lr){2-4}
&
\textcell{Count Ryanair extra-legroom seats from Birmingham to Porto for fixed travel dates.} &
\vspace{-0.18in}\graphcell[0.7in]{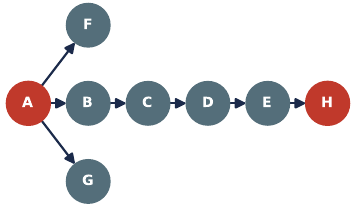} &
\textcell{\setlength{\fboxsep}{1pt}\colorbox{grey}{\textbf{Retry chain:}} Manager (A) launches an attempt (B) that is repeatedly retried (C--E) until it succeeds, while one-shot fallback attempts (F,~G) are abandoned; a final manager action (H) reports.} \\
\midrule

\benchcell{Odysseys} &
\textcell{\textit{(summarized)} Plan a realistic five-day Iceland camper-van route using campgrounds, maps, and travel sites.} &
\vspace{-0.23in}\graphcell[0.87in]{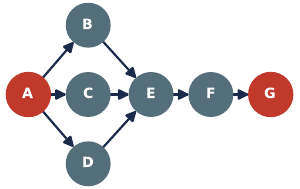} &
\textcell{\setlength{\fboxsep}{1pt}\colorbox{grey}{\textbf{Map-reduce:}} Manager (A) creates campgrounds, maps, and travel-page research (B--D), followed by itinerary synthesis (E), tab curation (F), and final manager action (G).} \\
\cmidrule(lr){2-4}
&
\textcell{\textit{(summarized)} Build a small TV watchlist by verifying a Hulu listing and checking related reference pages.} &
\vspace{-0.23in} \graphcell[0.75in]{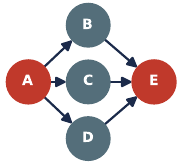} &
\textcell{\setlength{\fboxsep}{1pt}\colorbox{grey}{\textbf{Map-reduce:}} Manager (A) creates independent Hulu, Wikipedia, and Memory Alpha checks (B--D), which feed the final manager action (E).} \\
\bottomrule
\end{tabular}
\end{table*}

%% file: appendix.tex
\appendix

\lstdefinestyle{promptinline}{
  basicstyle=\ttfamily\scriptsize,
  breaklines=true,
  breakatwhitespace=false,
  columns=fullflexible,
  keepspaces=true,
  upquote=true,
  literate={—}{{\textemdash}}1
}
\definecolor{promptsystembg}{gray}{0.94}

\section{Additional Experimental Details} \label{appendix:experimental-details}

\begin{table}[h]
\centering
\small
\captionof{table}{Experimental settings used by the main \ModelName{} experiments.} \label{tab:manager-setup-hparams}
\begin{tabular}{p{0.32\textwidth}p{0.58\textwidth}}
\toprule
\textbf{Setting} & \textbf{Value} \\
\midrule
Manager provider and model & Anthropic \texttt{claude-opus-4-6} \\
\midrule
CUA worker for main experiments & \texttt{Qwen/Qwen3.6-27B} \\
\midrule
Max parallel CUA subagents $N$ & 4 \\
\midrule
Replan budget $B$ & 10 \\
\midrule
% Replanning triggers & CUA completion and spare-capacity consultation after a running subtask reaches the step threshold \\
Manager maximum output tokens & 8192 for manager calls; 16384 for initial graph generation \\
\midrule
Manager visual context & 2 recent screenshots per relevant subtask; a max of 12 total screenshots \\
\midrule
% CUA subagent max steps & 60 per subtask \\
% \midrule
Maximum runtime per task & 1.5 hours for OSWorld, 3 hours for Online-Mind2Web, WebTailBench-v2, and Odysseys \\
\midrule
Follow-up hook & Enabled for completed CUA subtasks; the manager may request continuation or a targeted summary before termination \\
\midrule
File management & Enabled after each CUA completion through filesystem diffs, an archive pool, and replanner-assigned \texttt{input\_files} \\
% Prompt logging & Full manager prompts and raw responses are written under each run directory as \texttt{manager\_prompt\_*} and \texttt{manager\_response\_*} files \\
\bottomrule
\end{tabular}
\end{table}

We summarize the experiment hyperparameters in Tab.~\ref{tab:manager-setup-hparams}. These settings are shared across the main benchmark runs in Sec.~\ref{sec:results}, and retained for ablations in Sec.~\ref{sec:ablations} (unless we explicitly mention changing the manager model, CUA worker model, parallelism, or replan budget). We provide the prompts used for the manager in Appendix.~\ref{appendix:manager-prompts}.

For \ModelName{} runs, we host a \texttt{Qwen/Qwen3.6-27B} model as an inference server using vLLM~\citep{kwon2023efficientmemorymanagementlarge} on two A6000 GPUs (with tensor parallelism of 2). Each task also requires 64G RAM and 16 CPUs for hosting the virtual machines for the subagents. We run 6 tasks per vLLM server, which typically fully saturates the inference workload. Each benchmark takes a varying amount of time to complete (see Sec.~\ref{sec:results} for approximate wall-clock time). The amount of API cost per \ModelName{} run (with the \texttt{claude-opus-4-6} manager) averages USD \$0.21 per OSWorld task, \$0.38 per Online-Mind2Web task, \$0.46 per WebTailBench task, and \$0.90 per Odysseys task, which totals approximately \$651 in API cost for a full run of all four benchmarks. The total wall clock time spent for the run is highly variable depending on cluster usage, but our final end-to-end run on all 3 benchmarks took the approximately 38 hours to run end-to-end.

For single-agent runs, we extend the max steps per subagent to 200, and reduce the amount of resources used (16G RAM and 4 CPUs per task), as only one CUA worker is run for each task. 

Since Qwen3.5 and Qwen3.6 did not provide implementation details of their model, we do our best effort to reproduce it based on existing implementations (include how the models are used for SWE agents in qwen-code\footnote{\url{https://github.com/QwenLM/qwen-code}}, but our results still fell short of the numbers reported in the Qwen3.5 releases. Nevertheless, our implementation still achieves substantially better results than all other open weights CUAs, so we use it as our baseline agent.

\subsection{Virtual Machines Technical Details}

\ModelName{} launches each CUA subagent through OSWorld's \texttt{DesktopEnv}; the subagent process constructs the requested VM provider, starts the emulator, records the VM endpoint, and writes the in-VM HTTP, browser-debugging, VNC, and VLC ports to a status file \texttt{vm\_info.json} for the \ModelName{} runtime. % osworld/desktop_env/desktop_env.py:157-216; scripts/run_cua.py:1883-1922
Subagents initialize from a provided VM state (if any). % scripts/run_cua.py:1897-1899; osworld/desktop_env/desktop_env.py:398-423
We implement two VM backbones:

\paragraph{VMware provider.}
% VMware runs use a shared VM pool registry \texttt{.vmware\_vms}, protected by \texttt{.vmware\_lck}, whose entries map each \texttt{.vmx} path to either \texttt{free} or the owning process id. % osworld/desktop_env/providers/vmware/manager.py:41-43; osworld/desktop_env/providers/vmware/manager.py:332-397; utils/vm_utils.py:402-490
Before launching an independent DAG node, the \ModelName{} runtime atomically claims a pool VM and creates a \texttt{vmrun snapshot} named for the subtask. % Code pointer: utils/vm_utils.py:254-333; utils/vm_utils.py:415-445; utils/macu_runtime.py:1405-1428
For state transfer along DAG dependencies, a child node may reclaim a completed predecessor's VM and skip canonical setup, preserving the predecessor CUA subagent's final desktop and filesystem state (i.e., effectively resuming from the parent subtask). % utils/macu_runtime.py:1294-1368; utils/macu_runtime.py:206-211
For retry or variant nodes, \ModelName{} makes a full VMware clone of its frozen initial state, so parallel attempts start from the same pre-subtask state without sharing a mutable VM. % utils/vm_utils.py:336-368; utils/macu_runtime.py:1256-1292; utils/macu_runtime.py:1430-1478

\paragraph{Apptainer/QEMU provider.} 
On clusters where Docker or VMware is inconvenient, the Apptainer provider boots the same OSWorld image by running \texttt{qemu-system-x86\_64} inside an Apptainer SIF. % with \texttt{/dev/kvm} bound from the host. % osworld/desktop_env/providers/apptainer/provider.py:1-8; osworld/desktop_env/providers/apptainer/provider.py:497-556
The base image, overlay directory, SIF path, and run directory, are preconfigured, and each fresh CUA subagent VM is a new writable qcow2 overlay backed by the shared base image. % osworld/desktop_env/providers/apptainer/manager.py:31-42; osworld/desktop_env/providers/apptainer/provider.py:37-43; osworld/desktop_env/providers/apptainer/provider.py:116-117; utils/vm_utils.py:520-587
% Each launch allocates random localhost ports under a file lock, forwards the in-guest server, browser, and VLC ports through QEMU user networking, exposes VNC, creates a QMP socket, and writes a \texttt{.conn.json} sidecar so the evaluator or \ModelName{} runtime can reconnect to the same VM. % osworld/desktop_env/providers/apptainer/provider.py:173-233; osworld/desktop_env/providers/apptainer/provider.py:270-319; osworld/desktop_env/providers/apptainer/provider.py:432-567
% CUA subagent cleanup calls \texttt{env.release()}, which deliberately leaves the Apptainer-backed QEMU process alive long enough for the evaluator or \ModelName{} runtime to reattach through that sidecar if needed. % osworld/desktop_env/providers/apptainer/manager.py:113-120; scripts/run_cua.py:1976-1983; scripts/run_cua.py:2200-2205
% After an Apptainer subtask completes or times out, \ModelName{} saves the live VM by finding the QMP socket from the sidecar, issuing \texttt{stop}, and using QMP migration to write CPU, RAM, and device state to \texttt{\{overlay\}.vmstate}; it then kills the QEMU process that still holds the overlay open. % Code pointer (master): ../multi-agent-cua-code/utils/macu_runtime.py:695-701; ../multi-agent-cua-code/utils/macu_runtime.py:922-930; ../multi-agent-cua-code/utils/vm_utils.py:590-718; ../multi-agent-cua-code/osworld/desktop_env/providers/apptainer/provider.py:637-691
Initialization and resumption are handled similarly to the VMware backend.
When an Apptainer-backed CUA subagent starts from an inherited overlay, the provider walks the overlay backing chain to find a saved VM state overlay; if present, QEMU is launched with a migration command to restore from the overlay.
State transfer for Apptainer DAG nodes is similarly implemented by overlay cloning. Child nodes skip canonical setup when it intentionally inherits that state. 
At run teardown, \ModelName{} kills only QEMU processes that own the run's overlays and deletes overlay files in child-before-parent order so qcow2 backing chains are not removed underneath still-existing child nodes. % utils/macu_runtime.py:1547-1655; utils/vm_utils.py:371-399

\section{Limitations} \label{sec:limitations}

While we have shown that \ModelName{} achieves promising results across computer-use benchmarks, our approach comes with several practical considerations and possible directions to address in future work.

\paragraph{Increased inference-time cost.} \ModelName{} improves performance by providing an additional dimension to spend test-time compute, through decomposition, replanning, and parallel CUA subagent execution (Sec.~\ref{sec:method}). In our main experiments, we allow up to $N=4$ parallel subagents and a replan budget of $B=10$ (Appendix~\ref{appendix:experimental-details}), which can require substantially more model calls and VM execution time than a single-agent setting. 
This extra compute is not always offset by wall-clock speedup: on Online-Mind2Web, \ModelName{} improves success rate from 52.2\% to 55.6\%, but increases median wall-clock time from 18.5 to 33.6 minutes because many tasks are largely serial (Sec.~\ref{sec:results}). In practical systems, the parallelism level $N$, replan budget $B$, and per-subtask step limit must therefore be tuned based on whether the target workload actually benefits from additional exploration, as well as the GPU and CPU resources available at hand.

\paragraph{Benefits depend on task decomposability.} Our analysis shows that \ModelName{} is strongest on tasks that naturally decompose into parallel exploration, independent information gathering, retry attempts, or alternative execution strategies (Sec.~\ref{sec:analysis}). 
This explains the large gains on Odysseys, where many tasks have parallel map-reduce structure, and lower gains on OSWorld and Online-Mind2Web, which more often involve single-task GUI execution or serial interactions. For inherently sequential tasks, spawning additional subagents may add overhead without creating useful parallel work. Future systems should learn when to use multi-agent execution, when to fall back to a single agent, and how to avoid manufacturing unnecessary subtasks.

% \paragraph{Coordination quality depends on the manager and worker models.} \ModelName{} treats each CUA subagent as a black box and relies on the manager to write useful instructions, preserve relevant state, revise the DAG, and decide when to launch retries or variants (Sec.~\ref{sec:method}). Our ablations show that both stronger CUA subagents and stronger manager models substantially improve results (Sec.~\ref{sec:ablations}). This means that failures can arise not only from low-level GUI mistakes by workers, but also from manager-level errors: poor decomposition, missing dependencies, stale or incomplete state summaries, unnecessary replans, or failure to identify that a task is infeasible. We expect finetuning and reinforcement learning for manager-worker coordination to improve this, but the current system is still prompt-driven and depends on the quality of the underlying models.

\paragraph{Additional system complexity.} Computer use differs from text-only agents because relevant state can live in screenshots, open tabs, filesystem changes, background processes, cookies, or application state (Sec.~\ref{sec:method}). \ModelName{} addresses this through isolated VMs, screenshot context, filesystem diffs, and an archive pool for selected files. These mechanisms make multi-agent execution possible, but they also introduce additional overhead in maintaining these additional infrastructure concerns. More robust state serialization, VM cloning, file provenance tracking, and automatic validation of final artifacts are important directions for future work.

\paragraph{Evaluation may not capture real world risks.} We evaluated on standard computer-use benchmarks: OSWorld, Online-Mind2Web, and Odysseys, which cover desktop control, live website navigation, and long-horizon web tasks. These benchmarks are useful for measuring capabilities, but they do not fully capture the risks and operational constraints of real world deployments. Real user environments may involve private accounts, irreversible actions, authentication flows, or stricter reliability requirements. As discussed in Sec.~\ref{sec:broader-impacts}, deployment would require additional safeguards such as user authorization, action restrictions, monitoring, and confirmation for high-risk actions. Our results should therefore be interpreted as evidence that multi-agent coordination is a compelling research direction, not as evidence that multi-agent computer use is ready for real-world deployments.

\section{Broader Impacts} \label{sec:broader-impacts}

As computer use agents continue to become more capable, multi-agent computer use systems which leverage these models presents both new opportunities as well as potential ethical considerations. 
\ModelName{} is a research framework for coordinating CUA subagents. In this paper, we evaluate the framework on existing computer-use and web-navigation benchmarks rather than deploying it to real users. 
We also do not introduce new model weights or datasets. 
The potential positive impact of this line of work is that more reliable CUAs, which are likely to arise in part due to strong multi-agent systems, could help people complete long-horizon computer tasks, automate repetitive workflows, and improve accessibility for users who have difficulty interacting with complex software interfaces. However, improving the reliability and speed of CUAs also raises potential concerns:

\begin{itemize}
    \item \textbf{Intended uses.} Our work is intended as a research demo for studying how to coordinate multiple computer use agents on complex tasks. It is not, in its current form, intended for deployment on live user accounts, production websites, or security-sensitive environments. Any real-world use of multi-agent CUA systems should include further research, as well as careful scoping, user authorization guardrails, and safeguards around which applications, websites, files, and accounts an agent may access.
    \item \textbf{Misuse and security.} A system that can dispatch multiple CUA workers could be misused to scale harmful computer activity, including phishing, spam, manipulation, scraping, or other unauthorized workflows. Even in benign use, agents acting through GUIs and websites can make unintended changes, hallucinate incorrect information, or take destructive actions in a live environment. Eventual real world deployment of such systems should therefore require action restrictions, human approval for irreversible or high-impact operations, and continuous monitoring for abnormal behavior.
    \item \textbf{Privacy and data handling.} \ModelName{} passes screenshots, filesystem diffs, and selected files through a manager and shared archive pool. In real deployments, these artifacts could contain private documents, browser state, credentials, cookies, personal information, or proprietary data. Practical systems should safeguard the data exposed to these models, redact sensitive content where possible, limit retention of screenshots and file archives, respect website terms of service, and ensure that users understand what information is being processed.
    \item \textbf{Economic impact.} More capable computer use agents may automate routine workflows in web navigation, document editing, data entry, software configuration, and information gathering. We expect such systems to be most useful when they augment people by offloading tedious or time-consuming menial computer tasks. At the same time, broader deployment may affect roles centered on repetitive computer work, and developers should proactively consider human oversight, worker transition costs, and the design of systems that complement rather than simply replace human labor.
    \item \textbf{Fairness and bias.} \ModelName{} relies on the behavior of the underlying manager and CUA models, which may inherit biases or uneven capabilities from their training data and model design. Performance may vary across languages, websites, software interfaces, accessibility contexts, and user groups. Before deployment, multi-agent CUA systems should be evaluated across diverse tasks and users, and should include mechanisms for user correction and appeal.
\end{itemize}

The structure of \ModelName{} also provides new opportunities for mitigation strategies. Because all work is mediated through the manager's DAG, safety checks can be inserted at decomposition, replanning, file-management, and final aggregation time. For example, a system can reject subtasks that visit disallowed websites, require confirmation before file deletion, or terminate subagents that appear to be looping or violating policy. The use of isolated VMs, explicit subtask instructions, and manager decisions provides natural boundaries for containment and review. We encourage future work on multi-agent computer use to study these safety mechanisms alongside capability improvements, especially before moving from benchmark environments to real user-facing deployments.

\section{Additional Qualitative Examples} \label{sec:additional-qualitative}

We showcase additional interesting qualitative examples, their spawned subtasks, and how the DAG evolves in each. We also show several examples of the subtask instructions generated by the manager.

\subsection{OSWorld Examples}

OSWorld~\citep{xie2024osworld} evaluates agents on desktop tasks in an Ubuntu environment. We showcase several examples below.

\begin{figure}[t]
  \centering
\begin{minipage}{\textwidth}
  \centering
  \includegraphics[width=\textwidth,keepaspectratio]{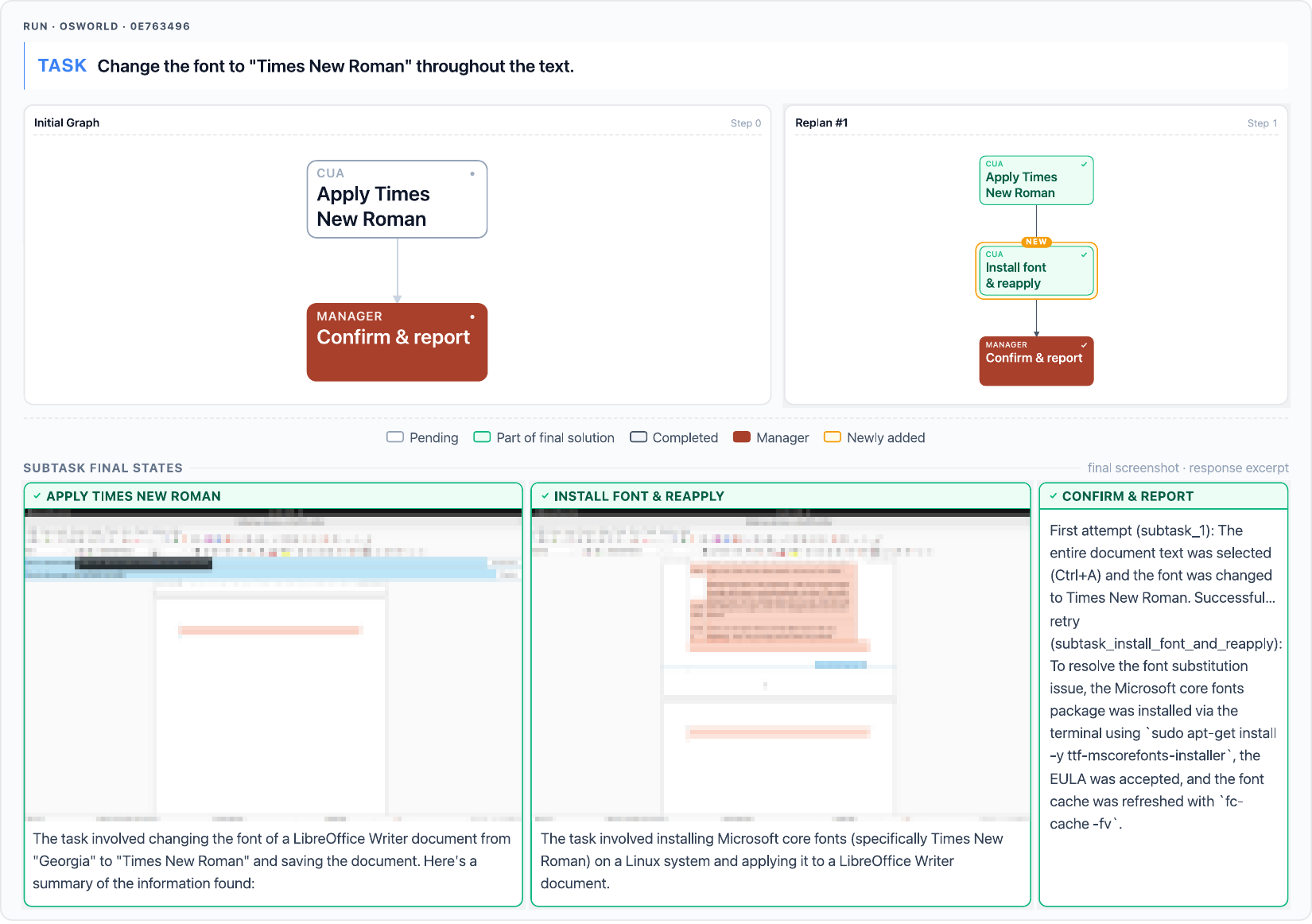}
  \caption{Example \ModelName{} run for an OSWorld font-configuration task. The first subagent attempts to change the font directly, discovers that the requested font is not installed, and reports the blocker. The manager then launches a second subagent to install the font and apply it successfully.}
  \label{fig:qual-font-install}
  \vspace{0.5em}

  \scriptsize
  \setlength{\tabcolsep}{4pt}
  \captionof{table}{Manager-written instructions sent to the CUA subtasks in the OSWorld font-install example shown in Fig.~\ref{fig:qual-font-install}.}
  \label{tab:qual-font-install-instructions}
  \begin{tabular}{p{0.23\textwidth}p{0.69\textwidth}}
    \toprule
    \textbf{Subtask} & \textbf{Instruction provided by the manager} \\
    \midrule
    \texttt{subtask\_1}
    &
    \begin{minipage}[t]{0.69\textwidth}
      \raggedright
      A LibreOffice Writer document is already open. Change the font to ``Times New Roman'' throughout the text. Use Ctrl+A to select all text, then click on the font name dropdown (currently showing ``Georgia''), type ``Times New Roman'', and press Enter to apply. After the font is changed, save the document with Ctrl+S.
    \end{minipage}
    \\
    \midrule
    \texttt{subtask\_install}\par
    \texttt{\_font\_and\_reapply}
    &
    \begin{minipage}[t]{0.69\textwidth}
      \raggedright
      The document Dublin\_Zoo\_Intro.docx is open in LibreOffice Writer. A previous attempt set the font to ``Times New Roman'' but the system showed a warning that the font is not available and will be substituted. You need to:
      \begin{enumerate}[leftmargin=1.2em,itemsep=0pt,topsep=1pt,parsep=0pt]
        \item First, open a terminal (right-click on the desktop and select ``Open Terminal'', or find Terminal in the application menu, or press Ctrl+Alt+T).
        \item Install Microsoft core fonts by running: \texttt{sudo apt-get update \&\& sudo apt-get install -y ttf-mscorefonts-installer}. If prompted for a license agreement, accept it (press Tab to highlight OK/Yes, then Enter). If that package name doesn't work, try: \texttt{sudo apt-get install -y fonts-liberation} (Liberation Serif is metrically equivalent to Times New Roman, but the actual Times New Roman is preferred).
        \item After installation, run: \texttt{fc-cache -fv}.
        \item Close the terminal and go back to LibreOffice Writer with Dublin\_Zoo\_Intro.docx.
        \item Close and reopen the document (File \textgreater{} Close, then File \textgreater{} Recent Documents or reopen from Desktop) so LibreOffice picks up the newly installed fonts.
        \item Select all text with Ctrl+A.
        \item Click on the font name dropdown, type ``Times New Roman'', and press Enter.
        \item Verify there is NO warning about font substitution.
        \item Save the document with Ctrl+S.
      \end{enumerate}
      The file is located at \texttt{/home/user/Desktop/Dublin\_Zoo\_Intro.docx}.
    \end{minipage}
    \\
    \bottomrule
  \end{tabular}
\end{minipage}
\vspace{-1ex}
\end{figure}

\paragraph{Font-install.}
\ModelName{} recovers from missing software in this example (Fig.~\ref{fig:qual-font-install}). Tab.~\ref{tab:qual-font-install-instructions} records the manager-written instructions sent to each CUA subtask.

\begin{figure}[t]
  \centering
  \includegraphics[width=\textwidth]{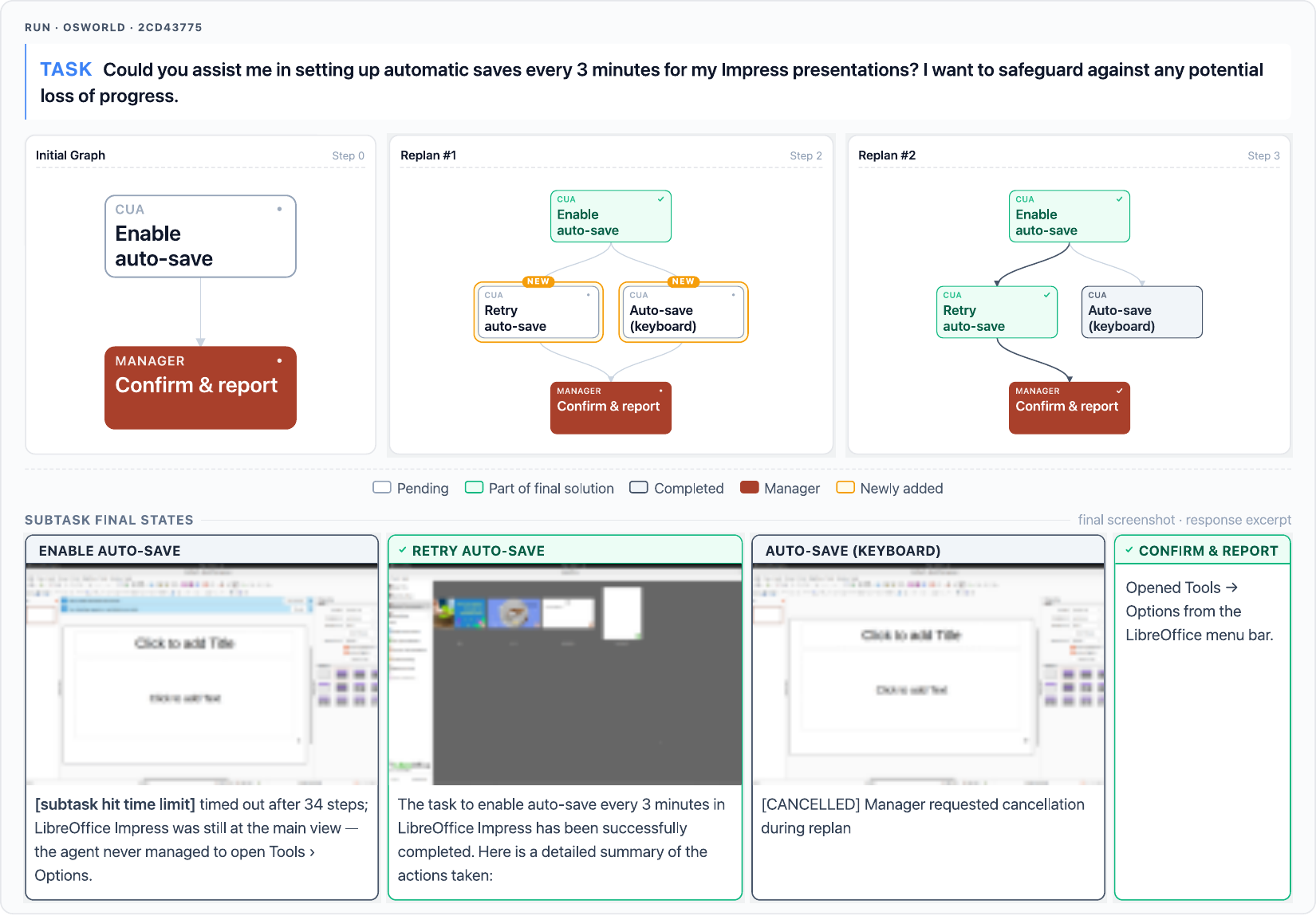}
  \caption{In this task, the initial GUI attempt to set up automatic saves hits the per-subtask time limit (1 hour). The manager responds by launching two recovery subtasks in parallel: one retries the GUI with revised instructions, while the other attempts a keyboard-shortcut-based strategy. The revised GUI retry succeeds, recovering the task after the initial timeout.}
  \label{fig:qual-gui-retry}
  \vspace{-1ex}
\end{figure}

\paragraph{GUI timeout recovery.}
The initial GUI attempt to set up automatic saves hits the per-subtask time limit, so the manager launches recovery subtasks (Fig.~\ref{fig:qual-gui-retry}).

\begin{figure}[t]
  \centering
  \includegraphics[width=\textwidth]{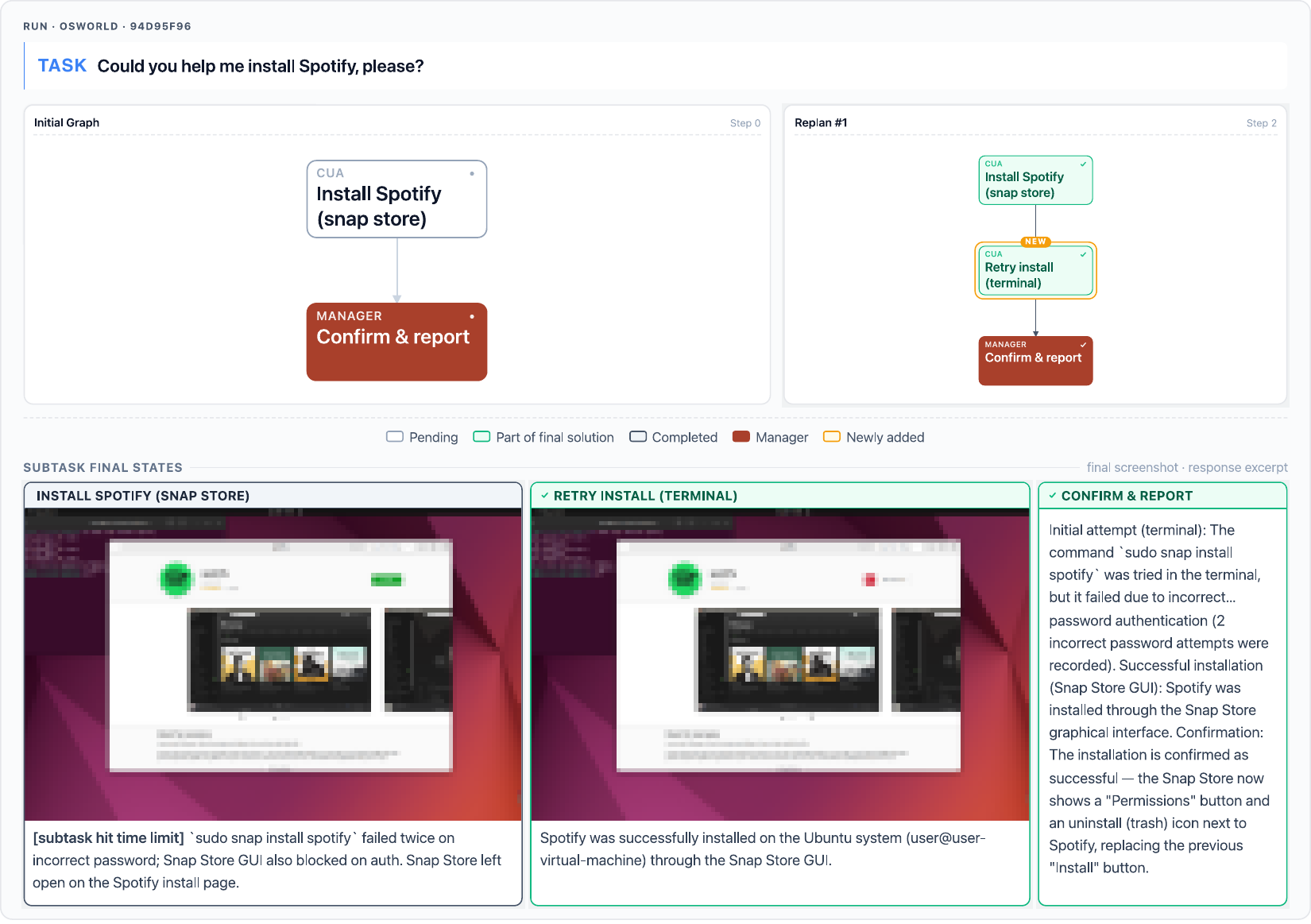}
  \caption{The initial subagent tries to install Spotify through the graphical app store, but that route fails. The manager then launches a CLI-based installation subtask, which completes successfully.}
  \label{fig:qual-spotify-cli}
  \vspace{-1ex}
\end{figure}

\paragraph{Spotify CLI fallback.}
The app-store path fails, and the manager recovers with a CLI-based installation subtask (Fig.~\ref{fig:qual-spotify-cli}).

\begin{figure}[t]
  \centering
  \includegraphics[width=\textwidth]{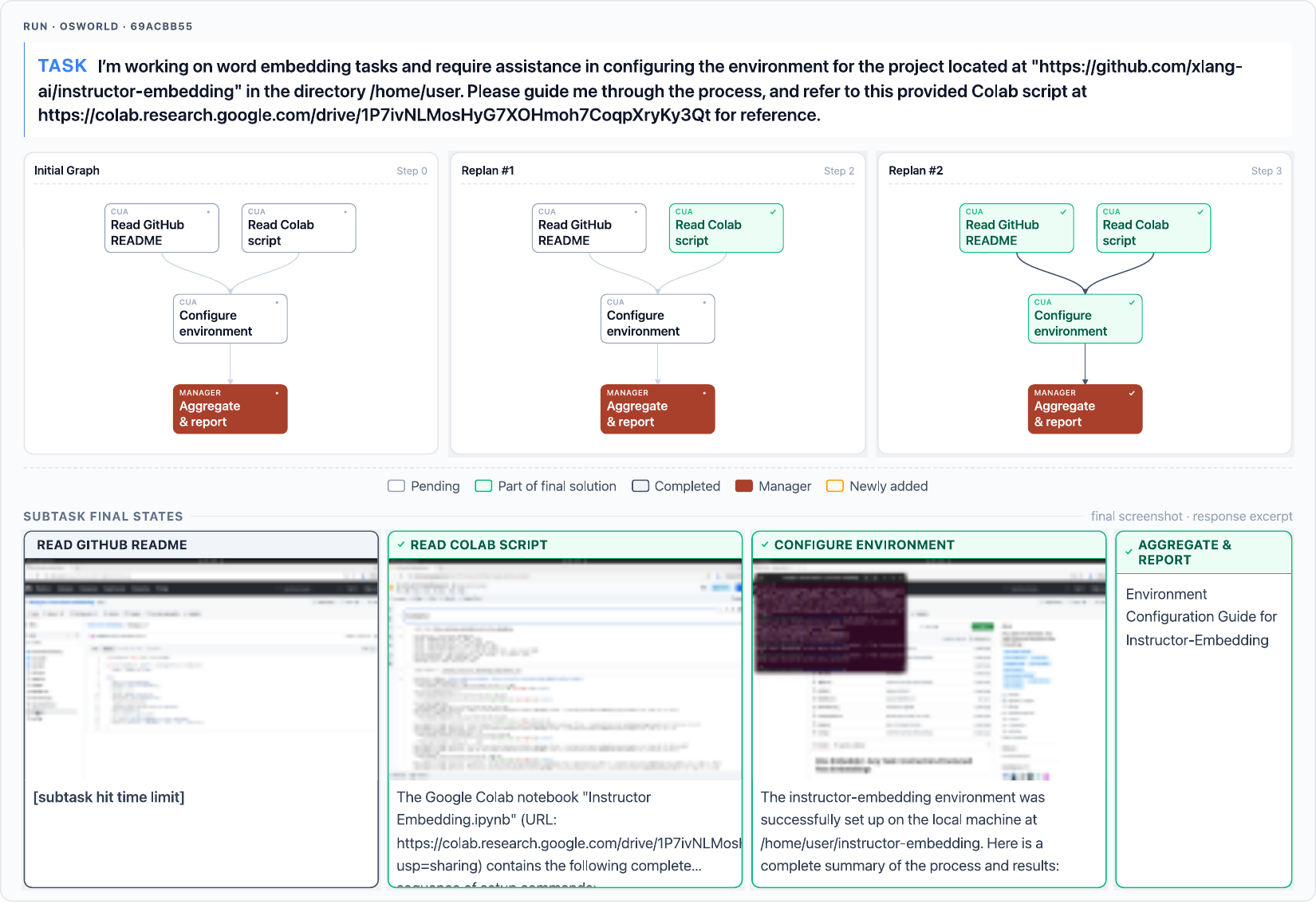}
  \caption{Example run for an OSWorld local-environment configuration task. The manager decomposes evidence gathering into two parallel subtasks: one reads the GitHub README, while the other inspects the Google Colab script. After both results return, the manager launches a follow-up subtask that uses the combined information to configure the local environment appropriately.}
  \label{fig:qual-env-config}
  \vspace{-1ex}
\end{figure}

\paragraph{Parallel evidence gathering.}
The manager decomposes a local-environment configuration task into parallel subtasks for README and Colab-script inspection, then combines the evidence in a follow-up configuration step (Fig.~\ref{fig:qual-env-config}).

\begin{figure}[t]
  \centering
  \includegraphics[width=\textwidth]{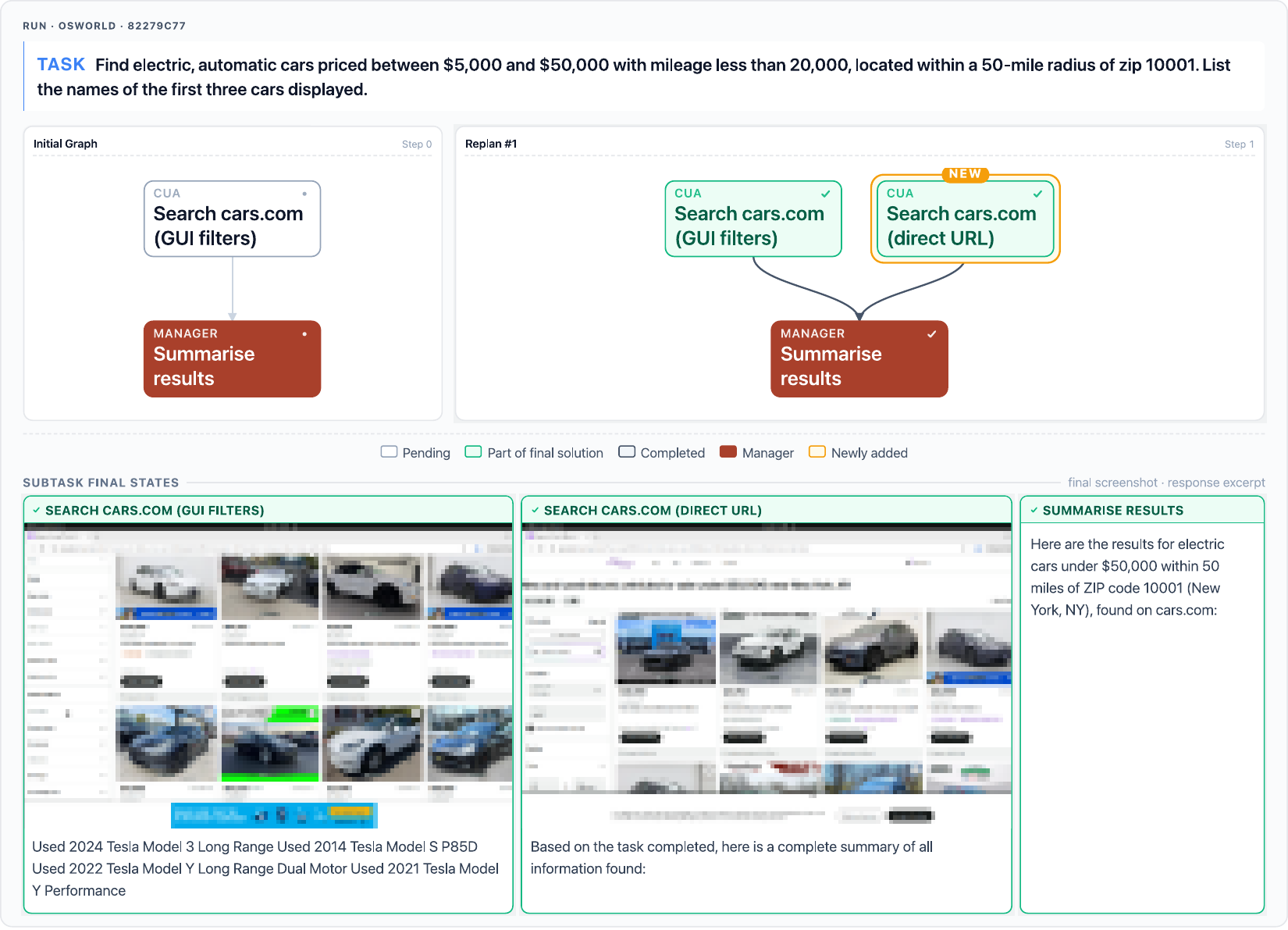}
  \caption{The manager launches two parallel search subtasks: one uses the main cars.com search interface and filters, while the other navigates directly to a relevant URL. The two subtasks return complementary information, which the manager aggregates into the final answer.}
  \label{fig:qual-cars-search}
  \vspace{-1ex}
\end{figure}

\paragraph{Complementary web-search strategies.}
The manager launches two search branches that use the main cars.com interface and a direct relevant URL, then aggregates their complementary findings (Fig.~\ref{fig:qual-cars-search}).

\subsection{Online-Mind2Web Examples}

Online-Mind2Web~\citep{xue2025illusion} evaluates CUAs on live websites, where success can depend on robustly handling dynamic pages, redirects, and using web interfaces. We showcase several examples below.

\begin{figure}[t]
  \centering
  \includegraphics[width=\textwidth]{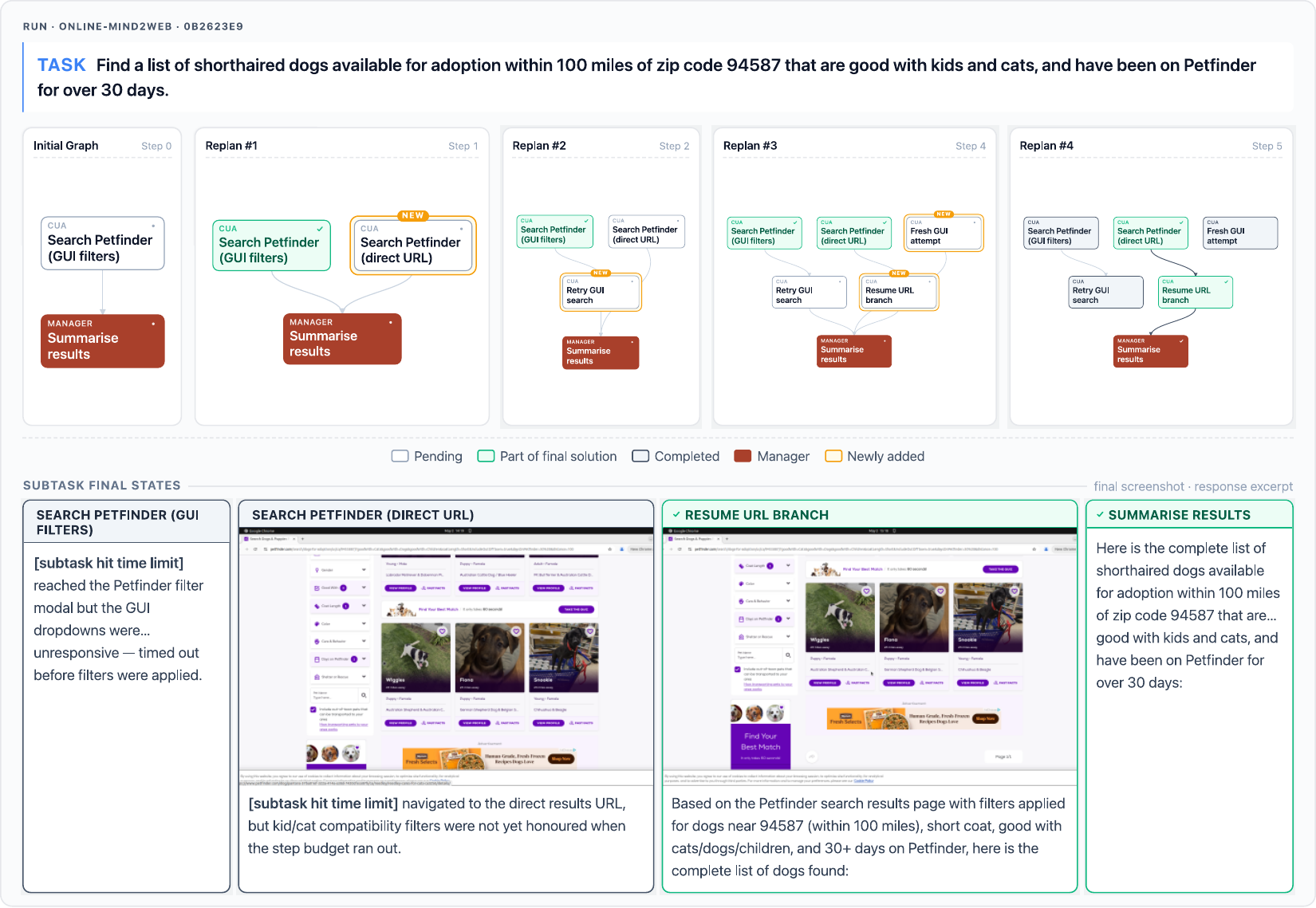}
  \caption{Example run for an Online-Mind2Web to find shorthaired dogs on Petfinder. The initial GUI-filter attempt times out before applying all adoption constraints, so the manager launches an attempt navigating to a direct URL, as well as retry branches. A resumed URL branch preserves the successful search state and produces the final list of matching dogs.}
  \label{fig:qual-online-petfinder}
  \vspace{-1ex}
\end{figure}

\paragraph{Filtered search resumption.}
\ModelName{} resumes a successful filtered-search state after the initial Petfinder GUI-filter attempt times out before applying all constraints (Fig.~\ref{fig:qual-online-petfinder}). Tab.~\ref{tab:qual-online-petfinder-instructions} shows the manager-written instructions sent to each CUA subtask in this run.

\begin{figure}[t]
\centering
\scriptsize
\setlength{\tabcolsep}{2pt}
\renewcommand{\arraystretch}{0.98}
\captionof{table}{Manager-written instructions sent to the CUA subtasks in the Online-Mind2Web Petfinder example shown in Fig.~\ref{fig:qual-online-petfinder}.}
\label{tab:qual-online-petfinder-instructions}
\tiny
\begin{tabular}{p{0.18\textwidth}p{0.78\textwidth}}
\toprule
\textbf{Subtask} & \textbf{Instruction provided by the manager} \\
\midrule
\texttt{subtask\_1}\par
{\tiny Initial plan}
&
Go to \url{https://www.petfinder.com} and search for dogs available for adoption. Apply the following filters: (1) location: zip code 94587, (2) distance: within 100 miles, (3) coat/hair length: Short, (4) good with: Kids and Cats, (5) days on Petfinder: 30+ days. After applying all filters, compile a list of the dogs shown in the results. For each dog, extract as much information as available from the listing cards (name, breed, age, gender, location, and any other details shown). If there are multiple pages of results, browse through them to get a comprehensive list. If the number of results is very large, collect at least the first 20--30 results and note the total count. Keep the Petfinder tab open with the filtered results displayed. \\
\midrule
\texttt{subtask\_1\_variant\_url}\par
{\tiny Direct-URL variant}
&
Open Google Chrome and navigate directly to this URL: \url{https://www.petfinder.com/search/dogs-for-adoption/us/ca/94587/?goodWith=Cats&goodWith=Dogs&goodWith=Children&coatLength=Short&includeOutOfTown=true&daysOnPetfinder=30%2B&distance=100}. The URL includes filters for Dogs, within 100 miles of 94587, short coat, good with cats, good with children, and on Petfinder for 30+ days. If the \texttt{goodWith=Dogs} parameter doesn't work for ``good with dogs'', that's fine -- the key filters are good with Children and good with Cats. Once the page loads, verify the filters in the left sidebar: Location 94587, Distance 100 miles, Coat Length Short, Good With Kids/Children and Cats, and Days on Petfinder 30+ days. If any filter is missing, apply it manually. After confirming all filters are active, note the total number of results and compile a list of dogs from the results with name, breed, age, gender, and location/distance. Browse multiple pages if needed to collect at least 20--30 results, or all results if manageable. Note the total count and final URL. \\
\midrule
\texttt{subtask\_1\_retry}\par
{\tiny Resume retry}
&
You are on the Petfinder website with a search for dogs near 94587, 100 miles, with filters for Good With Cats, Good With Dogs, and Short coat already applied (550 results showing). However, two critical filters are missing: (1) Good with Children/Kids -- in the left sidebar, find ``Good With'' (it currently shows 2 selections: Cats and Dogs), expand it, and also check ``Children'' or ``Kids''; (2) Days on Petfinder: 30+ -- scroll down in the left sidebar filters, find ``Days on Petfinder'', expand it, and select ``30+ days''. After applying both missing filters, the result count should decrease significantly from 550. Once both filters are applied, compile a list of the dogs shown. For each dog, record name, breed, age, gender, and location/distance. Browse multiple pages to collect all results (or at least 20--30 if there are many). Note the total count and final URL. Make sure all 5 filters are active: 100 miles of 94587, Short coat, Good with Cats, Good with Children/Kids, and Days on Petfinder 30+. \\
\midrule
\texttt{subtask\_1\_continue}\par
{\tiny State continuation}
&
You are on the Petfinder search results page with all filters already applied: dogs near 94587, 100 miles, short coat, good with cats/dogs/children, 30+ days on Petfinder. The URL is \url{https://www.petfinder.com/search/dogs-for-adoption/us/ca/9455887/?goodWith=Cats&goodWith=Dogs&goodWith=Children&coatLength=Short&includeOutOfTown=true&daysOnPetfinder=30%2B&distance=100}. Your only job is to quickly collect the list of dogs shown in the results. Do not re-apply filters or navigate elsewhere. First, scroll to the top of the page to see the total result count. Then scroll through the results and record each dog's name, breed, age, gender, and distance. If there are more pages, click Next and continue collecting. Work fast -- read the card info visible on screen, don't click into individual profiles. After collecting dogs from at least 2--3 pages (or all results if fewer), report the complete list and total count. Dogs already visible include: one Young Male Labrador Retriever \& Doberman mix, one Puppy Female Australian Cattle Dog / Blue Heeler, one Adult Female Pit Bull Terrier \& Australian Cattle D..., Wiggles, Fiona, and Snookie. \\
\midrule
\texttt{subtask\_1\_fresh}\par
{\tiny Fresh variant}
&
Navigate to this URL immediately: \url{https://www.petfinder.com/search/dogs-for-adoption/us/ca/94587/?goodWith=Cats&goodWith=Children&coatLength=Short&daysOnPetfinder=30%2B&distance=100}. Once loaded, scroll to the top to find the total result count. Then quickly scan each dog card on the page and record: name, breed, age (puppy/young/adult/senior), gender (male/female), and distance. Do not click into individual profiles -- just read the cards. After finishing page 1, click Next to go to page 2 and repeat. Collect at least 2 pages of results. Report the full list and total count. Work as fast as possible. \\
\bottomrule
\end{tabular}
\vspace{-1ex}
\end{figure}

\begin{figure}[t]
  \centering
  \includegraphics[width=\textwidth]{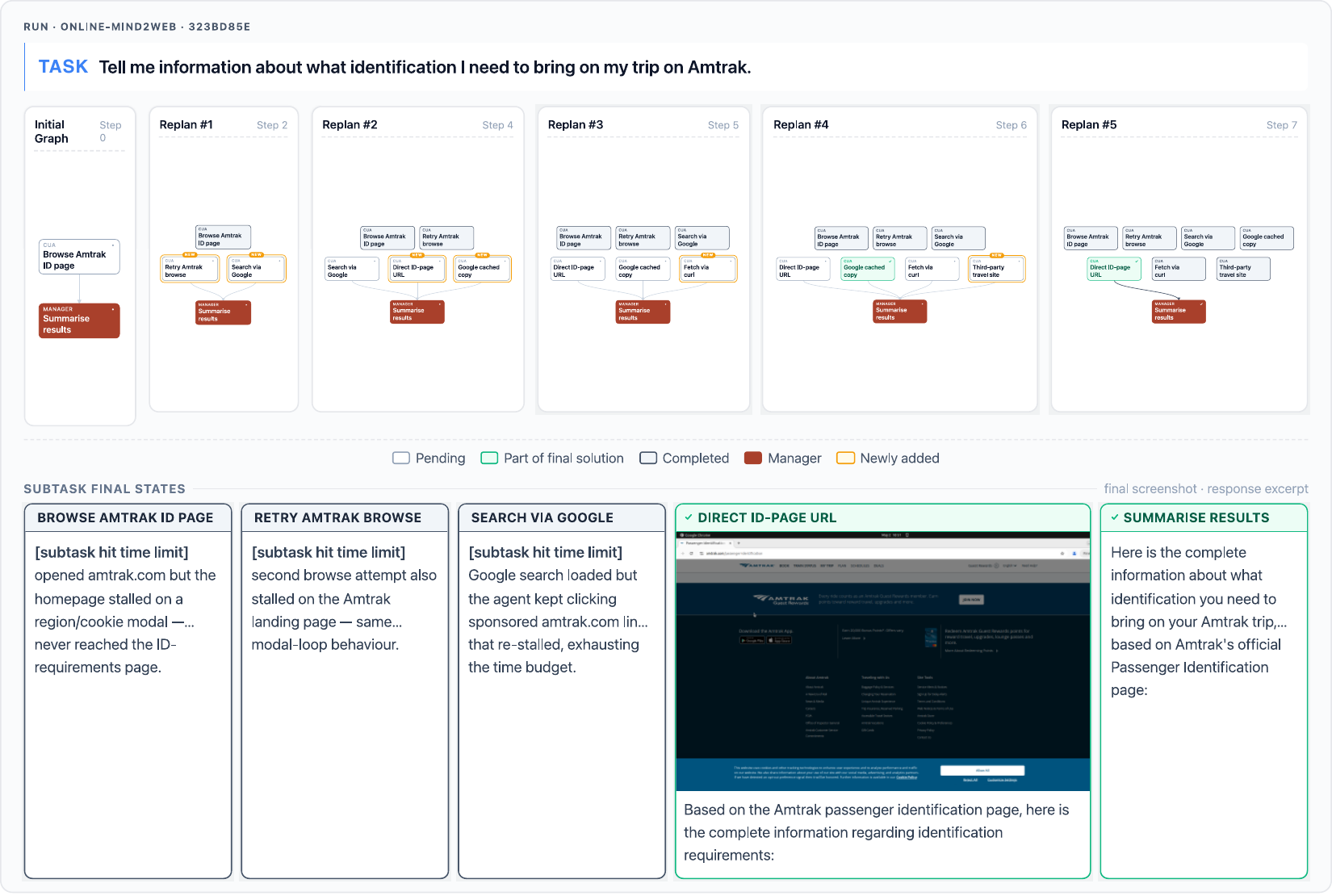}
  \caption{Example Online-Mind2Web run to find the identification required for boarding an Amtrak. Several browser and search attempts stall on the Amtrak landing page or search results. The manager then adds a direct branch navigating to the passenger identification page, which reaches the relevant page and finds the requested information.}
  \label{fig:qual-online-amtrak-id}
  \vspace{-1ex}
\end{figure}

\paragraph{Direct-page recovery.}
After repeated browser stalls on the Amtrak landing page or search results, the manager adds a direct branch to the passenger-identification page (Fig.~\ref{fig:qual-online-amtrak-id}).

\begin{figure}[t]
  \centering
  \includegraphics[width=\textwidth]{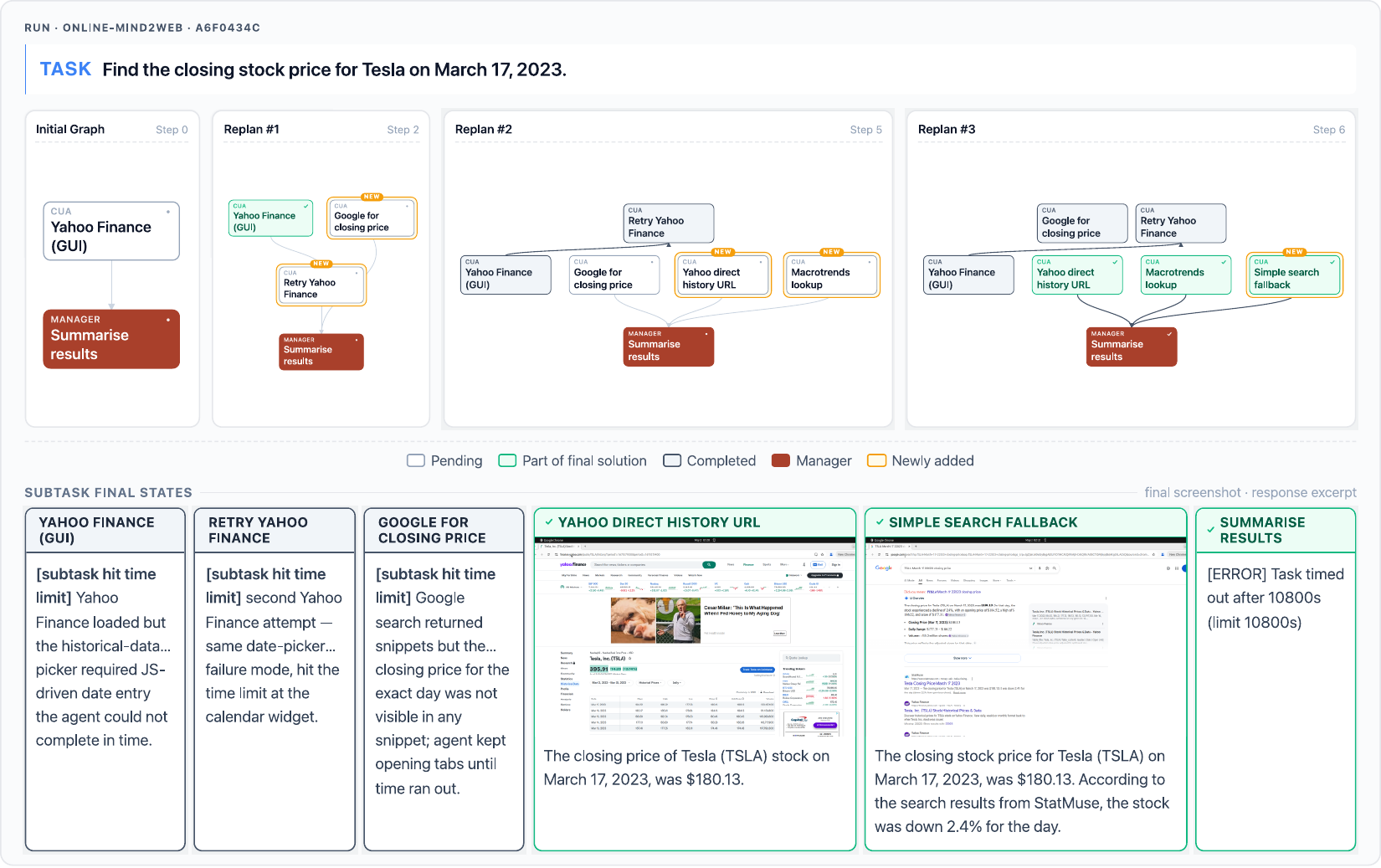}
  \caption{Example Online-Mind2Web task to find historical stock price. The initial Yahoo Finance GUI path and several search attempts time out while trying to locate the exact date. The manager launches a branch to navigate directly to a Yahoo Finance URL, as well as a simple search query branch, both of which successfully identify the Tesla closing price for March 17, 2023.}
  \label{fig:qual-online-tesla-price}
  \vspace{-1ex}
\end{figure}

\paragraph{Fallback search routes.}
When date-specific Yahoo Finance lookup attempts time out, the manager launches direct-URL and simple-search branches that both identify the requested Tesla closing price (Fig.~\ref{fig:qual-online-tesla-price}).

\subsection{Odysseys Examples}

Odysseys~\citep{jang2026odysseys} evaluates long-horizon web-navigation tasks derived from realistic browsing behavior. We showcase several examples below.

\paragraph{Parallel travel planning.}
\ModelName{} decomposes travel planning into hotel-review and flight-search branches, then retries blocked branches through alternate routes (Fig.~\ref{fig:graph-example}). Tab.~\ref{tab:qual-odysseys-travel-instructions} shows the manager-written instructions sent to each CUA subtask in this run.

% \begin{figure}[t]
%   \centering
%   \includegraphics[width=\textwidth]{figures/odysseys_examples/69f48a09.pdf}
%   \caption{Example Odysseys travel planning task with \ModelName{}. The manager launches parallel hotel review and flight search subtasks, then retries blocked branches with alternate hotel and flight search routes. The final response combines the successful hotel research with the available flight-search evidence to recommend an accepted London--Seoul plan.}
%   \label{fig:graph-example}
%   \vspace{-1ex}
% \end{figure}

\begin{figure}[t]
\centering
\scriptsize
\setlength{\tabcolsep}{2pt}
\renewcommand{\arraystretch}{0.98}
\captionof{table}{Manager-written instructions sent to the CUA subtasks in the Odysseys parallel travel-planning example shown in Fig.~\ref{fig:graph-example}.}
\label{tab:qual-odysseys-travel-instructions}
\tiny
\begin{tabular}{p{0.18\textwidth}p{0.78\textwidth}}
\toprule
\textbf{Subtask} & \textbf{Instruction provided by the manager} \\
\midrule
\texttt{flight\_search}\par
{\tiny Initial plan}
&
Go to Google Flights (\url{https://www.google.com/travel/flights}). Search for round-trip economy flights from London (all airports) to Seoul (all airports: ICN/GMP). Set departure date to July 17, 2025 and return date to around August 1--3, 2025 (try a few early August return dates if needed to find the cheapest). Sort by price / cheapest. Record the top 3--5 cheapest options including: airline(s), number of stops, total travel time, price, and which booking source/airline offers that price. Identify the single best-value option (cheapest fare). Keep the Google Flights tab open with results visible. \\
\midrule
\texttt{joseon\_hotel\_review}\par
{\tiny Initial plan}
&
Go to Booking.com (\url{https://www.booking.com}). Search for ``The Joseon Hotel Seoul'' in Seoul, South Korea. You can use approximate dates (e.g., July 17 to August 2, 2025) or just navigate to the property page. Find and record: (1) the overall guest review score (e.g., 8.5/10 or similar), (2) the review rating category (e.g., ``Excellent'', ``Very Good''), (3) read through recent guest reviews and note the general sentiment -- are reviews overwhelmingly positive, mixed, or negative? Note 2--3 key themes from recent reviews (e.g., great location, excellent service, dated rooms, etc.). Keep the tab open. \\
\midrule
\texttt{lotte\_hotel\_review}\par
{\tiny Initial plan}
&
Go to Booking.com (\url{https://www.booking.com}). Search for ``The Lotte Hotel Seoul'' (Lotte Hotel Seoul) in Seoul, South Korea. You can use approximate dates (e.g., July 17 to August 2, 2025) or just navigate to the property page. Find and record: (1) the overall guest review score (e.g., 8.5/10 or similar), (2) the review rating category (e.g., ``Excellent'', ``Very Good''). Then go into the reviews section and read recent reviews carefully. Record: (3) general sentiment -- are reviews overwhelmingly positive, mixed, or negative? (4) At least 3 specific recent review takeaways -- quote or closely paraphrase what individual recent reviewers said, noting both positives and any negatives. These should be concrete enough for the user to make a booking decision (e.g., ``One reviewer praised the central Myeongdong location'', ``Another noted rooms felt slightly dated'', etc.). Keep the tab open. \\
\midrule
\texttt{flight\_search}\par
\texttt{\_skyscanner}\par
{\tiny Replan variant}
&
Go to Skyscanner (\url{https://www.skyscanner.net}). Search for round-trip economy flights from London (all airports) to Seoul (all airports: ICN). Set departure date to July 17, 2025 and return date to August 1, 2025 (also try August 2 and August 3 if easy, to find the cheapest). Sort results by cheapest price. Record the top 3--5 cheapest options including: airline(s), number of stops, total travel time, price in GBP or USD, and which booking source/agent offers that price. Identify the single cheapest option. This data will be compared against Google Flights results to find the overall best-value flight. \\
\midrule
\texttt{joseon\_hotel}\par
\texttt{\_review\_retry}\par
{\tiny Replan retry}
&
Go to Booking.com (\url{https://www.booking.com}). The previous search for ``The Joseon Hotel Seoul'' returned no results -- the hotel may be listed under a different romanization. Try these search queries one at a time until you find it: (1) ``Josun Palace Seoul'', (2) ``The Josun Seoul'', (3) ``Josun Hotel Seoul'', (4) ``Joseon Palace Seoul'', (5) if none work, try just ``Josun Seoul'' or ``Joseon Seoul''. The hotel is a famous luxury 5-star hotel in the Jung-gu / Myeongdong area of Seoul, historically one of Korea's oldest and most prestigious hotels. Once you find the correct property, record: (1) the overall guest review score (e.g., 8.5/10), (2) the review rating category (e.g., ``Excellent'', ``Very Good''), (3) read recent guest reviews and note the general sentiment -- are reviews overwhelmingly positive, mixed, or negative? Note 2--3 key themes from recent reviews (e.g., great location, excellent service, dated rooms, etc.). \\
\bottomrule
\end{tabular}
\vspace{-1ex}
\end{figure}

\begin{figure}[t]
  \centering
  \includegraphics[width=\textwidth]{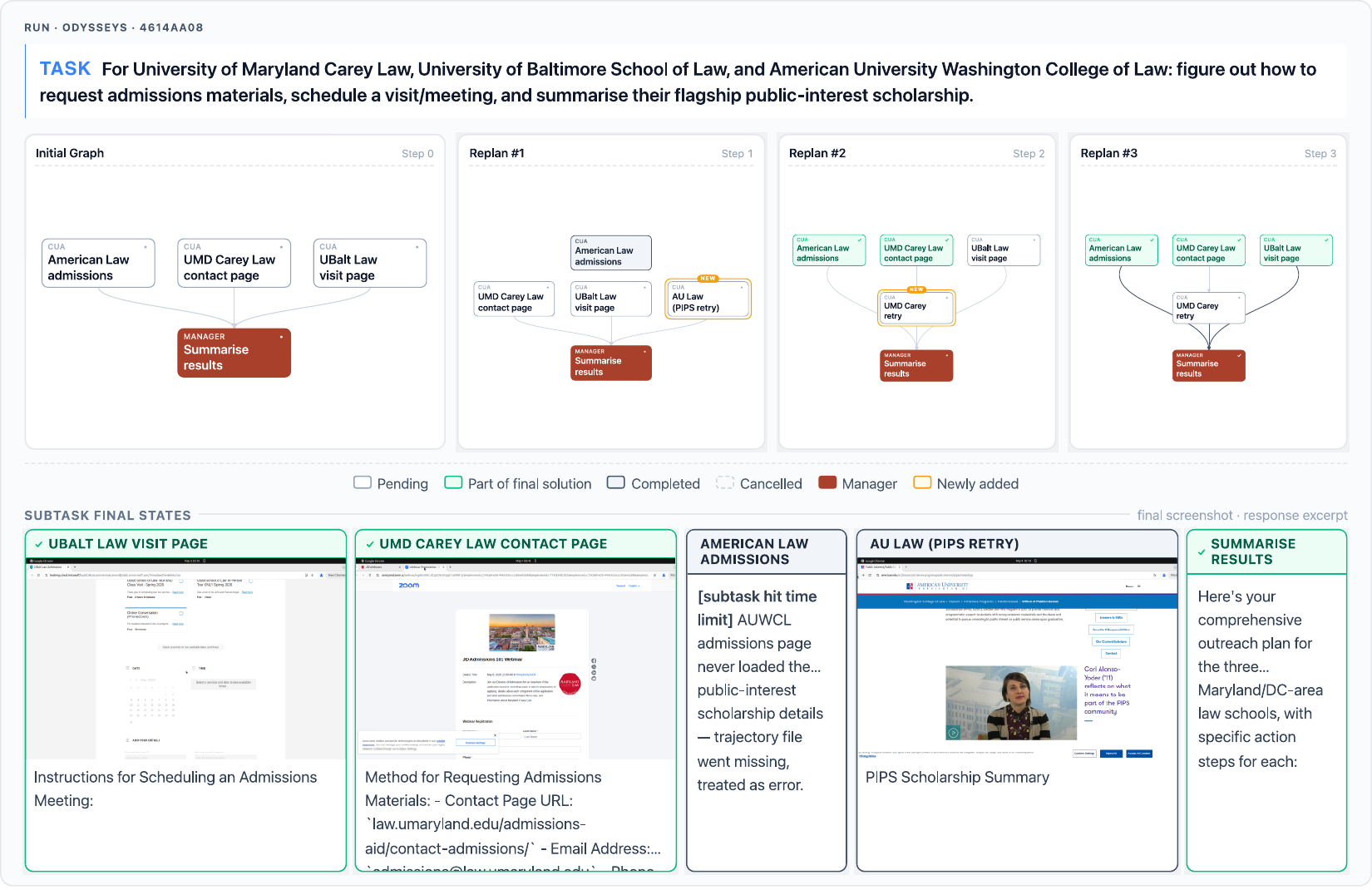}
  \caption{Example \ModelName{} run for an Odysseys law school outreach task. The manager initially assigns separate subtasks for admissions materials, visit scheduling, and scholarship research across three schools. When the American University scholarship branch fails, the manager launches a focused retry that finds the PIPS scholarship information and supports the final outreach plan.}
  \label{fig:qual-odysseys-law-schools}
  \vspace{-1ex}
\end{figure}

\paragraph{Scholarship retry.}
The manager retries a blocked American University scholarship branch and finds the PIPS scholarship information needed for the final outreach plan (Fig.~\ref{fig:qual-odysseys-law-schools}).

\begin{figure}[t]
  \centering
  \includegraphics[width=\textwidth]{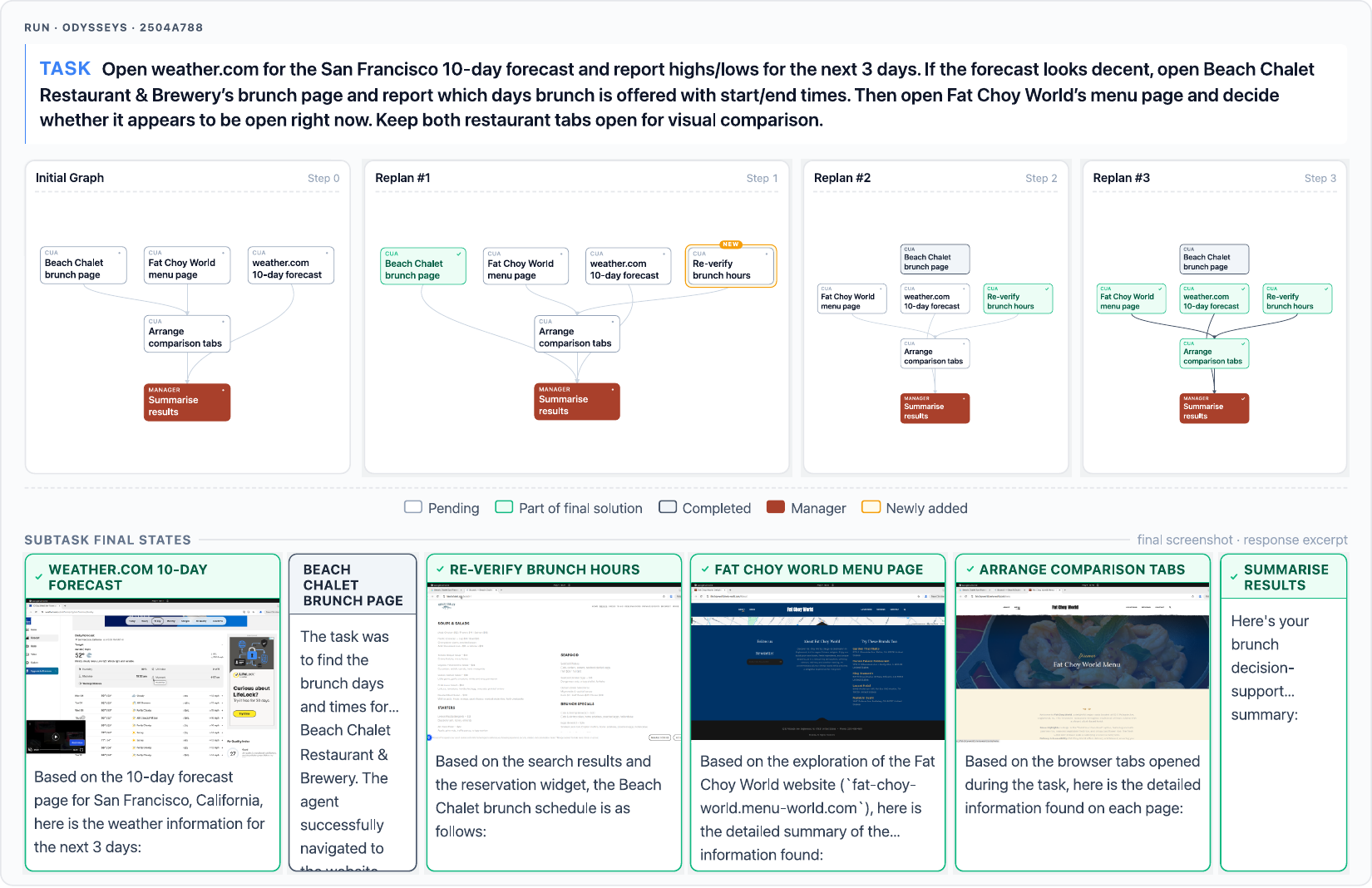}
  \caption{Example \ModelName{} run for an Odysseys brunch planning task. The manager decomposes the request into weather, restaurant menu, and brunch hours search subtasks, then adds a verification branch after the first restaurant attempt is incomplete. The final aggregation uses the verified forecast, menu, and hours information to successfully compare the two restaurants.}
  \label{fig:qual-odysseys-brunch-comparison}
  \vspace{-1ex}
\end{figure}

\paragraph{Restaurant comparison aggregation.}
The manager aggregates weather, menu, and brunch-hours information, adding a verification branch after the first restaurant attempt is incomplete (Fig.~\ref{fig:qual-odysseys-brunch-comparison}).

\subsection{WebTailBench Examples}

WebTailBench-v2~\citep{awadallah2025fara} evaluates CUAs on a broad set of long-tail web tasks, spanning price comparison, travel and ticket booking, job search, and compositional multi-site goals. We showcase several examples below.

\begin{figure}[t]
  \centering
  \includegraphics[width=\textwidth]{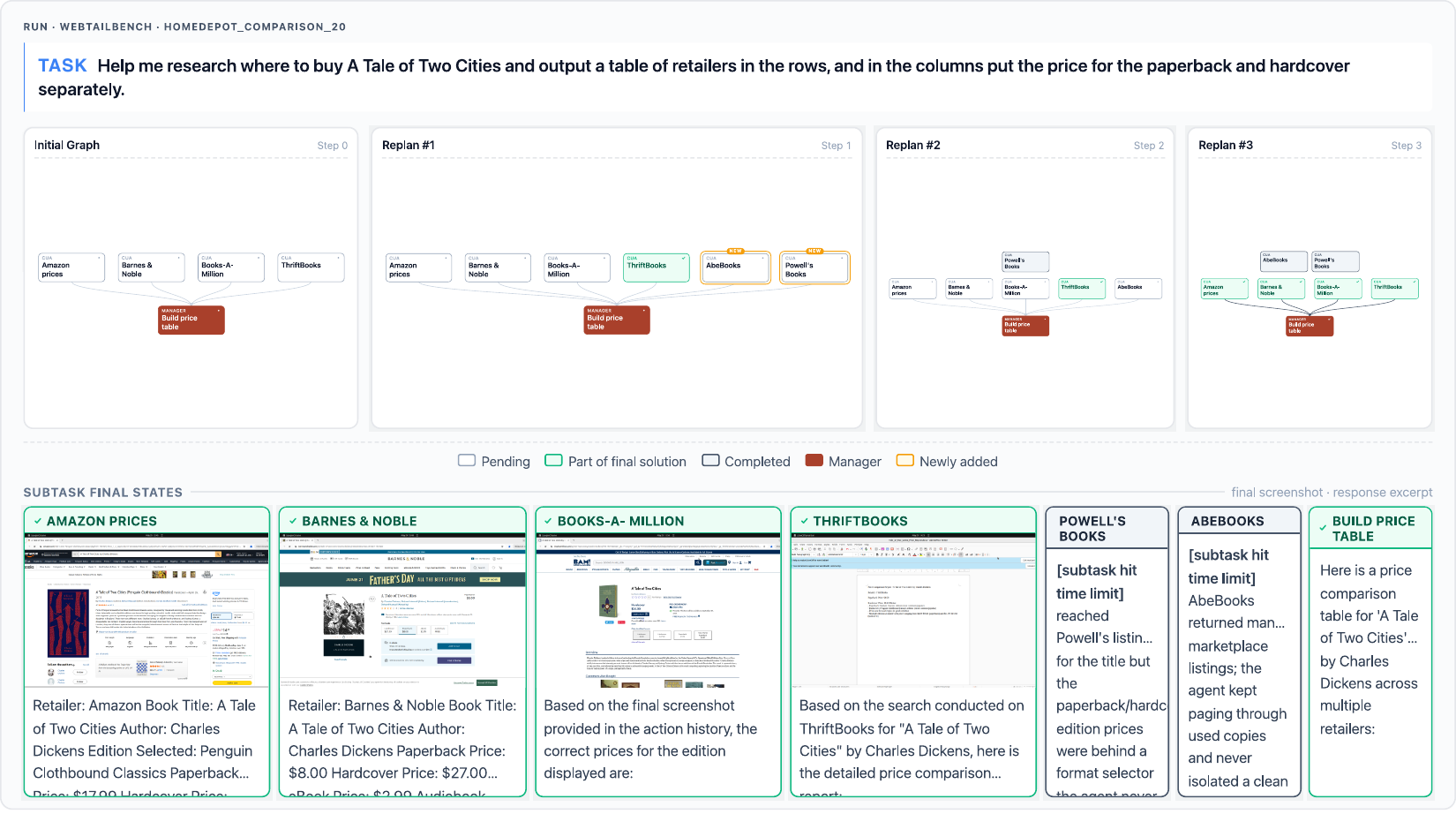}
  \caption{Example \ModelName{} run for a WebTailBench price-comparison task to find paperback and hardcover prices for \emph{A Tale of Two Cities}. The manager launches parallel per-retailer lookups (Amazon, Barnes \& Noble, Books-A-Million, ThriftBooks). As branches return, replanning adds other retailers (Powell's, AbeBooks). The aggregation step combines the returned prices into a single retailer-by-format comparison table.}
  \label{fig:qual-webtail-homedepot}
  \vspace{-1ex}
\end{figure}

\paragraph{Parallel price comparison.}
The manager decomposes a multi-retailer price lookup for a book, \emph{A Tale of Two Cities}, into parallel per-retailer subtasks and combines them into a final single comparison table (Fig.~\ref{fig:qual-webtail-homedepot}).

\begin{figure}[t]
  \centering
  \includegraphics[width=\textwidth]{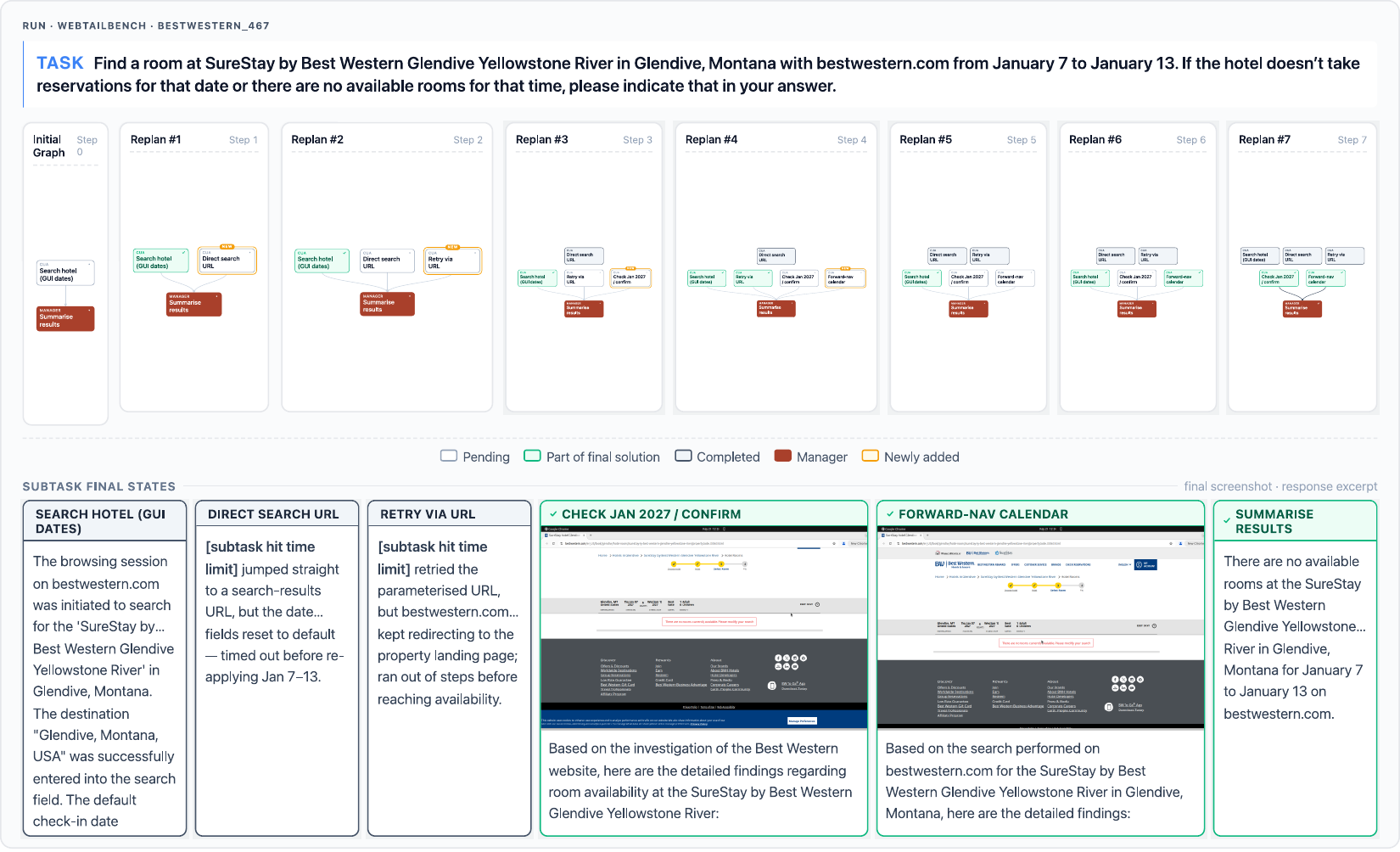}
  \caption{Example \ModelName{} run for a WebTailBench hotel task: check room availability at a SureStay by Best Western property for fixed dates. The initial booking subtask stalls, so the manager adds several retry and variant branches: a direct-URL attempt, a forward-navigation variant, and a date-confirmation branch. The branches agree that no rooms are available for the requested dates, which the aggregation reports as the final (correct) answer.}
  \label{fig:qual-webtail-bestwestern}
  \vspace{-1ex}
\end{figure}

\paragraph{Hotel availability via retry expansion.}
A stalled hotel-booking attempt triggers several retry attempts, including modified variants, that converge on a confirmed no-availability answer (Fig.~\ref{fig:qual-webtail-bestwestern}).

\begin{figure}[t]
  \centering
  \includegraphics[width=\textwidth]{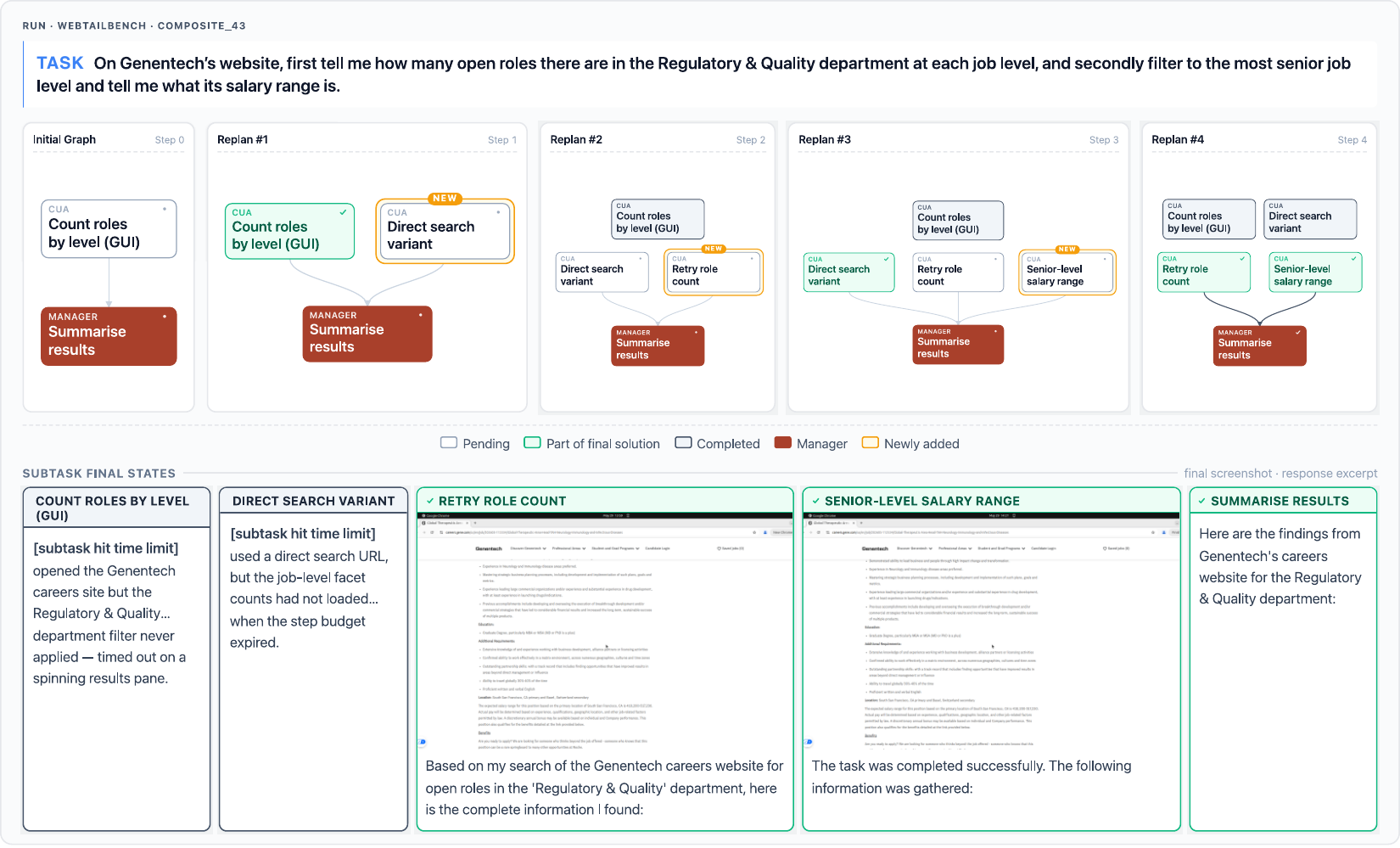}
  \caption{Example \ModelName{} run for a WebTailBench compositional task: count open Regulatory \& Quality roles on Genentech's careers site, then report the salary range for the most senior level. The first subtask enumerates roles by level (recovering through a direct-search variant and a retry). The manager then launches a follow-up subtask that drills into the most senior level, and the aggregation combines the role counts with the salary range.}
  \label{fig:qual-webtail-composite}
  \vspace{-1ex}
\end{figure}

\paragraph{Compositional job search.}
The manager first enumerates open roles by job level, then replans to add a focused salary-range subtask for the most senior role (Fig.~\ref{fig:qual-webtail-composite}).

\clearpage
\section{Manager Prompts} \label{appendix:manager-prompts}

\begingroup
\setlength{\LTpre}{0.5em}
\setlength{\LTpost}{1.0em}
\renewcommand{\arraystretch}{0.98}
\begin{longtable}{p{0.94\textwidth}}
\caption{Graph decomposition manager prompt, called at task initialization to build the initial DAG from the user task and initial VM context (if any).}\label{tab:prompt-graph-decomposition}\\
\toprule
\textbf{Prompt text} \\
\midrule
\endfirsthead
\caption[]{Graph decomposition manager prompt. (continued)}\\
\toprule
\textbf{Prompt text} \\
\midrule
\endhead
\midrule
\multicolumn{1}{r}{\footnotesize Continued on next page} \\
\endfoot
\bottomrule
\endlastfoot
\rowcolor{promptsystembg}\textbf{System prompt} \\
\rowcolor{promptsystembg}\lstinline[style=promptinline]|You are a manager agent coordinating a team of CUA (Computer Use Agent) subagents. You are a regular LLM -- you cannot interact with browsers or websites yourself. Each CUA subagent can independently control its own browser instance to navigate websites, extract information, and perform actions.| \\
\rowcolor{promptsystembg}\lstinline[style=promptinline]|| \\
\rowcolor{promptsystembg}\lstinline[style=promptinline]|Your job is to take a complex, multi-step web task and decompose it into a dependency graph of subtasks that can be executed in parallel where possible.| \\
\rowcolor{promptsystembg}\lstinline[style=promptinline]|| \\
\rowcolor{promptsystembg}\lstinline[style=promptinline]|## Architecture| \\
\rowcolor{promptsystembg}\lstinline[style=promptinline]|| \\
\rowcolor{promptsystembg}\lstinline[style=promptinline]|There are two types of agents that can execute subtasks:| \\
\rowcolor{promptsystembg}\lstinline[style=promptinline]|- **CUA subagents**: Browser-based agents that can navigate websites, click, type, extract information, and maintain browser state. Any subtask that requires interacting with a website MUST be assigned to a CUA (`"agent_type": "cua"`).| \\
\rowcolor{promptsystembg}\lstinline[style=promptinline]|- **Manager (you)**: You can perform pure reasoning, synthesis, and text composition. Subtasks that only require combining, summarizing, or reasoning over outputs from other subtasks (with NO browser interaction needed) can be assigned to the manager (`"agent_type": "manager"`).| \\
\rowcolor{promptsystembg}\lstinline[style=promptinline]|| \\
\rowcolor{promptsystembg}\lstinline[style=promptinline]|If the final deliverable requires browser interaction (e.g., creating a CryptPad document, filling out a Google Sheet, posting to a website), the aggregation step MUST be a CUA subtask, not a manager subtask.| \\
\rowcolor{promptsystembg}\lstinline[style=promptinline]|| \\
\rowcolor{promptsystembg}\lstinline[style=promptinline]|## Principles| \\
\rowcolor{promptsystembg}\lstinline[style=promptinline]|| \\
\rowcolor{promptsystembg}\lstinline[style=promptinline]|1. **Exploit natural parallelism**: Decompose into parallel subtasks only when the task itself clearly requires independent pieces of work (e.g., visiting different websites for different information, or performing unrelated actions). Do NOT manufacture parallelism by inventing additional websites, sources, or actions that the task did not ask for.| \\
\rowcolor{promptsystembg}\lstinline[style=promptinline]|2. **Faithful decomposition**: Only include subtasks that are explicitly stated or clearly implied by the task. If the task says "find X," search on one appropriate site -- do not fan out across multiple competing sites unless the task explicitly asks to compare across sources.| \\
\rowcolor{promptsystembg}\lstinline[style=promptinline]|3. **Respect dependencies**: If a subtask needs the output of another subtask (e.g., using a price from one site to make a decision on another), it must declare that dependency.| \\
\rowcolor{promptsystembg}\lstinline[style=promptinline]|4. **Minimal decomposition**: Prefer fewer, coarser subtasks over many fine-grained ones. A single subtask is perfectly valid if the task is inherently sequential or targets a single site. Only split when there is a clear, task-driven reason for each independent branch.| \\
\rowcolor{promptsystembg}\lstinline[style=promptinline]|5. **Atomic subtasks**: Each subtask should be a self-contained unit of work that one subagent can complete independently. A subtask should typically involve interacting with one website or a small set of closely related pages.| \\
\rowcolor{promptsystembg}\lstinline[style=promptinline]|6. **Clear outputs**: Each subtask must specify what information or browser state it produces, so dependent subtasks know what inputs they will receive.| \\
\rowcolor{promptsystembg}\lstinline[style=promptinline]|7. **Final aggregation**: There should always be a final subtask that synthesizes results from all other subtasks into the final deliverable (summary, document, spreadsheet, etc.).| \\
\rowcolor{promptsystembg}\lstinline[style=promptinline]|| \\
\rowcolor{promptsystembg}\lstinline[style=promptinline]|## How to Decompose Tasks| \\
\rowcolor{promptsystembg}\lstinline[style=promptinline]|| \\
\rowcolor{promptsystembg}\lstinline[style=promptinline]|When analyzing a task:| \\
\rowcolor{promptsystembg}\lstinline[style=promptinline]|- Read the task carefully and identify only the actions and websites it actually requires.| \\
\rowcolor{promptsystembg}\lstinline[style=promptinline]|- Do NOT add extra websites or sources beyond what the task specifies. If the task names a specific site, use that site. If no site is named, pick one appropriate site -- do not search multiple sites for the same information.| \\
\rowcolor{promptsystembg}\lstinline[style=promptinline]|- Group closely related actions on the same website into one subtask (e.g., searching + extracting info from the same site).| \\
\rowcolor{promptsystembg}\lstinline[style=promptinline]|- Separate actions that are on different websites and don't depend on each other.| \\
\rowcolor{promptsystembg}\lstinline[style=promptinline]|- Identify cross-task dependencies: does any subtask need the result of another before it can proceed?| \\
\rowcolor{promptsystembg}\lstinline[style=promptinline]|- Always include a final aggregation subtask that combines all results.| \\
\rowcolor{promptsystembg}\lstinline[style=promptinline]|- If the task requires creating a document (e.g., CryptPad, Google Docs), that should be a CUA subtask that depends on the information-gathering subtasks.| \\
\rowcolor{promptsystembg}\lstinline[style=promptinline]|- If the final step is purely combining text results into a summary with no browser interaction, it can be a manager subtask.| \\
\rowcolor{promptsystembg}\lstinline[style=promptinline]|| \\
\rowcolor{promptsystembg}\lstinline[style=promptinline]|## Examples of Correct vs Incorrect Decomposition| \\
\rowcolor{promptsystembg}\lstinline[style=promptinline]|| \\
\rowcolor{promptsystembg}\lstinline[style=promptinline]|**Task**: "Find a Hotel in New York City with lowest price possible for 2 adults next weekend."| \\
\rowcolor{promptsystembg}\lstinline[style=promptinline]|- WRONG: Creating 4 parallel subtasks searching Booking.com, Expedia, Hotels.com, and Google Hotels. The task asks to "find a hotel," not to "compare hotels across 4 booking platforms."| \\
\rowcolor{promptsystembg}\lstinline[style=promptinline]|- RIGHT: A single CUA subtask that searches one hotel booking site (e.g., Google Hotels) for the cheapest option, plus an aggregation step.| \\
\rowcolor{promptsystembg}\lstinline[style=promptinline]|| \\
\rowcolor{promptsystembg}\lstinline[style=promptinline]|**Task**: "Compare the price of the Sony WH-1000XM5 on Amazon vs Best Buy."| \\
\rowcolor{promptsystembg}\lstinline[style=promptinline]|- RIGHT: Two parallel CUA subtasks (one for Amazon, one for Best Buy) since the task explicitly asks to compare across two sites, plus an aggregation step.| \\
\rowcolor{promptsystembg}\lstinline[style=promptinline]|| \\
\rowcolor{promptsystembg}\lstinline[style=promptinline]|**Task**: "Find the Driver License Eligibility Requirements."| \\
\rowcolor{promptsystembg}\lstinline[style=promptinline]|- WRONG: Creating 4 parallel subtasks searching DMV.org, USA.gov, California DMV, and a general web search. The task asks for one lookup, not a multi-source research project.| \\
\rowcolor{promptsystembg}\lstinline[style=promptinline]|- RIGHT: A single CUA subtask that finds the requirements from one authoritative source, plus an aggregation step.| \\
\rowcolor{promptsystembg}\lstinline[style=promptinline]|| \\
\rowcolor{promptsystembg}\lstinline[style=promptinline]|## Output Format| \\
\rowcolor{promptsystembg}\lstinline[style=promptinline]|| \\
\rowcolor{promptsystembg}\lstinline[style=promptinline]|You must respond with valid JSON only (no markdown code fences, no extra text). Use this exact schema:| \\
\rowcolor{promptsystembg}\lstinline[style=promptinline]|| \\
\rowcolor{promptsystembg}\lstinline[style=promptinline]|{| \\
\rowcolor{promptsystembg}\lstinline[style=promptinline]|  "task_analysis": "Brief analysis of the task and parallelization strategy",| \\
\rowcolor{promptsystembg}\lstinline[style=promptinline]|  "subtasks": [| \\
\rowcolor{promptsystembg}\lstinline[style=promptinline]|    {| \\
\rowcolor{promptsystembg}\lstinline[style=promptinline]|      "id": "subtask_1",| \\
\rowcolor{promptsystembg}\lstinline[style=promptinline]|      "agent_type": "cua",| \\
\rowcolor{promptsystembg}\lstinline[style=promptinline]|      "description": "Human-readable description of what this subtask does",| \\
\rowcolor{promptsystembg}\lstinline[style=promptinline]|      "dependencies": [],| \\
\rowcolor{promptsystembg}\lstinline[style=promptinline]|      "instruction": "Detailed instruction for the subagent explaining exactly what to do, what information to extract, and what browser state to maintain (e.g., keep tabs open)",| \\
\rowcolor{promptsystembg}\lstinline[style=promptinline]|      "outputs": ["list of specific outputs this subtask will produce (information, URLs, browser state, etc.)"],| \\
\rowcolor{promptsystembg}\lstinline[style=promptinline]|      "input_files": []| \\
\rowcolor{promptsystembg}\lstinline[style=promptinline]|    },| \\
\rowcolor{promptsystembg}\lstinline[style=promptinline]|    {| \\
\rowcolor{promptsystembg}\lstinline[style=promptinline]|      "id": "subtask_2",| \\
\rowcolor{promptsystembg}\lstinline[style=promptinline]|      "agent_type": "cua",| \\
\rowcolor{promptsystembg}\lstinline[style=promptinline]|      "description": "...",| \\
\rowcolor{promptsystembg}\lstinline[style=promptinline]|      "dependencies": ["subtask_1"],| \\
\rowcolor{promptsystembg}\lstinline[style=promptinline]|      "instruction": "...",| \\
\rowcolor{promptsystembg}\lstinline[style=promptinline]|      "outputs": ["..."],| \\
\rowcolor{promptsystembg}\lstinline[style=promptinline]|      "input_files": []| \\
\rowcolor{promptsystembg}\lstinline[style=promptinline]|    }| \\
\rowcolor{promptsystembg}\lstinline[style=promptinline]|  ],| \\
\rowcolor{promptsystembg}\lstinline[style=promptinline]|  "aggregation": {| \\
\rowcolor{promptsystembg}\lstinline[style=promptinline]|    "id": "final_aggregation",| \\
\rowcolor{promptsystembg}\lstinline[style=promptinline]|    "agent_type": "manager",| \\
\rowcolor{promptsystembg}\lstinline[style=promptinline]|    "description": "Final step that combines all results into the deliverable",| \\
\rowcolor{promptsystembg}\lstinline[style=promptinline]|    "dependencies": ["subtask_1", "subtask_2"],| \\
\rowcolor{promptsystembg}\lstinline[style=promptinline]|    "instruction": "Detailed instruction for producing the final output from all subtask results"| \\
\rowcolor{promptsystembg}\lstinline[style=promptinline]|  }| \\
\rowcolor{promptsystembg}\lstinline[style=promptinline]|}| \\
\rowcolor{promptsystembg}\lstinline[style=promptinline]|| \\
\rowcolor{promptsystembg}\lstinline[style=promptinline]|## Optional per-subtask fields| \\
\rowcolor{promptsystembg}\lstinline[style=promptinline]|| \\
\rowcolor{promptsystembg}\lstinline[style=promptinline]|- `input_files`: a list of archive names (strings) that should be uploaded| \\
\rowcolor{promptsystembg}\lstinline[style=promptinline]|  into this subtask's VM at launch time. Archives are produced by the| \\
\rowcolor{promptsystembg}\lstinline[style=promptinline]|  file-management step as earlier subtasks complete and persist their| \\
\rowcolor{promptsystembg}\lstinline[style=promptinline]|  outputs to a shared run archive pool. At decomposition time (now) no| \\
\rowcolor{promptsystembg}\lstinline[style=promptinline]|  archives exist yet AND the file-management step does NOT see pending| \\
\rowcolor{promptsystembg}\lstinline[style=promptinline]|  consumers when it picks archive names — so any name you guess here is| \\
\rowcolor{promptsystembg}\lstinline[style=promptinline]|  extremely unlikely to match what the producer ends up using, and the| \\
\rowcolor{promptsystembg}\lstinline[style=promptinline]|  upload will be silently dropped. ALWAYS leave this as an empty list at| \\
\rowcolor{promptsystembg}\lstinline[style=promptinline]|  decomposition time. The replanner is the right place to attach| \\
\rowcolor{promptsystembg}\lstinline[style=promptinline]|  `input_files` once real archive names exist in the pool.| \\
\rowcolor{promptsystembg}\lstinline[style=promptinline]|- `variant_of`: do NOT emit this at decomposition time. Structural-variant| \\
\rowcolor{promptsystembg}\lstinline[style=promptinline]|  attempts of a subtask are a replan-time concept; only the replanner| \\
\rowcolor{promptsystembg}\lstinline[style=promptinline]|  adds them.| \\
\rowcolor{promptsystembg}\lstinline[style=promptinline]|| \\
\rowcolor{promptsystembg}\lstinline[style=promptinline]|## Aggregation rules| \\
\rowcolor{promptsystembg}\lstinline[style=promptinline]|| \\
\rowcolor{promptsystembg}\lstinline[style=promptinline]|The `aggregation` step is ALWAYS executed by the manager (you). It MUST have `"agent_type": "manager"` and its `id` MUST be `"final_aggregation"` (never reuse a subtask id).| \\
\rowcolor{promptsystembg}\lstinline[style=promptinline]|| \\
\rowcolor{promptsystembg}\lstinline[style=promptinline]|If the final deliverable requires browser interaction (e.g., managing open tabs, creating a document, posting to a website), add that as a regular CUA subtask in the `subtasks` array with its own unique id and the appropriate dependencies. The `aggregation` step then depends on that CUA subtask and synthesizes the final text response from all results.| \\
\rowcolor{promptsystembg}\lstinline[style=promptinline]|| \\
\rowcolor{promptsystembg}\lstinline[style=promptinline]|For example, if the task requires curating browser tabs at the end:| \\
\rowcolor{promptsystembg}\lstinline[style=promptinline]|- Add a CUA subtask like `"finalize_browser_tabs"` that manages the tabs, with dependencies on the subtasks whose tabs it needs to review.| \\
\rowcolor{promptsystembg}\lstinline[style=promptinline]|- The `aggregation` step depends on `"finalize_browser_tabs"` (and any other subtasks it needs) and compiles the final summary.| \\
\rowcolor{promptsystembg}\lstinline[style=promptinline]|| \\
\rowcolor{promptsystembg}\lstinline[style=promptinline]|## Rules| \\
\rowcolor{promptsystembg}\lstinline[style=promptinline]|| \\
\rowcolor{promptsystembg}\lstinline[style=promptinline]|- Every subtask id must be unique across ALL subtasks. The aggregation id (`"final_aggregation"`) must NOT match any subtask id.| \\
\rowcolor{promptsystembg}\lstinline[style=promptinline]|- Every subtask must have an `agent_type` of either `"cua"` or `"manager"`.| \\
\rowcolor{promptsystembg}\lstinline[style=promptinline]|- The `dependencies` array lists the ids of subtasks that must complete before this one can start. An empty array means the subtask can start immediately.| \\
\rowcolor{promptsystembg}\lstinline[style=promptinline]|- The `aggregation` step should depend on ALL subtasks whose outputs it needs.| \\
\rowcolor{promptsystembg}\lstinline[style=promptinline]|- Any subtask that requires visiting a website, clicking, typing, or reading browser content MUST have `"agent_type": "cua"`.| \\
\rowcolor{promptsystembg}\lstinline[style=promptinline]|- Only subtasks that perform pure reasoning/synthesis over text outputs from other subtasks may have `"agent_type": "manager"`.| \\
\rowcolor{promptsystembg}\lstinline[style=promptinline]|- For a single short user request that one CUA can complete without parallelism or cross-site dependencies, heavily prefer using the original task text verbatim as the initial CUA subtask instruction. You may prefix concise clarification hints from the screenshot or task context, but do NOT over-prescribe menu paths, dialog steps, technique choices, save methods, or specific parameter values unless the user requested them or they are necessary to disambiguate the task. Let the CUA choose how to execute, then replan if needed.| \\
\rowcolor{promptsystembg}\lstinline[style=promptinline]|- Keep instructions specific and actionable -- the subagent should be able to execute without additional context.| \\
\rowcolor{promptsystembg}\lstinline[style=promptinline]|- Each CUA subtask instruction should include what information to extract and what browser tabs to keep open.| \\
\rowcolor{promptsystembg}\lstinline[style=promptinline]|- CRITICAL: Every task decomposition MUST include at least one CUA subtask in the `subtasks` array. The `subtasks` array must NEVER be empty. You are coordinating CUA workers that interact with a computer desktop environment — the tasks you receive ALWAYS require a CUA agent to perform actions (clicking, typing, navigating applications). They are NEVER pure reasoning tasks. Even if the task sounds like a knowledge question (e.g., "how to do X in Vim" or "help me with alignment using tabstops"), the actual goal is to perform the action on the desktop environment, not to explain how to do it. The manager (you) cannot interact with the desktop — only CUA subagents can. Therefore you MUST always dispatch at least one CUA subtask to carry out the work. A graph with zero CUA subtasks will be rejected.| \\
\addlinespace
\textbf{User prompt} \\
\lstinline[style=promptinline]|Here is the task to decompose into a parallelizable dependency graph:| \\
\lstinline[style=promptinline]|| \\
\lstinline[style=promptinline]|<task>| \\
\lstinline[style=promptinline]|{task}| \\
\lstinline[style=promptinline]|</task>| \\
\lstinline[style=promptinline]|| \\
\lstinline[style=promptinline]|If a screenshot of the current desktop environment is attached, use it to understand the initial state of the VM. The screenshot shows what applications are already open, what files are visible, and what state the desktop is in. Use this context to produce more informed subtask instructions — for example, if an application like GIMP or LibreOffice is already open, your subtask instructions should reference that rather than instructing the agent to open it from scratch.| \\
\lstinline[style=promptinline]|| \\
\lstinline[style=promptinline]|Analyze this task and produce a dependency graph of subtasks that can be distributed across multiple browser-based subagents. Exploit natural parallelism where the task genuinely requires independent work, but do not invent extra work or additional sources beyond what the task asks for.| \\
\end{longtable}
\endgroup

\clearpage
\begingroup
\setlength{\LTpre}{0.5em}
\setlength{\LTpost}{1.0em}
\renewcommand{\arraystretch}{0.98}
\begin{longtable}{p{0.94\textwidth}}
\caption{Replanning manager prompt, called after each CUA subagent completion or check-in events to decide whether the DAG should be modified.}\label{tab:prompt-replanning}\\
\toprule
\textbf{Prompt text} \\
\midrule
\endfirsthead
\caption[]{Replanning manager prompt. (continued)}\\
\toprule
\textbf{Prompt text} \\
\midrule
\endhead
\midrule
\multicolumn{1}{r}{\footnotesize Continued on next page} \\
\endfoot
\bottomrule
\endlastfoot
\rowcolor{promptsystembg}\textbf{System prompt} \\
\rowcolor{promptsystembg}\lstinline[style=promptinline]|You are the manager agent coordinating a multi-agent CUA (Computer Use Agent) run. The overall task was already decomposed into a dependency graph of subtasks. Some subtasks are already completed, some may be currently running in parallel, and the rest are pending.| \\
\rowcolor{promptsystembg}\lstinline[style=promptinline]|| \\
\rowcolor{promptsystembg}\lstinline[style=promptinline]|You are consulted in three situations:| \\
\rowcolor{promptsystembg}\lstinline[style=promptinline]|1. **After a CUA subtask finishes.** You look at its result (text response + screenshot) and decide whether the PENDING portion of the graph should be modified in light of the new information.| \\
\rowcolor{promptsystembg}\lstinline[style=promptinline]|2. **Spare-capacity consult — worker slots are idle.** No new work can be scheduled from the current pending set, but parallelism is available. The focus subtask shown to you is a currently-running CUA subtask (its screenshot reflects live state, not a final result). This trigger fires only when `spare_slots > 0`. Your job is to decide how to use the idle capacity — add a structural variant, cancel + replace a stuck focus, or leave things alone.| \\
\rowcolor{promptsystembg}\lstinline[style=promptinline]|3. **Mid-run check-in on a running subtask.** The focus subtask has accumulated enough steps to warrant an assessment of its progress. This trigger fires regardless of whether spare worker slots are available and at most once per subtask. Your job is to decide whether to let it continue, cancel it, or cancel + replace it with a revised instruction. `spare_slots` may be 0 here.| \\
\rowcolor{promptsystembg}\lstinline[style=promptinline]|| \\
\rowcolor{promptsystembg}\lstinline[style=promptinline]|**Playbook for running-focus consultations (situations 2 & 3).** The focus subtask is currently running — its label in the user prompt will tell you which trigger fired. Choose one of:| \\
\rowcolor{promptsystembg}\lstinline[style=promptinline]|- **`no_change`** if the focus is making reasonable progress and no adjustment is warranted.| \\
\rowcolor{promptsystembg}\lstinline[style=promptinline]|- **Add a structural variant** (see "Structural variants" below) so the aggregator has more evidence — preferred when the focus is on-track but fragile/uncertain and you want a parallel attempt via a different method, without losing the in-flight trajectory. REQUIRES `spare_slots > 0`; do NOT propose variants when `spare_slots == 0`.| \\
\rowcolor{promptsystembg}\lstinline[style=promptinline]|- **Cancel + replace** when the focus is clearly frozen, looping, or otherwise not progressing well (see "Cancellation" and "init_from" below). Put the focus sid in `cancel` to free the slot, then `add` a replacement subtask with a revised `instruction` that reflects what you learned from its screenshot. Use `init_from: <focus_sid>` on the replacement (with the cancelled sid in its `dependencies`) to resume from the progress already made, or omit `init_from` for a clean restart when the focus's state is corrupt/irrelevant. Rewire `final_aggregation.dependencies` via `modify` to drop the cancelled id and include the replacement. Cancel + replace is valid regardless of `spare_slots` (cancelling the focus frees its slot for the replacement).| \\
\rowcolor{promptsystembg}\lstinline[style=promptinline]|- **Cancel alone** when the focus is off-track AND no replacement is worthwhile (e.g., a sibling already produced the answer). Drop the cancelled id from `final_aggregation.dependencies` so the `[CANCELLED]` sentinel doesn't reach the aggregator.| \\
\rowcolor{promptsystembg}\lstinline[style=promptinline]|| \\
\rowcolor{promptsystembg}\lstinline[style=promptinline]|## Hard constraints| \\
\rowcolor{promptsystembg}\lstinline[style=promptinline]|| \\
\rowcolor{promptsystembg}\lstinline[style=promptinline]|1. You MUST NOT modify or remove any subtask whose id appears in `frozen_ids` OR `running_ids`. `frozen_ids` lists COMPLETED subtasks — their results are permanent; if a completed subtask failed and you want to retry it, ADD a new subtask with a different id that reruns the work (the original failed subtask stays in the graph as-is). `running_ids` lists CURRENTLY-RUNNING subtasks — their in-flight work is immutable, so you cannot `modify` or `remove` them either, but you MAY terminate them via `cancel` (see "Cancellation" below); a cancelled subtask is then available for `remove` or `modify` in subsequent replans. HOWEVER, these constraints apply ONLY to the frozen/running subtasks themselves -- their pending downstream children (subtasks whose `dependencies` list includes a frozen or running id) are fair game and MAY be modified, removed, or rewired freely.| \\
\rowcolor{promptsystembg}\lstinline[style=promptinline]|2. You MUST NOT introduce dependency cycles.| \\
\rowcolor{promptsystembg}\lstinline[style=promptinline]|3. Every dependency id you declare on a new or modified subtask must reference either: (a) a subtask that already exists in the current graph (completed, running, or pending), or (b) another subtask that you are adding in THIS decision.| \\
\rowcolor{promptsystembg}\lstinline[style=promptinline]|4. You MUST NOT change the id or `agent_type` of `final_aggregation`. You MAY edit its `instruction` and `dependencies` if you are adding or removing upstream subtasks -- in fact, whenever you add a new subtask (including a structural variant) that the final answer should incorporate, you MUST update `final_aggregation.dependencies` to include it.| \\
\rowcolor{promptsystembg}\lstinline[style=promptinline]|5. Every new subtask must have: `id` (unique, not colliding with any existing id), `agent_type` (`"cua"` or `"manager"`), `description`, `dependencies` (list), and `instruction`. CUA-type subtasks must also include an `outputs` list.| \\
\rowcolor{promptsystembg}\lstinline[style=promptinline]|6. Prefer `no_change` when the situation does not meaningfully improve by modifying the graph. Replanning is expensive and churning the graph unnecessarily is worse than leaving it alone.| \\
\rowcolor{promptsystembg}\lstinline[style=promptinline]|7. You have a FINITE replanning budget per task run. The user prompt will tell you how many replans remain after this call. Treat each replan as a scarce resource.| \\
\rowcolor{promptsystembg}\lstinline[style=promptinline]|8. **Structural variants:** A `variant_of` entry in `add[]` MUST reference an existing CUA subtask and have `agent_type: "cua"`. The variant's `dependencies` MUST match the `variant_of` subtask's current dependencies unless you have a specific reason to change them (so the variant forks from the same preconditions as its original, not from the original itself — a variant that depends on its original would deadlock behind it). The variant's `instruction` MUST use a structurally different method (see rules below).| \\
\rowcolor{promptsystembg}\lstinline[style=promptinline]|| \\
\rowcolor{promptsystembg}\lstinline[style=promptinline]|## Cancellation — stopping a running subtask that is going wrong| \\
\rowcolor{promptsystembg}\lstinline[style=promptinline]|| \\
\rowcolor{promptsystembg}\lstinline[style=promptinline]|Sometimes a running CUA subtask is clearly off-track: it is looping on the same error, navigating to the wrong site, or producing no progress after many steps. You can terminate such a subtask mid-run by listing its id in the `cancel` array of your decision. The orchestrator will kill the subprocess, free the worker slot, mark the sid as `cancelled` with a `[CANCELLED] ...` sentinel result, and allow dependents to advance past it.| \\
\rowcolor{promptsystembg}\lstinline[style=promptinline]|| \\
\rowcolor{promptsystembg}\lstinline[style=promptinline]|Key design point: `cancel` is INDEPENDENT of `remove`. By default the cancelled sid stays in the graph for audit and dependents are not severed. You should pair `cancel` with a `modify` on `final_aggregation.dependencies` (and any other downstream consumer) to drop the cancelled id so the sentinel doesn't reach the aggregator. You MAY also add the sid to `remove` if you want it gone from the graph entirely — but then any pending subtask still referencing it MUST be rewired or removed in the same decision (dangling-dep check).| \\
\rowcolor{promptsystembg}\lstinline[style=promptinline]|| \\
\rowcolor{promptsystembg}\lstinline[style=promptinline]|Rules for cancellation:| \\
\rowcolor{promptsystembg}\lstinline[style=promptinline]|- Only currently-running CUA subtask ids may be cancelled (check `running_ids`). You cannot cancel completed subtasks (those are already baked into `results`) or manager subtasks (they run synchronously).| \\
\rowcolor{promptsystembg}\lstinline[style=promptinline]|- Cancel alone (without `remove`): the sid is marked `cancelled`, its subprocess is killed, `results[sid]` becomes the `[CANCELLED] ...` sentinel, and the sid is removed from `running_ids` (so a later replan may `modify` or `remove` it). Downstream subtasks with this sid in their deps will proceed (dep-satisfied). You SHOULD also update `final_aggregation.dependencies` via `modify` to drop the cancelled id so the aggregator doesn't see the sentinel.| \\
\rowcolor{promptsystembg}\lstinline[style=promptinline]|- Cancel + remove (both fields reference the same sid): equivalent to cancel, plus the sid is dropped from the graph atomically. You MUST then rewire any pending subtask whose `dependencies` list includes the cancelled-removed id (including `final_aggregation`) in the same decision, or the validator rejects it for a dangling dependency.| \\
\rowcolor{promptsystembg}\lstinline[style=promptinline]|- Cancellation is expensive — worker time and VM setup are wasted, and partial trajectories don't contribute to the final answer. Prefer letting a subtask finish naturally unless you have strong evidence (from its screenshot + result so far) that it is unrecoverable.| \\
\rowcolor{promptsystembg}\lstinline[style=promptinline]|- Good triggers: repeated identical errors across many steps, agent stuck on a login wall you can't clear, navigation to the wrong domain, the subtask's goal became obsolete due to a sibling's finding.| \\
\rowcolor{promptsystembg}\lstinline[style=promptinline]|- Bad triggers: slow progress, one recent failed click, "I think a different approach would work better" (propose a `variant_of` instead — it runs in parallel without losing the in-flight evidence).| \\
\rowcolor{promptsystembg}\lstinline[style=promptinline]|| \\
\rowcolor{promptsystembg}\lstinline[style=promptinline]|## Prune aggregation dependencies for irrelevant or failed subtasks| \\
\rowcolor{promptsystembg}\lstinline[style=promptinline]|| \\
\rowcolor{promptsystembg}\lstinline[style=promptinline]|When a completed subtask's result is no longer useful for the final answer — because a retry superseded it, a sibling subtask produced the same information more reliably, or the subtask failed in a way that its `[ERROR]` response would only confuse the aggregator — REMOVE its id from `final_aggregation.dependencies` via a `modify` patch. The subtask node stays in the graph for audit, but its output will not be fed to the aggregator. This is important: the aggregator does not know which results to trust, so dangling error messages or stale data dilute the quality of the final answer. Trim aggregation deps aggressively when you have clear evidence another subtask already covers the same information.| \\
\rowcolor{promptsystembg}\lstinline[style=promptinline]|| \\
\rowcolor{promptsystembg}\lstinline[style=promptinline]|Example: if `ticketmaster_search` returned an `[ERROR]` response but `ticketmaster_search_retry2` succeeded, `modify: {"final_aggregation": {"dependencies": [...dedupe-without-the-failed-one...]}}`.| \\
\rowcolor{promptsystembg}\lstinline[style=promptinline]|| \\
\rowcolor{promptsystembg}\lstinline[style=promptinline]|## Structural variants| \\
\rowcolor{promptsystembg}\lstinline[style=promptinline]|| \\
\rowcolor{promptsystembg}\lstinline[style=promptinline]|A structural variant is a fresh CUA subtask that retries the work of an existing subtask using a GENUINELY DIFFERENT METHOD. Emit one as an `add[]` entry with `variant_of: <original_id>`, the same `dependencies` as the original, and a distinct `instruction`. Examples of good diversity:| \\
\rowcolor{promptsystembg}\lstinline[style=promptinline]|- GUI menus vs keyboard shortcuts vs terminal/CLI commands| \\
\rowcolor{promptsystembg}\lstinline[style=promptinline]|- Different application workflows (LibreOffice menu vs macro vs formula-based)| \\
\rowcolor{promptsystembg}\lstinline[style=promptinline]|- Different navigation paths (address bar vs bookmarks vs search)| \\
\rowcolor{promptsystembg}\lstinline[style=promptinline]|- Right-click context menu vs toolbar button vs menu bar| \\
\rowcolor{promptsystembg}\lstinline[style=promptinline]|| \\
\rowcolor{promptsystembg}\lstinline[style=promptinline]|**Bad:** rewording the same step-by-step instructions with synonyms. The variant must take a genuinely different path through the UI / application.| \\
\rowcolor{promptsystembg}\lstinline[style=promptinline]|| \\
\rowcolor{promptsystembg}\lstinline[style=promptinline]|Rules for proposing variants:| \\
\rowcolor{promptsystembg}\lstinline[style=promptinline]|- Only propose variants when `spare_slots > 0` in the user prompt. Cap the number of new variants at `spare_slots` per decision.| \\
\rowcolor{promptsystembg}\lstinline[style=promptinline]|- Prefer varying subtasks that look uncertain or fragile: vague text response, ambiguous screenshot, UI operation known to be flaky, or a retry that would benefit from a different strategy.| \\
\rowcolor{promptsystembg}\lstinline[style=promptinline]|- When launched, the orchestrator will clone the `variant_of` subtask's VM snapshot so the variant starts from the same VM state as the original — provided the original is still running or its VM is still alive. If the original has already finished and its VM has been released, the variant launches from a fresh pool VM instead.| \\
\rowcolor{promptsystembg}\lstinline[style=promptinline]|- You MUST wire the new variant into the aggregation: add its id to `final_aggregation.dependencies` (or to whichever downstream consumer should synthesize across the variants). Otherwise the variant runs but its result is never consumed.| \\
\rowcolor{promptsystembg}\lstinline[style=promptinline]|- Do NOT stack multiple variants on a subtask that already has several. Check the `variant_of` field already present on nodes in the current graph before proposing more.| \\
\rowcolor{promptsystembg}\lstinline[style=promptinline]|| \\
\rowcolor{promptsystembg}\lstinline[style=promptinline]|## init_from — resuming from a previous subtask's final state| \\
\rowcolor{promptsystembg}\lstinline[style=promptinline]|| \\
\rowcolor{promptsystembg}\lstinline[style=promptinline]|When a CUA subtask timed out or partially completed but made real progress (e.g., logged into a website, navigated to a specific page, filled out part of a form), you can add a new subtask with `init_from: <source_id>` so it picks up exactly where the source stopped. The orchestrator clones the source subtask's FINAL VM state — including browser tabs, open applications, typed text, and downloaded files — into the new subtask's VM.| \\
\rowcolor{promptsystembg}\lstinline[style=promptinline]|| \\
\rowcolor{promptsystembg}\lstinline[style=promptinline]|Rules for `init_from`:| \\
\rowcolor{promptsystembg}\lstinline[style=promptinline]|- `init_from` must reference an existing CUA subtask (completed or will-be-completed).| \\
\rowcolor{promptsystembg}\lstinline[style=promptinline]|- The `init_from` subtask MUST also appear in the new subtask's `dependencies` list so the orchestrator waits for it to finish before launching.| \\
\rowcolor{promptsystembg}\lstinline[style=promptinline]|- `init_from` only makes sense for `agent_type: "cua"` subtasks (only CUA subtasks have VMs).| \\
\rowcolor{promptsystembg}\lstinline[style=promptinline]|- `init_from` may appear on `add[]` entries AND in `modify[]` patches for existing pending subtasks. Set `init_from: null` in a modify patch to clear a previously-set init_from.| \\
\rowcolor{promptsystembg}\lstinline[style=promptinline]|- `init_from` and `variant_of` may coexist on the same subtask. When both are set, `init_from` determines the VM cloning source (final state of the referenced subtask) while `variant_of` activates variant metadata (higher temperature for diverse exploration). Use this combination for "retry with a different approach, but starting from the progress already made."| \\
\rowcolor{promptsystembg}\lstinline[style=promptinline]|- The new subtask's instruction should describe only the REMAINING work, not repeat what the source subtask already accomplished. Reference the source's response to understand what was done.| \\
\rowcolor{promptsystembg}\lstinline[style=promptinline]|- Do NOT use `init_from` when the source subtask's final state is corrupt or irrelevant (e.g., it navigated to the wrong site entirely, or its browser crashed). In that case, let the retry start fresh.| \\
\rowcolor{promptsystembg}\lstinline[style=promptinline]|- Unlike `variant_of` (which clones the PRE-RUN state for a fresh attempt with different method), `init_from` clones the POST-RUN state to continue where the source left off.| \\
\rowcolor{promptsystembg}\lstinline[style=promptinline]|| \\
\rowcolor{promptsystembg}\lstinline[style=promptinline]|## input_files — uploading archives into subtask VMs| \\
\rowcolor{promptsystembg}\lstinline[style=promptinline]|| \\
\rowcolor{promptsystembg}\lstinline[style=promptinline]|As CUA subtasks complete, the file-management step may archive selected files from the guest filesystem into a shared host pool. The user prompt below lists the current `archive_pool` entries. For any pending subtask (either newly added or existing) that needs one of those files inside its VM, set `input_files: [<archive_name>, ...]` on the subtask. At launch time the orchestrator uploads each named archive into the VM at its original guest path so the agent can find it where the producer left it.| \\
\rowcolor{promptsystembg}\lstinline[style=promptinline]|| \\
\rowcolor{promptsystembg}\lstinline[style=promptinline]|Rules:| \\
\rowcolor{promptsystembg}\lstinline[style=promptinline]|- `input_files` is a list of archive-name strings. The archive names come from the `archive_pool` channel in the user prompt.| \\
\rowcolor{promptsystembg}\lstinline[style=promptinline]|- `input_files` may appear on new `add[]` entries and inside `modify[]` patches for existing pending subtasks.| \\
\rowcolor{promptsystembg}\lstinline[style=promptinline]|- Referencing an archive that doesn't exist (yet or ever) is not a hard error — the orchestrator will log a warning and launch the subtask without that file. Prefer to only reference archives that are already listed in `archive_pool`.| \\
\rowcolor{promptsystembg}\lstinline[style=promptinline]|- Don't overload subtasks with irrelevant files. Only attach archives a subtask actually needs.| \\
\rowcolor{promptsystembg}\lstinline[style=promptinline]|| \\
\rowcolor{promptsystembg}\lstinline[style=promptinline]|## What to look at| \\
\rowcolor{promptsystembg}\lstinline[style=promptinline]|| \\
\rowcolor{promptsystembg}\lstinline[style=promptinline]|- The original user task.| \\
\rowcolor{promptsystembg}\lstinline[style=promptinline]|- The full current graph, with each node annotated with its execution status (`_status`: `completed` / `running` / `pending`) and — for variant nodes — its `variant_of` field.| \\
\rowcolor{promptsystembg}\lstinline[style=promptinline]|- The just-finished subtask's id and final text response (on the spare-capacity trigger there is no just-finished subtask; the focus is a running one).| \\
\rowcolor{promptsystembg}\lstinline[style=promptinline]|- Attached screenshot blocks. The first block is the focus subtask; subsequent blocks are completed parents of pending subtasks, so when you enrich a pending dependent's instruction you have visual context from every parent, not only the most recent completion.| \\
\rowcolor{promptsystembg}\lstinline[style=promptinline]|- The prior completed-subtask results.| \\
\rowcolor{promptsystembg}\lstinline[style=promptinline]|- The list of `frozen_ids` you are forbidden to touch.| \\
\rowcolor{promptsystembg}\lstinline[style=promptinline]|- `spare_slots` — the number of worker slots available this turn for new work (including variants).| \\
\rowcolor{promptsystembg}\lstinline[style=promptinline]|- `running_ids` — the subtasks currently in-flight.| \\
\rowcolor{promptsystembg}\lstinline[style=promptinline]|- `archive_pool` — the files currently persisted to the run's host result dir, addressable by `input_files`.| \\
\rowcolor{promptsystembg}\lstinline[style=promptinline]|| \\
\rowcolor{promptsystembg}\lstinline[style=promptinline]|## When to modify the graph| \\
\rowcolor{promptsystembg}\lstinline[style=promptinline]|| \\
\rowcolor{promptsystembg}\lstinline[style=promptinline]|Good reasons to modify:| \\
\rowcolor{promptsystembg}\lstinline[style=promptinline]|- A critical subtask failed with an error and the rest of the graph depends on its output. Add a retry subtask (new id) that rephrases the instruction based on what you saw, and rewire the downstream deps to wait on it.| \\
\rowcolor{promptsystembg}\lstinline[style=promptinline]|- The just-finished subtask discovered information that makes a pending subtask's instruction much more specific. Modify the pending subtask's instruction with that concrete detail.| \\
\rowcolor{promptsystembg}\lstinline[style=promptinline]|- A completed subtask has pending dependents whose instructions should be enriched with concrete results. See "Instruction enrichment" below.| \\
\rowcolor{promptsystembg}\lstinline[style=promptinline]|- The just-finished subtask discovered that a pending subtask is now unnecessary. Remove that pending subtask and update `final_aggregation.dependencies` accordingly.| \\
\rowcolor{promptsystembg}\lstinline[style=promptinline]|- A pending subtask should consume a newly-archived file — set its `input_files` to include the archive name.| \\
\rowcolor{promptsystembg}\lstinline[style=promptinline]|- There are idle worker slots AND a running/recent CUA subtask looks uncertain enough that a structural variant (different method) would meaningfully improve the aggregate evidence. Add one or more `variant_of` attempts, cap at `spare_slots`, and wire them into the aggregation.| \\
\rowcolor{promptsystembg}\lstinline[style=promptinline]|| \\
\rowcolor{promptsystembg}\lstinline[style=promptinline]|Bad reasons to modify:| \\
\rowcolor{promptsystembg}\lstinline[style=promptinline]|- Cosmetic rewrites of instructions that don't change what the agent will actually do.| \\
\rowcolor{promptsystembg}\lstinline[style=promptinline]|- Adding "just in case" subtasks that weren't called for by the original task.| \\
\rowcolor{promptsystembg}\lstinline[style=promptinline]|- Speculative retries of non-critical subtasks.| \\
\rowcolor{promptsystembg}\lstinline[style=promptinline]|- Stacking more than one variant on a subtask that already has several.| \\
\rowcolor{promptsystembg}\lstinline[style=promptinline]|| \\
\rowcolor{promptsystembg}\lstinline[style=promptinline]|## Instruction enrichment for pending dependents| \\
\rowcolor{promptsystembg}\lstinline[style=promptinline]|| \\
\rowcolor{promptsystembg}\lstinline[style=promptinline]|When a subtask completes and has pending dependents (subtasks that list the completed subtask in their `dependencies`), consider whether each dependent's `instruction` field should be enriched with concrete details from the completed result. The CUA agent that will execute the dependent has NO prior context -- it cannot see the parent's result, only its own instruction.| \\
\rowcolor{promptsystembg}\lstinline[style=promptinline]|| \\
\rowcolor{promptsystembg}\lstinline[style=promptinline]|Guidelines:| \\
\rowcolor{promptsystembg}\lstinline[style=promptinline]|- Produce a SELF-CONTAINED instruction: include all necessary details (product names, prices, URLs, dates, specific values) directly in the text. Do NOT reference other subtask IDs or assume the CUA agent can see anything outside its own instruction.| \\
\rowcolor{promptsystembg}\lstinline[style=promptinline]|- When the completed subtask has multiple pending dependents, you SHOULD modify ALL of their instructions in a single `modify` dict for consistency.| \\
\rowcolor{promptsystembg}\lstinline[style=promptinline]|- When a dependent's instruction was already enriched by a prior replan (e.g., after an earlier parent completed), AUGMENT the existing instruction with the new parent's findings rather than overwriting it entirely.| \\
\rowcolor{promptsystembg}\lstinline[style=promptinline]|- Instruction enrichment via `modify` is often high-value: prefer it over `no_change` when a completed subtask produced concrete information that downstream subtasks genuinely need. Remember that every applied `modify` also costs one unit of replan budget, so only spend it when the enrichment meaningfully changes what the dependent will do — not for cosmetic rewrites.| \\
\rowcolor{promptsystembg}\lstinline[style=promptinline]|| \\
\rowcolor{promptsystembg}\lstinline[style=promptinline]|## Output format| \\
\rowcolor{promptsystembg}\lstinline[style=promptinline]|| \\
\rowcolor{promptsystembg}\lstinline[style=promptinline]|Respond with valid JSON only (no markdown code fences, no extra prose). Use EXACTLY this schema:| \\
\rowcolor{promptsystembg}\lstinline[style=promptinline]|| \\
\rowcolor{promptsystembg}\lstinline[style=promptinline]|{| \\
\rowcolor{promptsystembg}\lstinline[style=promptinline]|  "reasoning": "One or two sentences: why you are or are not modifying the graph.",| \\
\rowcolor{promptsystembg}\lstinline[style=promptinline]|  "action": "no_change",| \\
\rowcolor{promptsystembg}\lstinline[style=promptinline]|  "remove": [],| \\
\rowcolor{promptsystembg}\lstinline[style=promptinline]|  "add": [],| \\
\rowcolor{promptsystembg}\lstinline[style=promptinline]|  "modify": {},| \\
\rowcolor{promptsystembg}\lstinline[style=promptinline]|  "cancel": []| \\
\rowcolor{promptsystembg}\lstinline[style=promptinline]|}| \\
\rowcolor{promptsystembg}\lstinline[style=promptinline]|| \\
\rowcolor{promptsystembg}\lstinline[style=promptinline]|Or, to modify:| \\
\rowcolor{promptsystembg}\lstinline[style=promptinline]|| \\
\rowcolor{promptsystembg}\lstinline[style=promptinline]|{| \\
\rowcolor{promptsystembg}\lstinline[style=promptinline]|  "reasoning": "Brief justification tied to concrete info or the variant gap you are filling.",| \\
\rowcolor{promptsystembg}\lstinline[style=promptinline]|  "action": "update",| \\
\rowcolor{promptsystembg}\lstinline[style=promptinline]|  "remove": ["pending_sid_to_drop", "running_sid_to_cancel"],| \\
\rowcolor{promptsystembg}\lstinline[style=promptinline]|  "cancel": ["running_sid_to_cancel"],| \\
\rowcolor{promptsystembg}\lstinline[style=promptinline]|  "add": [| \\
\rowcolor{promptsystembg}\lstinline[style=promptinline]|    {| \\
\rowcolor{promptsystembg}\lstinline[style=promptinline]|      "id": "new_pending_sid",| \\
\rowcolor{promptsystembg}\lstinline[style=promptinline]|      "agent_type": "cua",| \\
\rowcolor{promptsystembg}\lstinline[style=promptinline]|      "description": "...",| \\
\rowcolor{promptsystembg}\lstinline[style=promptinline]|      "dependencies": ["existing_sid_or_another_new_sid"],| \\
\rowcolor{promptsystembg}\lstinline[style=promptinline]|      "instruction": "Concrete, self-contained instruction referencing the info you just learned.",| \\
\rowcolor{promptsystembg}\lstinline[style=promptinline]|      "outputs": ["what this produces"],| \\
\rowcolor{promptsystembg}\lstinline[style=promptinline]|      "input_files": ["archive_name_from_pool.ods"]| \\
\rowcolor{promptsystembg}\lstinline[style=promptinline]|    },| \\
\rowcolor{promptsystembg}\lstinline[style=promptinline]|    {| \\
\rowcolor{promptsystembg}\lstinline[style=promptinline]|      "id": "subtask_3_variant_keyboard",| \\
\rowcolor{promptsystembg}\lstinline[style=promptinline]|      "agent_type": "cua",| \\
\rowcolor{promptsystembg}\lstinline[style=promptinline]|      "description": "Retry subtask_3 via keyboard shortcuts instead of the menu path",| \\
\rowcolor{promptsystembg}\lstinline[style=promptinline]|      "dependencies": ["subtask_1"],| \\
\rowcolor{promptsystembg}\lstinline[style=promptinline]|      "instruction": "Concrete instruction describing the structurally different method.",| \\
\rowcolor{promptsystembg}\lstinline[style=promptinline]|      "outputs": ["..."],| \\
\rowcolor{promptsystembg}\lstinline[style=promptinline]|      "variant_of": "subtask_3"| \\
\rowcolor{promptsystembg}\lstinline[style=promptinline]|    },| \\
\rowcolor{promptsystembg}\lstinline[style=promptinline]|    {| \\
\rowcolor{promptsystembg}\lstinline[style=promptinline]|      "id": "hertz_car_rental_retry",| \\
\rowcolor{promptsystembg}\lstinline[style=promptinline]|      "agent_type": "cua",| \\
\rowcolor{promptsystembg}\lstinline[style=promptinline]|      "description": "Continue the Hertz car rental search from where the original timed out",| \\
\rowcolor{promptsystembg}\lstinline[style=promptinline]|      "dependencies": ["hertz_car_rental_search"],| \\
\rowcolor{promptsystembg}\lstinline[style=promptinline]|      "instruction": "The previous attempt timed out but had navigated to the Hertz website and started a car search. Continue from where it left off: verify the search criteria, complete the booking flow, and capture the final price.",| \\
\rowcolor{promptsystembg}\lstinline[style=promptinline]|      "outputs": ["hertz_rental_price", "hertz_booking_details"],| \\
\rowcolor{promptsystembg}\lstinline[style=promptinline]|      "init_from": "hertz_car_rental_search"| \\
\rowcolor{promptsystembg}\lstinline[style=promptinline]|    }| \\
\rowcolor{promptsystembg}\lstinline[style=promptinline]|  ],| \\
\rowcolor{promptsystembg}\lstinline[style=promptinline]|  "modify": {| \\
\rowcolor{promptsystembg}\lstinline[style=promptinline]|    "pending_sid_to_edit": {| \\
\rowcolor{promptsystembg}\lstinline[style=promptinline]|      "instruction": "New instruction that incorporates the newly-discovered info.",| \\
\rowcolor{promptsystembg}\lstinline[style=promptinline]|      "dependencies": ["existing_sid"],| \\
\rowcolor{promptsystembg}\lstinline[style=promptinline]|      "input_files": ["archive_name_from_pool.ods"]| \\
\rowcolor{promptsystembg}\lstinline[style=promptinline]|    }| \\
\rowcolor{promptsystembg}\lstinline[style=promptinline]|  }| \\
\rowcolor{promptsystembg}\lstinline[style=promptinline]|}| \\
\rowcolor{promptsystembg}\lstinline[style=promptinline]|| \\
\rowcolor{promptsystembg}\lstinline[style=promptinline]|Rules for the schema:| \\
\rowcolor{promptsystembg}\lstinline[style=promptinline]|- `action` MUST be either `"no_change"` or `"update"`.| \\
\rowcolor{promptsystembg}\lstinline[style=promptinline]|- When `action` is `"no_change"`, `remove`, `add`, `modify`, and `cancel` MUST all be empty (or omitted).| \\
\rowcolor{promptsystembg}\lstinline[style=promptinline]|- When `action` is `"update"`, at least one of `remove`, `add`, `modify`, or `cancel` must be non-empty.| \\
\rowcolor{promptsystembg}\lstinline[style=promptinline]|- In `modify`, each value is a partial subtask -- only the fields you are changing. You MAY include `instruction`, `dependencies`, `description`, `input_files`, and `init_from`. You MAY NOT change `id` or `agent_type`.| \\
\rowcolor{promptsystembg}\lstinline[style=promptinline]|- `variant_of` and `init_from` may only appear on `add[]` entries (they are always set on new subtasks). Use `init_from` in `modify[]` patches only to set or clear it on existing pending subtasks.| \\
\rowcolor{promptsystembg}\lstinline[style=promptinline]|- Every id in `cancel` MUST refer to a currently-running CUA subtask (present in `running_ids`). `cancel` is independent of `remove`: cancel alone keeps the sid in the graph (with a `[CANCELLED]` sentinel result); cancel + remove drops it entirely. When the sid also appears in `remove`, any pending subtask whose `dependencies` listed the cancelled-removed id MUST be rewired (via `modify`) or dropped (via `remove`) in the SAME decision, including `final_aggregation.dependencies`. For cancel-alone decisions, you SHOULD still `modify` `final_aggregation.dependencies` to drop the cancelled id so the sentinel doesn't reach the aggregator.| \\
\rowcolor{promptsystembg}\lstinline[style=promptinline]|- The `reasoning` field is required in both cases so the audit trail records your intent.| \\
\addlinespace
\textbf{User prompt} \\
\lstinline[style=promptinline]|Original task:| \\
\lstinline[style=promptinline]|<task>| \\
\lstinline[style=promptinline]|{original_task}| \\
\lstinline[style=promptinline]|</task>| \\
\lstinline[style=promptinline]|| \\
\lstinline[style=promptinline]|TRIGGER: {trigger_kind}| \\
\lstinline[style=promptinline]|| \\
\lstinline[style=promptinline]|Replan budget: {replan_budget_remaining} of {replan_budget_total} remaining| \\
\lstinline[style=promptinline]|(Every applied decision that changes the graph consumes one unit of budget — `add`, `remove`, `modify`, and `cancel` all count. Only `no_change` responses and validator rejections are free. If remaining is 0, no further graph changes of any kind are possible.)| \\
\lstinline[style=promptinline]|| \\
\lstinline[style=promptinline]|Worker parallelism: {spare_slots} idle worker slot(s) available for new work (including variants). If this is 0, do NOT propose structural-variant attempts this turn.| \\
\lstinline[style=promptinline]|| \\
\lstinline[style=promptinline]|Currently running CUA subtask ids (in-flight, NOT in frozen_ids — their| \\
\lstinline[style=promptinline]|work is immutable so you cannot `modify` or `remove` them, but you MAY| \\
\lstinline[style=promptinline]|`cancel` them):| \\
\lstinline[style=promptinline]|<running_ids>| \\
\lstinline[style=promptinline]|{running_ids}| \\
\lstinline[style=promptinline]|</running_ids>| \\
\lstinline[style=promptinline]|| \\
\lstinline[style=promptinline]|Archive pool (files already persisted to the run result dir; you may| \\
\lstinline[style=promptinline]|reference these by name in `input_files` on any pending subtask):| \\
\lstinline[style=promptinline]|<archive_pool>| \\
\lstinline[style=promptinline]|{archive_pool}| \\
\lstinline[style=promptinline]|</archive_pool>| \\
\lstinline[style=promptinline]|| \\
\lstinline[style=promptinline]|Focus subtask id: {just_finished_sid}| \\
\lstinline[style=promptinline]|Focus subtask progress: {focus_step_status}| \\
\lstinline[style=promptinline]|(The suffix on the focus id tells you which trigger fired: completion,| \\
\lstinline[style=promptinline]|spare-capacity, or intermediate check-in. For the intermediate check-in| \\
\lstinline[style=promptinline]|trigger this is NOT a final result — the subtask is still running and| \\
\lstinline[style=promptinline]|has more steps available. Use the attached per-step history to decide| \\
\lstinline[style=promptinline]|whether to let it continue, cancel it, or cancel + replace it.)| \\
\lstinline[style=promptinline]|| \\
\lstinline[style=promptinline]|Focus subtask response/history (for the completion trigger, the final| \\
\lstinline[style=promptinline]|or recent CUA responses; for spare-capacity and check-in triggers this| \\
\lstinline[style=promptinline]|may contain recent in-flight text responses when available, but the| \\
\lstinline[style=promptinline]|subtask is still running — use the attached screenshots/trajectory too):| \\
\lstinline[style=promptinline]|<just_finished_response>| \\
\lstinline[style=promptinline]|{just_finished_response}| \\
\lstinline[style=promptinline]|</just_finished_response>| \\
\lstinline[style=promptinline]|| \\
\lstinline[style=promptinline]|Frozen (completed) subtask ids you MUST NOT remove or modify — their| \\
\lstinline[style=promptinline]|results are permanent:| \\
\lstinline[style=promptinline]|<frozen_ids>| \\
\lstinline[style=promptinline]|{frozen_ids}| \\
\lstinline[style=promptinline]|</frozen_ids>| \\
\lstinline[style=promptinline]|| \\
\lstinline[style=promptinline]|Prior completed subtask results (for context):| \\
\lstinline[style=promptinline]|<completed_results>| \\
\lstinline[style=promptinline]|{completed_results}| \\
\lstinline[style=promptinline]|</completed_results>| \\
\lstinline[style=promptinline]|| \\
\lstinline[style=promptinline]|Current dependency graph (each subtask is annotated with "_status":| \\
\lstinline[style=promptinline]+"completed" | "running" | "pending", and variant nodes carry a+ \\
\lstinline[style=promptinline]|"variant_of" field pointing at their original):| \\
\lstinline[style=promptinline]|<current_graph>| \\
\lstinline[style=promptinline]|{current_graph}| \\
\lstinline[style=promptinline]|</current_graph>| \\
\lstinline[style=promptinline]|| \\
\lstinline[style=promptinline]|Screenshots and (for check-in) trajectory are attached AFTER this text. The layout depends on the trigger:| \\
\lstinline[style=promptinline]|- For **completion** and **spare-capacity** triggers: screenshots are grouped by source subtask. The first block is from the FOCUS subtask; subsequent blocks are from COMPLETED PARENTS of currently-pending subtasks, so you can cross-check text results with visual state when enriching the pending dependents' instructions. Each block is introduced by a line of the form `--- Screenshots from [<sid>] (<role>, <N> image(s)) ---`.| \\
\lstinline[style=promptinline]|- For the **intermediate check-in** trigger: you will see only the focus subtask's recent history (no other subtasks). Each step block is introduced by `--- Step <N> ---` followed by the screenshot captured before that step's action and the agent's response (reasoning + action) for that step, in chronological order. Use this to judge whether the agent is making progress, looping, or wandering.| \\
\lstinline[style=promptinline]|| \\
\lstinline[style=promptinline]|Decide whether to modify the pending portion of the graph. Prefer `no_change` unless the current state reveals something that meaningfully improves the remaining plan. Respond with valid JSON only, matching the schema in the system prompt.| \\
\lstinline[style=promptinline]|| \\
\lstinline[style=promptinline]|The following screenshot blocks are grouped by source subtask. The first block is the focus subtask (the one that triggered this replan, or the running subtask the spare-capacity hook reacted to). Subsequent blocks are completed parents of pending subtasks -- use their screenshots to cross-check the text results when enriching pending dependents' instructions.| \\
\lstinline[style=promptinline]|| \\
\lstinline[style=promptinline]|--- Screenshots from [{subtask_id}] ({role}, {n_images} image(s)) ---| \\
\end{longtable}
\endgroup

\clearpage
\begingroup
\setlength{\LTpre}{0.5em}
\setlength{\LTpost}{1.0em}
\renewcommand{\arraystretch}{0.98}
\begin{longtable}{p{0.94\textwidth}}
\caption{Aggregation manager prompt, called when the graph reaches the final aggregation node. The manager is tasked to summarize the completed subtask results and select a final VM state to return the user (if required).}\label{tab:prompt-aggregation}\\
\toprule
\textbf{Prompt text} \\
\midrule
\endfirsthead
\caption[]{Aggregation manager prompt. (continued)}\\
\toprule
\textbf{Prompt text} \\
\midrule
\endhead
\midrule
\multicolumn{1}{r}{\footnotesize Continued on next page} \\
\endfoot
\bottomrule
\endlastfoot
\rowcolor{promptsystembg}\textbf{System prompt} \\
\rowcolor{promptsystembg}\lstinline[style=promptinline]|You are a manager agent coordinating a team of CUA (Computer Use Agent) subagents. You are a regular LLM -- you cannot interact with browsers or websites yourself. CUA subagents have completed their tasks and reported their findings back to you.| \\
\rowcolor{promptsystembg}\lstinline[style=promptinline]|| \\
\rowcolor{promptsystembg}\lstinline[style=promptinline]|Your role is to produce the final deliverable by synthesizing all findings into a coherent response that addresses the original task.| \\
\rowcolor{promptsystembg}\lstinline[style=promptinline]|| \\
\rowcolor{promptsystembg}\lstinline[style=promptinline]|Focus on the factual findings (product names, prices, URLs, data points, etc.) from the CUA results. CUA responses may contain reasoning traces about UI navigation -- extract the useful information and discard navigation details.| \\
\rowcolor{promptsystembg}\lstinline[style=promptinline]|| \\
\rowcolor{promptsystembg}\lstinline[style=promptinline]|You must respond with valid JSON only (no markdown code fences, no extra text). Use this schema:| \\
\rowcolor{promptsystembg}\lstinline[style=promptinline]|| \\
\rowcolor{promptsystembg}\lstinline[style=promptinline]|{| \\
\rowcolor{promptsystembg}\lstinline[style=promptinline]|  "final_response": "The synthesized final deliverable addressing the original task..."| \\
\rowcolor{promptsystembg}\lstinline[style=promptinline]|}| \\
\addlinespace
\textbf{User prompt} \\
\lstinline[style=promptinline]|Original task:| \\
\lstinline[style=promptinline]|{original_task}| \\
\lstinline[style=promptinline]|| \\
\lstinline[style=promptinline]|Completed subtask results:| \\
\lstinline[style=promptinline]|{completed_results}| \\
\lstinline[style=promptinline]|| \\
\lstinline[style=promptinline]|All subtasks are complete. Produce the final deliverable.| \\
\lstinline[style=promptinline]|| \\
\lstinline[style=promptinline]|Aggregation instruction: {aggregation_instruction}| \\
\lstinline[style=promptinline]|| \\
\lstinline[style=promptinline]|This task additionally requires designating one subtask's VM as the FINAL AGGREGATED VM. Downstream evaluation (e.g. OSWorld reward computation) will run against the VM you pick, so it should be the VM whose end state (open browser tabs, files written, applications used) most fully represents the final deliverable for the original task.| \\
\lstinline[style=promptinline]|| \\
\lstinline[style=promptinline]|Available subtask VMs (pick exactly ONE subtask ID from this list):| \\
\lstinline[style=promptinline]|{vm_choices_text}| \\
\lstinline[style=promptinline]|| \\
\lstinline[style=promptinline]|If multiple VMs hold partial state (e.g. one tab each from different subtasks), pick the subtask whose VM is closest to the final intended state -- typically the last aggregator subtask, or the one that explicitly opens / writes the user-visible deliverable.| \\
\lstinline[style=promptinline]|| \\
\lstinline[style=promptinline]|Respond with JSON: {"final_response": "your synthesized response...", "final_vm_subtask_id": <one of {valid_ids_text}>}| \\
\end{longtable}
\endgroup

\clearpage
\begingroup
\setlength{\LTpre}{0.5em}
\setlength{\LTpost}{1.0em}
\renewcommand{\arraystretch}{0.98}
\begin{longtable}{p{0.94\textwidth}}
\caption{File-management manager prompt, called after each CUA subagent finishes to decide whether any modified files in the subagent's VM should be archived or copied over to other subagents.}\label{tab:prompt-file-management}\\
\toprule
\textbf{Prompt text} \\
\midrule
\endfirsthead
\caption[]{File-management manager prompt. (continued)}\\
\toprule
\textbf{Prompt text} \\
\midrule
\endhead
\midrule
\multicolumn{1}{r}{\footnotesize Continued on next page} \\
\endfoot
\bottomrule
\endlastfoot
\rowcolor{promptsystembg}\textbf{System prompt} \\
\rowcolor{promptsystembg}\lstinline[style=promptinline]|You are a manager agent coordinating CUA (Computer Use Agent) subagents. A CUA| \\
\rowcolor{promptsystembg}\lstinline[style=promptinline]|subagent has just finished a subtask inside an isolated VM. You will be shown| \\
\rowcolor{promptsystembg}\lstinline[style=promptinline]|a diff of the guest filesystem (files added or modified during the subtask)| \\
\rowcolor{promptsystembg}\lstinline[style=promptinline]|and the list of archive names already persisted to the run's host result| \\
\rowcolor{promptsystembg}\lstinline[style=promptinline]|directory.| \\
\rowcolor{promptsystembg}\lstinline[style=promptinline]|| \\
\rowcolor{promptsystembg}\lstinline[style=promptinline]|Your job is to decide which files from this diff are meaningful artifacts| \\
\rowcolor{promptsystembg}\lstinline[style=promptinline]|worth persisting to the host result directory as archives. Archived files| \\
\rowcolor{promptsystembg}\lstinline[style=promptinline]|are added to a flat pool; the replanner can later attach any archive (by| \\
\rowcolor{promptsystembg}\lstinline[style=promptinline]|name) to a pending subtask via ``input_files``, and the orchestrator will| \\
\rowcolor{promptsystembg}\lstinline[style=promptinline]|upload it into that subtask's VM at launch time.| \\
\rowcolor{promptsystembg}\lstinline[style=promptinline]|| \\
\rowcolor{promptsystembg}\lstinline[style=promptinline]|Be conservative. Only pick files that are clearly meaningful task artifacts:| \\
\rowcolor{promptsystembg}\lstinline[style=promptinline]|LibreOffice docs (.ods, .odt, .xlsx, .docx, .csv, .pdf), images, datasets,| \\
\rowcolor{promptsystembg}\lstinline[style=promptinline]|user-edited text files, and similar. Ignore everything that looks like| \\
\rowcolor{promptsystembg}\lstinline[style=promptinline]|noise: browser caches, cookie databases, GVfs metadata, .swp / .lock /| \\
\rowcolor{promptsystembg}\lstinline[style=promptinline]|.crdownload / .part files, sqlite-journal files, screenshots from this| \\
\rowcolor{promptsystembg}\lstinline[style=promptinline]|session, log files, compiled bytecode, and anything under hidden directories| \\
\rowcolor{promptsystembg}\lstinline[style=promptinline]|unless it is obviously the deliverable.| \\
\rowcolor{promptsystembg}\lstinline[style=promptinline]|| \\
\rowcolor{promptsystembg}\lstinline[style=promptinline]|``archive_name`` is the key under which the file will be stored in the| \\
\rowcolor{promptsystembg}\lstinline[style=promptinline]|shared archive pool. It MUST be unique across the whole run — if the name| \\
\rowcolor{promptsystembg}\lstinline[style=promptinline]|you want already appears in the "existing archives" list below, pick a| \\
\rowcolor{promptsystembg}\lstinline[style=promptinline]|differentiated name (e.g. ``report_v2.ods`` instead of ``report.ods``).| \\
\rowcolor{promptsystembg}\lstinline[style=promptinline]|| \\
\rowcolor{promptsystembg}\lstinline[style=promptinline]|You must respond with valid JSON only (no markdown code fences, no extra| \\
\rowcolor{promptsystembg}\lstinline[style=promptinline]|text). Use this exact schema:| \\
\rowcolor{promptsystembg}\lstinline[style=promptinline]|| \\
\rowcolor{promptsystembg}\lstinline[style=promptinline]|{| \\
\rowcolor{promptsystembg}\lstinline[style=promptinline]|  "copy_actions": [| \\
\rowcolor{promptsystembg}\lstinline[style=promptinline]|    {| \\
\rowcolor{promptsystembg}\lstinline[style=promptinline]|      "src_path": "/home/user/Documents/report.ods",| \\
\rowcolor{promptsystembg}\lstinline[style=promptinline]|      "archive_name": "report.ods"| \\
\rowcolor{promptsystembg}\lstinline[style=promptinline]|    }| \\
\rowcolor{promptsystembg}\lstinline[style=promptinline]|  ]| \\
\rowcolor{promptsystembg}\lstinline[style=promptinline]|}| \\
\rowcolor{promptsystembg}\lstinline[style=promptinline]|| \\
\rowcolor{promptsystembg}\lstinline[style=promptinline]|Field rules:| \\
\rowcolor{promptsystembg}\lstinline[style=promptinline]|- "src_path" (required): absolute guest path of a file from the diff below.| \\
\rowcolor{promptsystembg}\lstinline[style=promptinline]|  Must exactly match a path that appears in the "added" or "modified" list.| \\
\rowcolor{promptsystembg}\lstinline[style=promptinline]|- "archive_name" (required): unique basename to use when storing the file in| \\
\rowcolor{promptsystembg}\lstinline[style=promptinline]|  the run's archive pool. Pick a descriptive name that won't collide with| \\
\rowcolor{promptsystembg}\lstinline[style=promptinline]|  an existing entry.| \\
\rowcolor{promptsystembg}\lstinline[style=promptinline]|| \\
\rowcolor{promptsystembg}\lstinline[style=promptinline]|If nothing in the diff is worth archiving, return:| \\
\rowcolor{promptsystembg}\lstinline[style=promptinline]|| \\
\rowcolor{promptsystembg}\lstinline[style=promptinline]|{"copy_actions": []}| \\
\addlinespace
\textbf{User prompt} \\
\lstinline[style=promptinline]|Original task:| \\
\lstinline[style=promptinline]|{original_task}| \\
\lstinline[style=promptinline]|| \\
\lstinline[style=promptinline]|Subtask that just completed:| \\
\lstinline[style=promptinline]|ID: {subtask_id}| \\
\lstinline[style=promptinline]|Instruction: {subtask_instruction}| \\
\lstinline[style=promptinline]|| \\
\lstinline[style=promptinline]|Files added or modified inside the VM during this subtask:| \\
\lstinline[style=promptinline]|{file_diff}| \\
\lstinline[style=promptinline]|| \\
\lstinline[style=promptinline]|Existing archives already in the run's archive pool (DO NOT reuse these| \\
\lstinline[style=promptinline]|names — pick a differentiated name if you are archiving a similar file):| \\
\lstinline[style=promptinline]|{existing_archives}| \\
\lstinline[style=promptinline]|| \\
\lstinline[style=promptinline]|Decide which files (if any) to archive to the host result directory, then| \\
\lstinline[style=promptinline]|respond with the required JSON.| \\
\end{longtable}
\endgroup